\documentclass[
 reprint,
 superscriptaddress,
preprintnumbers,
 amsmath,amssymb,
 aps,
 prl,
]{revtex4-2}

\usepackage{graphicx}
\usepackage{gensymb}
\usepackage{siunitx}
\usepackage[italic]{hepnames}
\usepackage{verbatim}
\usepackage{hyperref}
\usepackage{upgreek}
\usepackage{tikz}
\usepackage{multirow}

\usepackage[mathlines]{lineno}

\newcommand{\nuewithbar}{\stackrel{\scalebox{0.7}{\textbf{\fontsize{1pt}{0pt}\selectfont(---)}}}{\nu}_{e}}
\newcommand{\nuebar}{\stackrel{\scalebox{0.7}{\textbf{\fontsize{1pt}{0pt}\selectfont---}}}{\nu}_{e}}
\newcommand{\numuwithbar}{\stackrel{\scalebox{0.7}{\textbf{\fontsize{1pt}{0pt}\selectfont(---)}}}{\nu}_{\mu}}
\newcommand{\numubar}{\stackrel{\scalebox{0.7}{\textbf{\fontsize{1pt}{0pt}\selectfont---}}}{\nu}_{\mu}}

\begin{document}

\null\hfill\begin{tabular}[t]{l@{}}
\small{FERMILAB-PUB-25-0130-PPD}
\end{tabular}
    
\title{First measurement of $\nu_e$ + $\bar{\nu}_e$ charged current single charged pion production differential cross sections on argon using the MicroBooNE detector}

\newcommand{\ANL}{Argonne National Laboratory (ANL), Lemont, IL, 60439, USA}
\newcommand{\Bern}{Universit{\"a}t Bern, Bern CH-3012, Switzerland}
\newcommand{\BNL}{Brookhaven National Laboratory (BNL), Upton, NY, 11973, USA}
\newcommand{\UCSB}{University of California, Santa Barbara, CA, 93106, USA}
\newcommand{\Cambridge}{University of Cambridge, Cambridge CB3 0HE, United Kingdom}
\newcommand{\CIEMAT}{Centro de Investigaciones Energ\'{e}ticas, Medioambientales y Tecnol\'{o}gicas (CIEMAT), Madrid E-28040, Spain}
\newcommand{\Chicago}{University of Chicago, Chicago, IL, 60637, USA}
\newcommand{\Cincinnati}{University of Cincinnati, Cincinnati, OH, 45221, USA}
\newcommand{\CSU}{Colorado State University, Fort Collins, CO, 80523, USA}
\newcommand{\Columbia}{Columbia University, New York, NY, 10027, USA}
\newcommand{\Edinburgh}{University of Edinburgh, Edinburgh EH9 3FD, United Kingdom}
\newcommand{\FNAL}{Fermi National Accelerator Laboratory (FNAL), Batavia, IL 60510, USA}
\newcommand{\Granada}{Universidad de Granada, Granada E-18071, Spain}
\newcommand{\IIT}{Illinois Institute of Technology (IIT), Chicago, IL 60616, USA}
\newcommand{\ICL}{Imperial College London, London SW7 2AZ, United Kingdom}
\newcommand{\Indiana}{Indiana University, Bloomington, IN 47405, USA}
\newcommand{\Kansas}{The University of Kansas, Lawrence, KS, 66045, USA}
\newcommand{\KSU}{Kansas State University (KSU), Manhattan, KS, 66506, USA}
\newcommand{\Lancaster}{Lancaster University, Lancaster LA1 4YW, United Kingdom}
\newcommand{\LANL}{Los Alamos National Laboratory (LANL), Los Alamos, NM, 87545, USA}
\newcommand{\Louisiana}{Louisiana State University, Baton Rouge, LA, 70803, USA}
\newcommand{\Manchester}{The University of Manchester, Manchester M13 9PL, United Kingdom}
\newcommand{\MIT}{Massachusetts Institute of Technology (MIT), Cambridge, MA, 02139, USA}
\newcommand{\Michigan}{University of Michigan, Ann Arbor, MI, 48109, USA}
\newcommand{\MSU}{Michigan State University, East Lansing, MI 48824, USA}
\newcommand{\Minnesota}{University of Minnesota, Minneapolis, MN, 55455, USA}
\newcommand{\Nankai}{Nankai University, Nankai District, Tianjin 300071, China}
\newcommand{\NMSU}{New Mexico State University (NMSU), Las Cruces, NM, 88003, USA}
\newcommand{\Oxford}{University of Oxford, Oxford OX1 3RH, United Kingdom}
\newcommand{\Pitt}{University of Pittsburgh, Pittsburgh, PA, 15260, USA}
\newcommand{\QMUL}{Queen Mary University of London, London E1 4NS, United Kingdom}
\newcommand{\Rutgers}{Rutgers University, Piscataway, NJ, 08854, USA}
\newcommand{\SLAC}{SLAC National Accelerator Laboratory, Menlo Park, CA, 94025, USA}
\newcommand{\SDSMT}{South Dakota School of Mines and Technology (SDSMT), Rapid City, SD, 57701, USA}
\newcommand{\Maine}{University of Southern Maine, Portland, ME, 04104, USA}
\newcommand{\TelAviv}{Tel Aviv University, Tel Aviv, Israel, 69978}
\newcommand{\UTA}{University of Texas, Arlington, TX, 76019, USA}
\newcommand{\Tufts}{Tufts University, Medford, MA, 02155, USA}
\newcommand{\VTech}{Center for Neutrino Physics, Virginia Tech, Blacksburg, VA, 24061, USA}
\newcommand{\Warwick}{University of Warwick, Coventry CV4 7AL, United Kingdom}

\affiliation{\ANL}
\affiliation{\Bern}
\affiliation{\BNL}
\affiliation{\UCSB}
\affiliation{\Cambridge}
\affiliation{\CIEMAT}
\affiliation{\Chicago}
\affiliation{\Cincinnati}
\affiliation{\CSU}
\affiliation{\Columbia}
\affiliation{\Edinburgh}
\affiliation{\FNAL}
\affiliation{\Granada}
\affiliation{\IIT}
\affiliation{\ICL}
\affiliation{\Indiana}
\affiliation{\Kansas}
\affiliation{\KSU}
\affiliation{\Lancaster}
\affiliation{\LANL}
\affiliation{\Louisiana}
\affiliation{\Manchester}
\affiliation{\MIT}
\affiliation{\Michigan}
\affiliation{\MSU}
\affiliation{\Minnesota}
\affiliation{\Nankai}
\affiliation{\NMSU}
\affiliation{\Oxford}
\affiliation{\Pitt}
\affiliation{\QMUL}
\affiliation{\Rutgers}
\affiliation{\SLAC}
\affiliation{\SDSMT}
\affiliation{\Maine}
\affiliation{\TelAviv}
\affiliation{\UTA}
\affiliation{\Tufts}
\affiliation{\VTech}
\affiliation{\Warwick}

\author{P.~Abratenko} \affiliation{\Tufts}
\author{D.~Andrade~Aldana} \affiliation{\IIT}
\author{L.~Arellano} \affiliation{\Manchester}
\author{J.~Asaadi} \affiliation{\UTA}
\author{A.~Ashkenazi}\affiliation{\TelAviv}
\author{S.~Balasubramanian}\affiliation{\FNAL}
\author{B.~Baller} \affiliation{\FNAL}
\author{A.~Barnard} \affiliation{\Oxford}
\author{G.~Barr} \affiliation{\Oxford}
\author{D.~Barrow} \affiliation{\Oxford}
% ...

\collaboration{The MicroBooNE Collaboration}
\thanks{microboone\_info@fnal.gov}\noaffiliation

\begin{abstract}
Understanding electron neutrino interactions is crucial for measurements of neutrino oscillations and searches for new physics in neutrino experiments. We present the first measurement of the flux-averaged $\nu_e$ + $\bar{\nu}_e$ charged current single charged pion production cross section on argon using the MicroBooNE detector and data from the NuMI neutrino beam. The total cross section is measured to be (0.93 $\pm$ 0.13 (stat.) $\pm$ 0.27 (syst.))$\times \num{e-39}$ cm$^2$/nucleon at a mean $\nu_e$ + $\bar{\nu}_e$ energy of 730\,MeV. Differential cross sections are also reported in electron energy, electron and pion angles, and electron-pion opening angle.
\end{abstract}

\maketitle

{\bf \textit{Introduction}}.--- The next generation of accelerator neutrino oscillation experiments will seek to address multiple open questions in neutrino physics through precision measurement of electron neutrino appearance in muon neutrino beams. These include the presence and scale of charge-parity violation in the neutrino sector, the neutrino mass ordering, and the resolution of anomalies observed at short baselines~\cite{SajjadAthar:2021prg}. Several of these experiments will make use of the liquid argon time projection chamber (LArTPC) detector technology~\cite{Machado:2019oxb, DUNE:2020jqi}. As such, in order to facilitate these measurements, precise understanding of $\nu_e$ interactions on argon is essential. 

Cross section modeling of $\nu_e$ interactions is typically extrapolated from $\nu_\mu$ measurements. However, uncertainties on the $\nu_e/\nu_\mu$ interaction cross section ratio arising from the different lepton masses and subsequent radiative corrections limit their constraining power~\cite{Day:2012gb, Martini:2016eec, Nikolakopoulos:2019qcr, Tomalak:2021hec}. Understanding these effects through measurements of $\nu_e$ interactions is crucial for precision neutrino oscillation measurements, rare-event searches, and benchmarking theoretical models used in neutrino experiments.
  
Reconstructing and measuring $\nu_e$ interactions also presents unique challenges compared to $\nu_\mu$ interactions due to the electromagnetic cascades produced by final-state electrons. These are complicated to reconstruct and sensitive to detector calibration uncertainties. Measurements of $\nu_e$ interactions allow algorithms targeting these topologies to be evaluated and improved reducing systematic uncertainties and enhancing detector performance. 

Existing measurements of $\nu_e$ cross sections on argon are limited and consist of several inclusive measurements~\cite{ArgoNeuT:2020kir, MicroBooNE:2021gfj, MicroBooNE:2021ppm} and a measurement without final state pions~\cite{MicroBooNE:2022tdd}. The Deep Underground Neutrino Experiment (DUNE) will be exposed to a neutrino flux peaking at energies of a few GeV~\cite{DUNE:2020jqi}. At these energies, one of the dominant neutrino interaction modes leads to the excitation of baryon resonances that subsequently decay producing pions. This process has never previously been measured for $\nu_e$ interactions in argon. 

This work presents the first measurement of the flux-averaged $\nu_e$ + $\nuebar$ charged-current (CC) single charged pion production cross section on argon using MicroBooNE. The final-state topology considered consists of an electron (or positron), a single charged pion, zero neutral pions (or heavier mesons), and any number of protons or neutrons:  
\begin{equation}
\nuewithbar+\mathrm{Ar} \rightarrow e^{\pm} + 1 \pi^{\pm} + 0 \pi^{0} + X,
\end{equation}
where $X$ represents the residual nucleus and any outgoing nucleons. These interactions will subsequently be referred to as $\nuewithbar \mathrm{CC}\,1\pi^\pm$ for simplicity.

The MicroBooNE detector is an 85 tonne LArTPC that collected data between 2015 and 2020. It consists of an instrumented argon volume of ($2.56\times2.32\times10.36$)\,m$^3$ (drift, vertical, beam direction). Ionization charge produced by charged particles resulting from neutrino interactions is drifted towards three planes of readout wires, orientated vertically and at $\pm60\degree$ to the vertical, by an electric field of 273\,V/cm. Additionally, scintillation light is collected by an array of 32 photomultiplier tubes located behind the readout planes~\cite{MicroBooNE:2016pwy}. 

MicroBooNE collected data from two neutrino beams: the on-axis 8\,GeV Booster Neutrino Beam and the approximately $8\degree$ off-axis 120\,GeV Neutrinos at the Main Injector (NuMI) beam~\cite{Adamson:2015dkw}. In this work, data from the NuMI beam operating in a mixture of forward-horn-current (FHC) neutrino mode and reverse-horn-current (RHC) anti-neutrino mode are used. The NuMI flux at MicroBooNE is shown for each horn-current mode in the Supplemental Material~\cite{supplemental}. The flux contains a significant fraction of $\nuewithbar$ ($2.5$\%) due to the high beam energy and the significantly off-axis position of the detector. This makes it particularly useful for $\nuewithbar$ cross-section measurements. The average $\nuewithbar$ energies incident on MicroBooNE are $715$\,MeV in FHC mode and $744$\,MeV in RHC mode. The integrated exposure is \num{8.9e20} protons-on-target (POT) in FHC mode and \num{11.1e20} POT in RHC mode. This corresponds to the full NuMI dataset accumulated by MicroBooNE and used for the first time in this analysis. 

At these neutrino energies, the dominant $\nuewithbar \mathrm{CC}\,1\pi^\pm$ production mechanism is through the $\Delta(1232)$ resonance that subsequently decays to a pion and a nucleon. In addition, there are subdominant contributions from coherent pion production and pion production as a result of deep inelastic scattering. The observed topology is also impacted by final-state interactions that can lead to pion production or absorption as the particles produced in the neutrino interaction leave the argon nucleus. 

Although pion production has not previously been measured in $\nu_e$ interactions on argon, it has been probed in $\nu_\mu$ interactions. ArgoNeuT has performed measurements of $\nu_\mu$ and $\numubar$ induced charged pion production on argon~\cite{ArgoNeuT:2018und}. MicroBooNE has performed several measurements of $\nu_\mu$ induced neutral pion production on argon for both CC~\cite{MicroBooNE:2018neo, MicroBooNE:2024bnl} and neutral-current (NC) interactions~\cite{MicroBooNE:2022zhr, MicroBooNE:2024sec}. This work presents the first measurement of $\nuewithbar \mathrm{CC}\,1\pi^\pm$ on argon. The total interaction cross section is reported along with differential cross sections in electron energy, electron and pion angles, and electron-pion opening angle. These measurements are complementary to the existing suite of $\numuwithbar$ measurements.

{\bf \textit{Simulation and reconstruction}}.--- The NuMI neutrino flux is simulated with \texttt{GEANT4 v4.10.4}~\cite{GEANT4:2002zbu, Allison:2006ve, Allison:2016lfl, flux_public_note} constrained with available hadron production data using the \texttt{PPFX} software package~\cite{MINERvA:2016iqn}. The integrated flux of $\nuewithbar$ with energy over 60 MeV is \num{1.86e-11}/cm$^2$/POT in FHC mode and \num{1.69e-11}/cm$^2$/POT in RHC mode. The combined FHC and RHC integrated flux is calculated by weighting the contributions from each horn-current mode according to the accumulated POT in that mode. The resulting combined integrated $\nuewithbar$ flux is \num{1.77e-11}/cm$^2$/POT.

The \texttt{LArSoft} software framework~\cite{Snider:2017wjd} is used to perform simulation and reconstruction. Neutrino interactions are modeled using the \texttt{GENIE v3.0.6 G18\_10a\_02\_11a} event generator~\cite{Andreopoulos:2009rq} with the MicroBooNE tune applied~\cite{MicroBooNE:2021ccs}. In particular, resonant pion production is simulated using the \texttt{Kuzmin-Lyubushkin-Naumov Berger-Sehgal} model~\cite{Nowak:2009se, Kuzmin:2003ji, Berger:2007rq, Graczyk:2007bc} and coherent pion production using the \texttt{Berger-Sehgal} model~\cite{Berger:2008xs}. Propagation of the final state particles through the detector is then performed using \texttt{GEANT4 v4\_10\_3\_p03c} with the \texttt{QGSP\_BERT} physics list~\cite{GEANT4:2002zbu, Allison:2006ve, Allison:2016lfl}. This is followed by simulation of the produced ionization electrons and scintillation light and the subsequent detector response~\cite{MicroBooNE:2018swd, MicroBooNE:2018vro}. Simulated neutrino interactions are overlaid onto data collected while the beam is off providing data-based modeling of cosmic-ray induced interactions and detector noise. The Monte-Carlo (MC) prediction consists of simulated neutrino interactions within the detector, simulated interactions upstream of the detector (out-of-cryostat), and data collected with the beam off to model beam spills where no neutrino interaction occurs (EXT).

Reconstruction is performed using the \texttt{Pandora} pattern-recognition toolkit. This uses a multi-algorithm approach to identify neutrino interactions from cosmic-ray-induced backgrounds and to reconstruct each resulting particle. Each particle associated with the neutrino interaction is categorized as a track (muons, pions, protons) or an electromagnetic shower (electrons, photons). A detailed description of the performance of the Pandora reconstruction in MicroBooNE can be found in Ref.~\cite{MicroBooNE:2017xvs}. 
Following the Pandora reconstruction, particle identification based on calorimetric and topological information~\cite{MicroBooNE:2019efx, MicroBooNE:2019rgx} is performed. Finally, energy reconstruction is performed using particle range for tracks and through calorimetry for showers.

\begin{figure}
\centering
\includegraphics[width=0.4\textwidth]{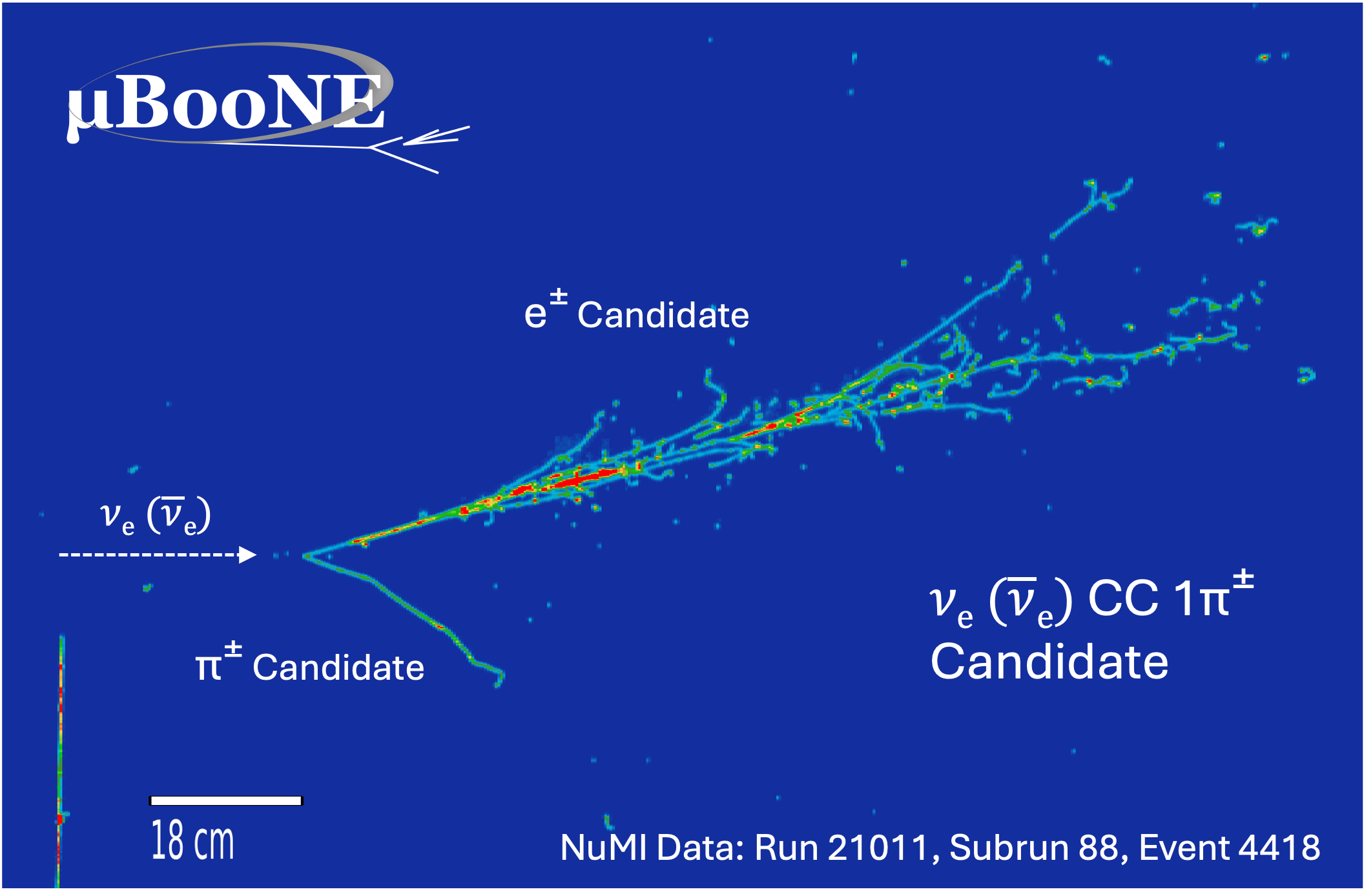}
\caption{Event display of a candidate $\nuewithbar \mathrm{CC}\,1\pi^\pm$ interaction.}
\label{fig:eventDisplay}
\end{figure}

{\bf \textit{Signal, selection and observables}}.--- Signal events are defined as $\nu_e$ or $\nuebar$ CC interactions that contain an outgoing electron or positron with kinetic energy KE$_{e^\pm} > 30$\,MeV, a single charged pion with KE$_{\pi^\pm} > 40$\,MeV, zero neutral pions or heavier mesons, and any number of outgoing protons or neutrons. The signal definition thresholds are guided by the reconstruction thresholds for each particle type. In addition, the opening-angle between the electron and charged pion is required to be $\theta_{e\pi^\pm} < 170\degree$. This removes a region of phase-space, containing less than 0.5\% of predicted signal events, for which the reconstruction performance is poor. Figure~\ref{fig:eventDisplay} shows a candidate $\nuewithbar \mathrm{CC}\,1\pi^\pm$ interaction matching the signal definition that is selected in MicroBooNE data. A high energy electromagnetic shower is visible along with a single track that is consistent with a charged pion.

The event selection expands upon tools developed in previous MicroBooNE $\nuewithbar$ analyses~\cite{MicroBooNE:2021gfj, MicroBooNE:2021ppm, MicroBooNE:2021wad, MicroBooNE:2022tdd, MicroBooNE:2024ryw}. Charged current $\nuewithbar$ interactions are first identified through the presence of an electromagnetic shower consistent with an electron. Tracks present in the event are then assumed to have originated from either protons or charged pions. These are distinguished based on their differing ionization profiles or the presence of pion re-interactions. The focus on first identifying charged current $\nuewithbar$ interactions avoids the need to distinguish between final state muons and charged pions that are otherwise more challenging to separate. 

Well-reconstructed candidate events are first identified that have at least one electromagnetic shower and at least one track contained within the instrumented volume. Next, a set of simple cuts are used to remove obvious background events. Cosmic-ray-induced interactions are rejected through \texttt{Pandora} classification based on topological characteristics. Electromagnetic showers resulting from neutral pion decays to photons are rejected by applying cuts on three features: the fraction of energy in the leading shower compared to all showers; the shower start position relative to the interaction vertex; and the transverse spread of the shower. Finally, $\nuewithbar$ interactions containing protons but not pions (referred to subsequently as $\nuewithbar$ CC N$p$) are rejected by requiring at least one track has an ionization profile ($dE/dx$) that is inconsistent with a stopping proton Bragg peak hypothesis~\cite{MicroBooNE:2021ddy}. 

After these initial cuts, more sophisticated methods are applied to refine the event selection further focusing on background suppression through the use of two Boosted Decision Trees (BDTs) trained with \texttt{XGBoost}~\cite{xgboost}. The BDTs are trained separately for FHC and RHC modes accounting for differences in neutrino composition and energies.

\begin{figure}
\centering
\begin{tikzpicture} 
\draw (0, 0) node[inner sep=0] {\includegraphics[width=0.495\textwidth]{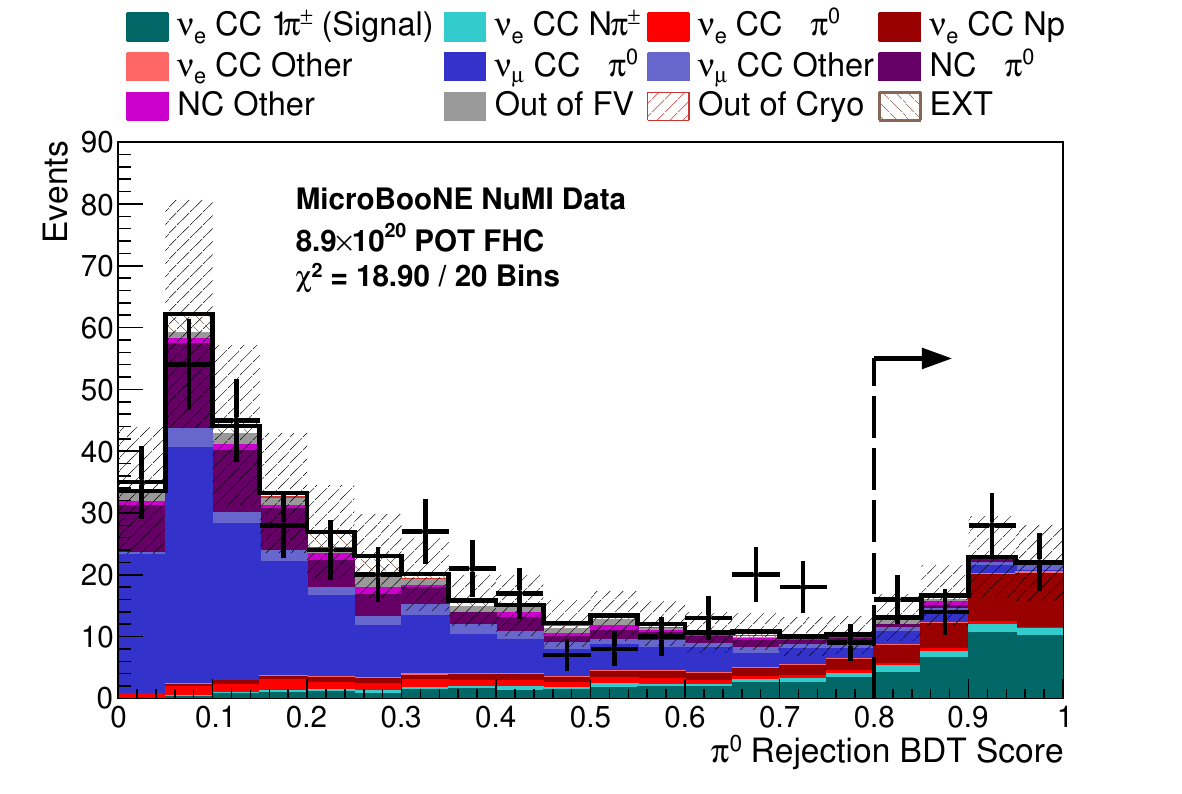}};
\draw (2.5, 1.5) node {\textbf{(a)}};
\end{tikzpicture}
\begin{tikzpicture} 
\draw (0, 0) node[inner sep=0] {\includegraphics[width=0.495\textwidth]{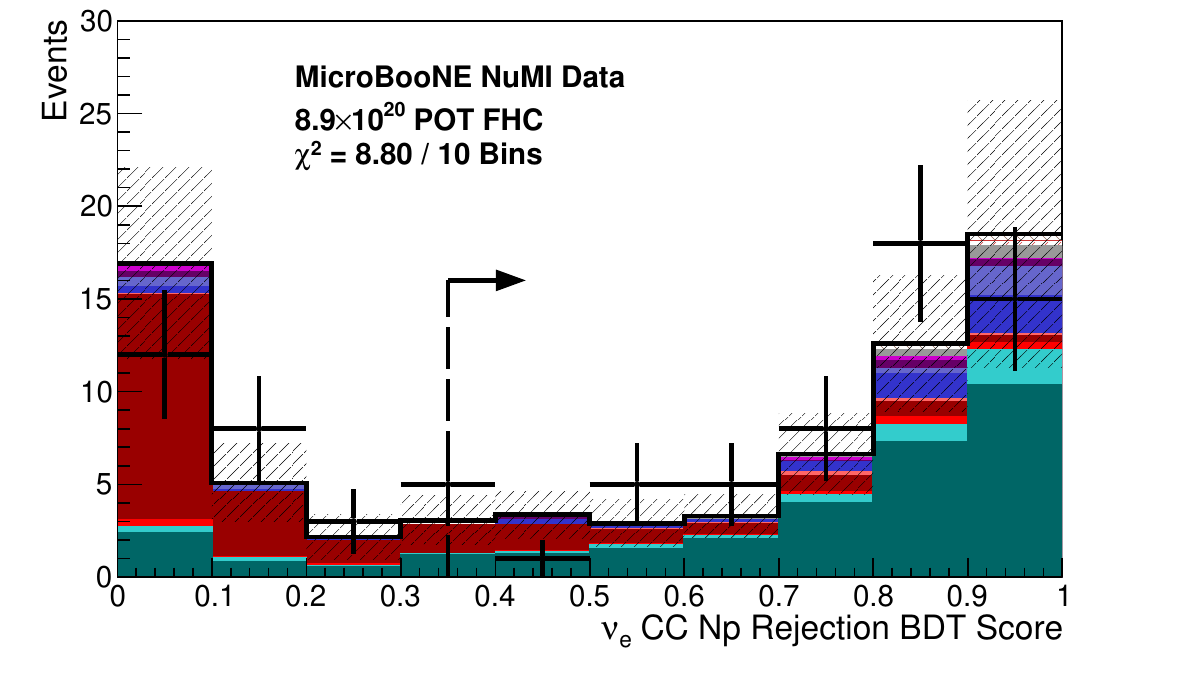}};
\draw (2.5, 1.9) node {\textbf{(b)}};
\end{tikzpicture}
\caption{Distribution of scores for (a) the $\pi^0$-rejection BDT and (b) the $\nuewithbar$ CC N$p$-rejection BDT compared with FHC mode data. The shaded band shows the systematic and statistical uncertainty on the MC prediction and the black points show the data with statistical uncertainties. The dashed lines show the cuts applied, where events to the right are selected.}
\label{fig:BDTsWithData}
\end{figure}

The first BDT focuses on the shower information to distinguish between electrons and photons from neutral pion backgrounds ($\nuewithbar$ CC $\pi^0$, $\numuwithbar$ CC $\pi^0$ and NC $\pi^0$). It uses both calorimetric and topological information about the primary shower, any secondary showers if present, and any shower-like clusters of charge identified close to the neutrino interaction.

The second BDT focuses on the track information and aims to distinguish between charged pions and protons, in particular targeting remaining $\nuewithbar$ CC N$p$ interactions where the proton does not have a clear Bragg peak and, hence, is not rejected at the previous stage. This BDT uses the $dE/dx$ of the track compared to different particle hypotheses along with topological information focusing on the ends of tracks to identify pion re-interactions or decays. Interactions are selected where only one track is identified as a charged pion candidate, with any number of additional proton candidates allowed.

Figure~\ref{fig:BDTsWithData} shows the distribution of scores for the $\pi^0$-rejection BDT and $\nuewithbar$ CC N$p$-rejection BDT compared with data. The $\nuewithbar$ CC N$p$-rejection BDT is shown after the cut on the $\pi^0$-rejection BDT has been applied. Both are shown for FHC mode; RHC mode can be found in the Supplemental Material~\cite{supplemental}. Both BDTs achieve good signal-background separation for their target topologies. For the $\pi^0$ rejection, 99.2\% of $\pi^0$ induced photon showers are removed while keeping 42.0\% of $\nuewithbar$ induced electron showers. For the $\nuewithbar$ CC N$p$ rejection, 94.3\% of $\nuewithbar$ CC N$p$ interactions are rejected while keeping 67.1\% of $\nuewithbar$ CC $\pi^\pm$ interactions. In addition, excellent data--MC agreement within statistical and systematic uncertainties is seen across the full BDT score distributions. A full description of the BDT input variables along with data--MC comparisons used can be found in the Supplemental Material~\cite{supplemental}.

Candidate $\nuewithbar$ CC $\pi^\pm$ interactions are selected with efficiencies of 6.2\% and 5.8\% and purities of 57.3\% and 58.5\% for the FHC and RHC beam periods, respectively. Further details on the selection performance can be found in the Supplemental Material~\cite{supplemental}. In total, 116 candidate events pass the selection in the combined \num{2.0e21} POT data sample. The selected signal events are predicted to be 78\% $\nu_e$ interactions and 22\% $\nuebar$ interactions. They are dominated by resonant production (85\%), with smaller contributions from deep-inelastic scattering (11\%), quasi-elastic interactions (2\%) and coherent interactions (2\%). 

The observables considered are: the total rate; the electron energy, $E_e$; the electron and pion angles, $\theta_e$ and $\theta_\pi$, with respect to the neutrino direction; and the electron-pion opening angle, $\theta_{e\pi}$. Since MicroBooNE is off-axis to the NuMI beam, the exact neutrino direction is not known in data. To reconstruct angles relative to the neutrino direction, neutrinos are assumed to be produced at the NuMI target position. This gives an effective approximation of the neutrino direction with 95\% of signal events having an approximated direction within $3\degree$ of the true direction. In simulation, the true direction is known and hence the smearing arising from this approximation can be accounted for during unfolding~\cite{MicroBooNE:2021ppm}. Details on the performance of this approximation can be found in the Supplemental Material~\cite{supplemental}. The reconstruction resolution is approximately 17\% for $E_e$, 10\% for $\theta_e$, 10\% for $\theta_\pi$ and 9\% for $\theta_{e\pi}$. 

Each observable is binned such that there are five bins with approximately ten expected signal events each. In the case of $E_e$, the highest energy bin also serves as an overflow bin. For each observable, good data--MC agreement is observed within uncertainties. The selected event distributions can be found in the Supplemental Material~\cite{supplemental}

The largest backgrounds are interactions producing neutral pions (13.0\% of passing events) and $\nuewithbar$ CC N$p$ interactions (11.7\% of passing events). Two sidebands are constructed to assess the agreement between data and simulation for these backgrounds. A $\pi^0$-rich sideband is constructed by reversing the $\pi^0$-rejection BDT cut. This results in a sample with 68.8\% purity of $\pi^0$-containing interactions with a mixture of $\nuewithbar$ CC $\pi^0$, $\numuwithbar$ CC $\pi^0$ and NC $\pi^0$ topologies. A $\nuewithbar$ CC N$p$-rich sideband is constructed by reversing both the $\nuewithbar$ CC N$p$-rejection BDT cut and the proton Bragg peak cut. This results in a sample with 77.6\% purity of $\nuewithbar$ CC N$p$ interactions. The level of agreement with data is assessed across each observable considered for both sidebands. Good data--MC agreement is seen across all distributions indicating the background modeling is sufficient to proceed with cross section extraction. The sideband selected event distributions can be found in the Supplemental Material~\cite{supplemental}.  

{\bf \textit{Cross section extraction and uncertainties}}.--- The flux-averaged total cross section and differential cross sections as a function of true kinematic variables are reported. 
The cross sections are extracted using the Wiener singular value decomposition unfolding technique~\cite{Tang:2017rob} using a first-derivative regularization term. The impact of the regularization is encoded in a regularization matrix that can be applied to generator predictions to allow direct comparison with the extracted cross sections in the regularized truth space. A block-wise approach to unfolding is used allowing correlations between bins in different variables to be evaluated and reported~\cite{Gardiner:2024gdy}.  

Uncertainties on the extracted cross sections are assessed from a variety of sources. The statistical and systematic uncertainties are encoded in a covariance matrix using a block-wise formalism~\cite{Gardiner:2024gdy}. The total covariance matrix is constructed by summing the covariance matrices of each individual uncertainty. 

Systematic uncertainties are considered on: the neutrino flux from hadron production and beam-line geometry modeling~\cite{MINERvA:2016iqn, flux_public_note}; the neutrino interaction cross section modeling with \texttt{GENIE}~\cite{MicroBooNE:2021ccs}; secondary particle re-interactions~\cite{Calcutt:2021zck}; detector response modeling including the scintillation light yield, recombination model, space charge effects~\cite{MicroBooNE:2020kca}, and ionization signal response~\cite{MicroBooNE:2021roa}; out-of-cryostat interaction modeling; the number of argon targets; and POT counting. In the unfolded results, the dominant systematic uncertainty arises from the neutrino flux modeling ($\sim$20--30\%) due to the challenges in simulating the $8\degree$ off-axis NuMI beam. This is followed by detector response modeling ($\sim$15\%), predominantly from recombination modeling due to the reliance on calorimetric variables to identify charged pions; and cross section modeling ($\sim$10\%). The other sources of systematic uncertainty are subdominant. 

Statistical uncertainties on the data and simulation are evaluated using Poisson uncertainties. Data statistical uncertainties are subdominant for the total cross-section measurement at around 10\% but are comparable with systematic uncertainties or dominant for the differential cross-section measurements at around 30\%. The total covariance and correlation matrices are reported in the Supplemental Material~\cite{supplemental}.

The robustness of the unfolding procedure and regularization is assessed using fake data produced by the \texttt{GENIE} and \texttt{NuWro 19.02.2}~\cite{Golan:2012wx} generator models. These fake-data tests motivated expanding the cross section modeling uncertainty by treating the \texttt{NuWro} sample as an additional systematic universe~\cite{MicroBooNE:2023cmw}.

\begin{table}
\begin{ruledtabular}
\begin{tabular}{lcc}
\textbf{Generator} & \textbf{$\sigma$} [$\num{e-39}$cm$^2$/nucleon] & \textbf{$\chi^2/\mathrm{n_{bins}}$} \\[2pt]
\hline\hline 
\\[-6pt]
\textbf{Unfolded Data} & \multicolumn{2}{l}{\textbf{0.93 $\pm$ 0.13 (stat.) $\pm$ 0.27 (syst.)}} \\[2pt]
\hline\hline
\\[-7pt]
\texttt{NuWro 21.09.2} & 0.76 & 0.30/1 \\
\texttt{NEUT 5.4.0.1} & 0.83  & 0.11/1 \\
\texttt{GiBUU 2025} & 0.74 & 0.42/1 \\
\texttt{GENIE 3.4.0 AR23} & 0.62 & 1.11/1 \\
\texttt{GENIE 3.0.6 G18 $\upmu$B} & 0.68 & 0.68/1
\end{tabular}
\end{ruledtabular}
\caption{Extracted total cross section compared with predictions from generators.}
\label{tab:UnfoldedResultsTotal}
\end{table}

\begin{figure*}
\centering
\begin{tikzpicture} 
\draw (0, 0) node[inner sep=0] {\includegraphics[width=0.495\textwidth]{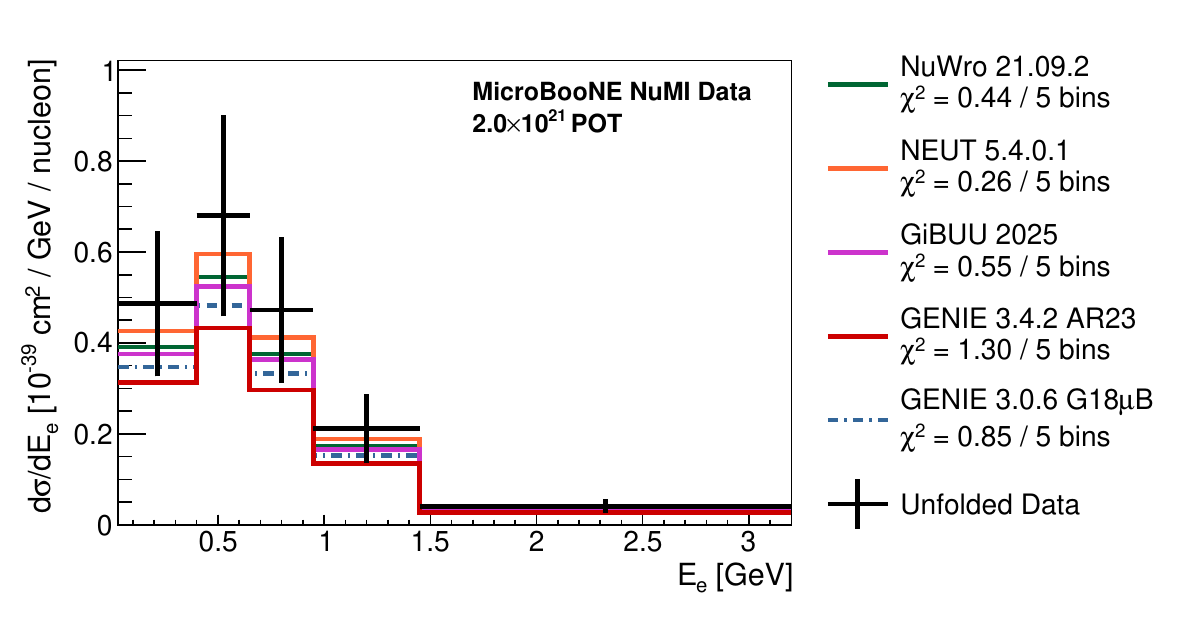}};
\draw (0.25, 1.0) node {\textbf{(a)}};
\end{tikzpicture}
\begin{tikzpicture} 
\draw (0, 0) node[inner sep=0] {\includegraphics[width=0.495\textwidth]{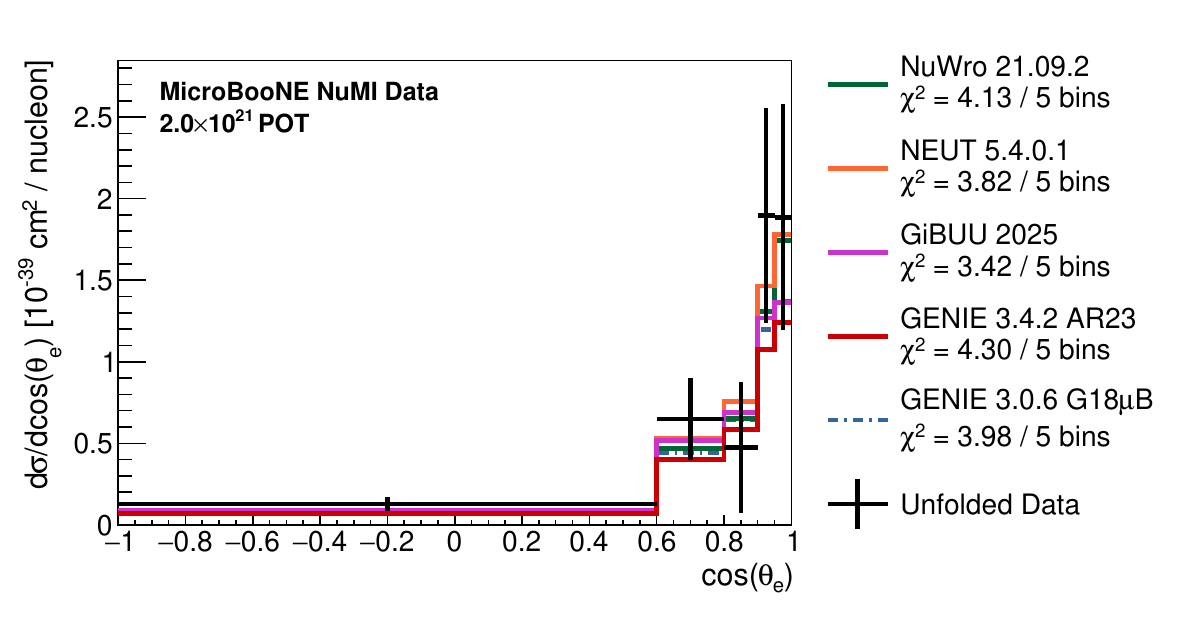}};
\draw (0.25, 1.5) node {\textbf{(b)}};
\end{tikzpicture}
\begin{tikzpicture} 
\draw (0, 0) node[inner sep=0] {\includegraphics[width=0.495\textwidth]{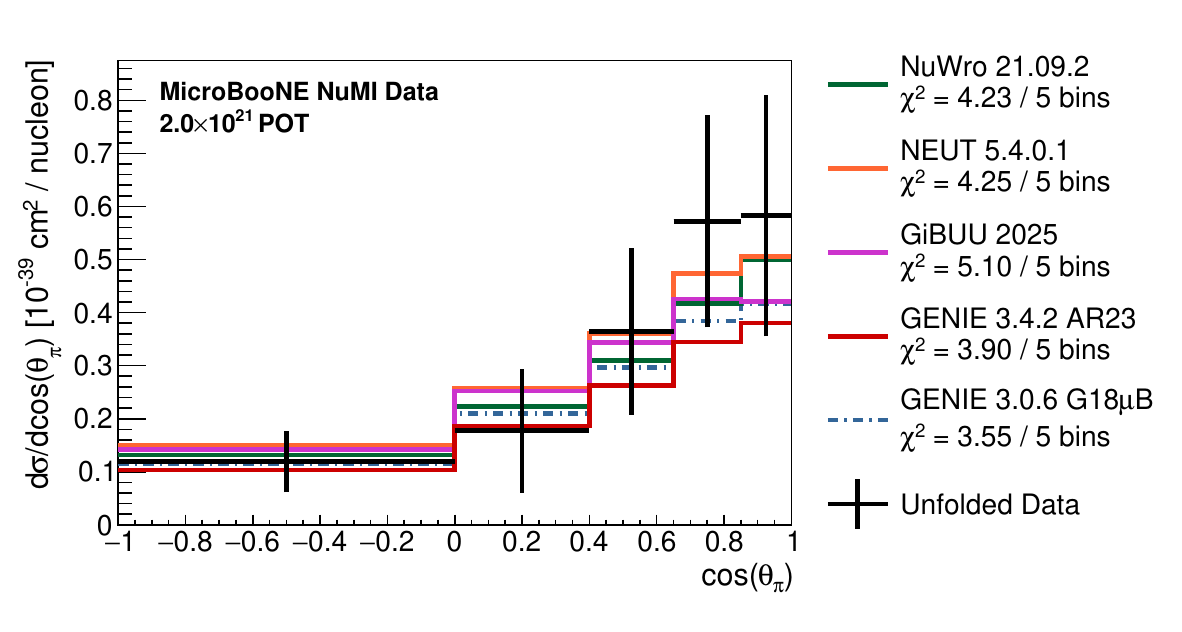}};
\draw (0.25, 1.5) node {\textbf{(c)}};
\end{tikzpicture}
\begin{tikzpicture} 
\draw (0, 0) node[inner sep=0] {\includegraphics[width=0.495\textwidth]{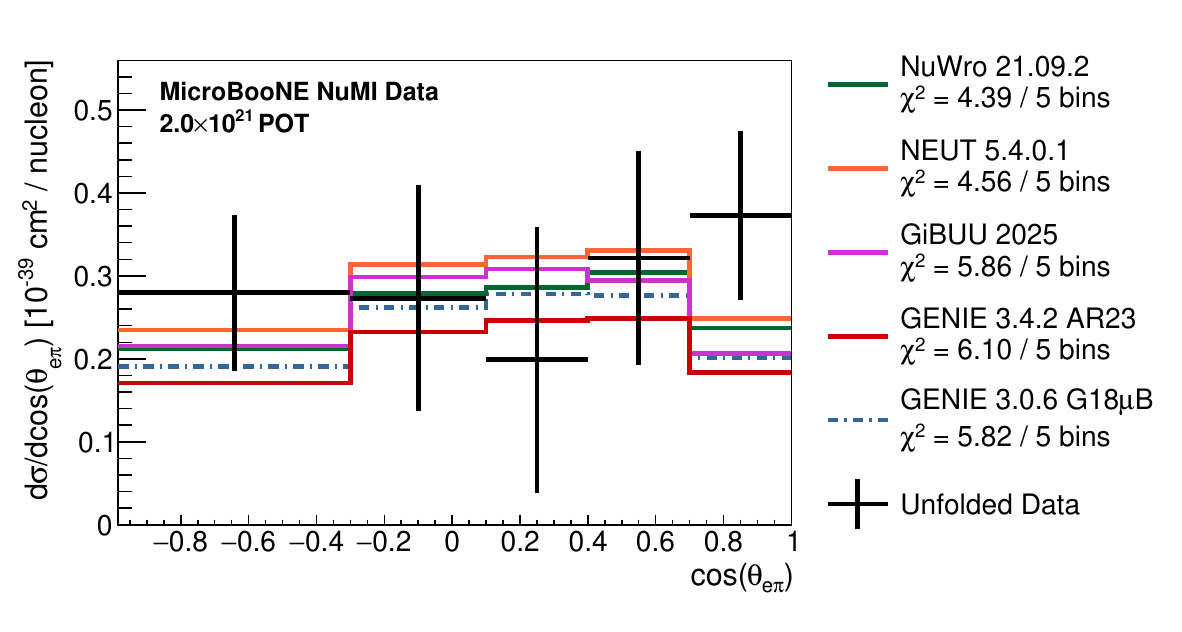}};
\draw (0.25, 1.5) node {\textbf{(d)}};
\end{tikzpicture}
\caption{Extracted differential cross sections in (a) electron energy, (b) electron angle, (c) pion angle and (d) electron-pion opening angle compared with generator predictions. The unfolded data points show both statistical and systematic uncertainties.}
\label{fig:UnfoldedResultsDifferential}
\end{figure*}

{\bf \textit{Results}}.--- The extracted total cross section is shown in Table~\ref{tab:UnfoldedResultsTotal} and differential cross sections in $E_e$, $\theta_e$, $\theta_\pi$ and $\theta_{e\pi}$ are shown in Fig.~\ref{fig:UnfoldedResultsDifferential}. The differential cross sections are presented in regularized truth space described by the regularization matrix available in the Supplemental Material~\cite{supplemental}. The unfolded data is compared with generator predictions from \texttt{NuWro 21.09.2}~\cite{Golan:2012wx}, \texttt{NEUT 5.4.0.1}~\cite{Hayato:2021heg}, \texttt{GiBUU 2025}~\cite{Buss:2011mx}, \texttt{GENIE 3.0.6 G18\_10a\_02\_11a MicroBooNE tune}~\cite{MicroBooNE:2021ccs} (labeled \texttt{GENIE 3.0.6 G18$\upmu$B}), and \texttt{GENIE 3.4.2 AR23\_20i\_00\_000}~\cite{Andreopoulos:2009rq} (labeled \texttt{GENIE 3.4.2 AR23}). Resonant pion production is simulated using the \texttt{Kuzmin-Lyubushkin-Naumov Berger-Sehgal} model~\cite{Nowak:2009se, Kuzmin:2003ji, Berger:2007rq, Graczyk:2007bc} in \texttt{GENIE}, the \texttt{Adler-Rarita-Schwinger} formalism~\cite{Graczyk:2009qm} in \texttt{NuWro}, and the \texttt{Rein-Sehgal} model in \texttt{NEUT}~\cite{Rein:1980wg}. Coherent pion production is simulated with the \texttt{Berger-Sehgal}~\cite{Berger:2008xs} model in \texttt{GENIE}, \texttt{NuWro} and \texttt{NEUT}. \texttt{GiBUU} models resonant pion production following the \texttt{MAID} analysis~\cite{Mosel:2019vhx} and does not simulate coherent pion production.

The total cross section is measured to be (0.93 $\pm$ 0.13 (stat.) $\pm$ 0.27 (syst.)) $\times \num{e-39}$ cm$^2$/nucleon. This is consistent with the predictions from each of the generators considered. The data slightly prefers the higher cross sections predicted by \texttt{NEUT}, \texttt{NuWro} and \texttt{GiBUU} compared with the lower cross sections predicted by \texttt{GENIE}. This could suggest a slight preference for the treatment of resonant pion production or nuclear medium effects in these models. However, the sensitivity of the measurement is limited by the large systematic uncertainties on the flux modeling. 

The extracted differential cross sections in $E_e$, $\theta_e$, $\theta_\pi$ and $\theta_{e\pi}$ are all consistent with the generator predictions. The preference for higher cross sections results in lower $\chi^2$ for \texttt{NEUT}, \texttt{NuWro} and \texttt{GiBUU} in $E_e$, $\theta_e$ and $\theta_{e\pi}$; whereas these models have a higher $\chi^2$ for $\theta_\pi$ hinting at possible shape disagreement in this variable. The smallest $\chi^2$ are seen for $E_e$ where there is minimal shape difference between the data and the generator predictions. The largest $\chi^2$ are seen for $\theta_{e\pi}$ driven by the smallest opening angle bin where, in particular, \texttt{GENIE} and \texttt{GiBUU} underpredict the cross section. However, due to the large statistical and flux uncertainties, all models lie within or close to 1$\sigma$ of the data suggesting in general good modeling of this process within the sensitivity of this measurement. To achieve greater model separation, future measurements would require greater statistics, potentially through improved reconstruction, and improved flux modeling. In particular, additional hadron production data to constrain the off-axis NuMI flux could significantly reduce the associated uncertainties.   

{\bf \textit{Conclusions}}.--- We have presented the first measurement of the flux-averaged $\nuewithbar$ charged current single charged pion production cross section on argon using the MicroBooNE detector. The full NuMI beam data set accumulated by MicroBooNE is used for the first time corresponding to a total exposure of \num{2.0e21} POT with mean $\nuewithbar$ energy of 730\,MeV. The total cross section is measured to be (0.93 $\pm$ 0.13 (stat.) $\pm$ 0.27 (syst.))$\times \num{e-39}$ cm$^2$/nucleon. Differential cross sections are also reported as functions of electron energy, the electron and pion angles with respect to the neutrino direction, and the electron-pion opening angle. These are found to be in good agreement with generator predictions within uncertainties. This is the first measurement of pion production in $\nuewithbar$ interactions on argon, one of the dominant interaction modes at the energies of the DUNE neutrino flux. It is essential to understand this process in order to facilitate the DUNE physics program, and direct measurements will have a critical impact to motivate modeling improvements. While this is the first time this process has been measured, the measurement is limited by statistical uncertainties and the challenges arising from the off-axis flux; future experiments in the SBN program~\cite{Machado:2019oxb} and the DUNE near detector~\cite{DUNE:2021tad} are expected to be able to improve on this. 

{\bf \textit{Acknowledgments}}.--- This document was prepared by the MicroBooNE collaboration using the resources of the Fermi National Accelerator Laboratory (Fermilab), a U.S. Department of Energy, Office of Science, Office of High Energy Physics HEP User Facility. Fermilab is managed by Fermi Forward Discovery Group, LLC, acting under Contract No. 89243024CSC000002. MicroBooNE is supported by the following: the U.S. Department of Energy, Office of Science, Offices of High Energy Physics and Nuclear Physics; the U.S. National Science Foundation; the Swiss National Science Foundation; the Science and Technology Facilities Council (STFC), part of the United Kingdom Research and Innovation; the Royal Society (United Kingdom); the UK Research and Innovation (UKRI) Future Leaders Fellowship; and the NSF AI Institute for Artificial Intelligence and Fundamental Interactions. Additional support for the laser calibration system and cosmic ray tagger was provided by the Albert Einstein Center for Fundamental Physics, Bern, Switzerland. We also acknowledge the contributions of technical and scientific staff to the design, construction, and operation of the MicroBooNE detector as well as the contributions of past collaborators to the development of MicroBooNE analyses, without whom this work would not have been possible. For the purpose of open access, the authors have applied a Creative Commons Attribution (CC BY) public copyright license to any Author Accepted Manuscript version arising from this submission.

\bibliography{main.bib}

\end{document}


\title{Supplemental material}
\maketitle

\section{NuMI flux at MicroBooNE}

\begin{figure}[htb]
\centering
\begin{tikzpicture} 
\draw (0, 0) node[inner sep=0] {\includegraphics[width=0.495\textwidth]{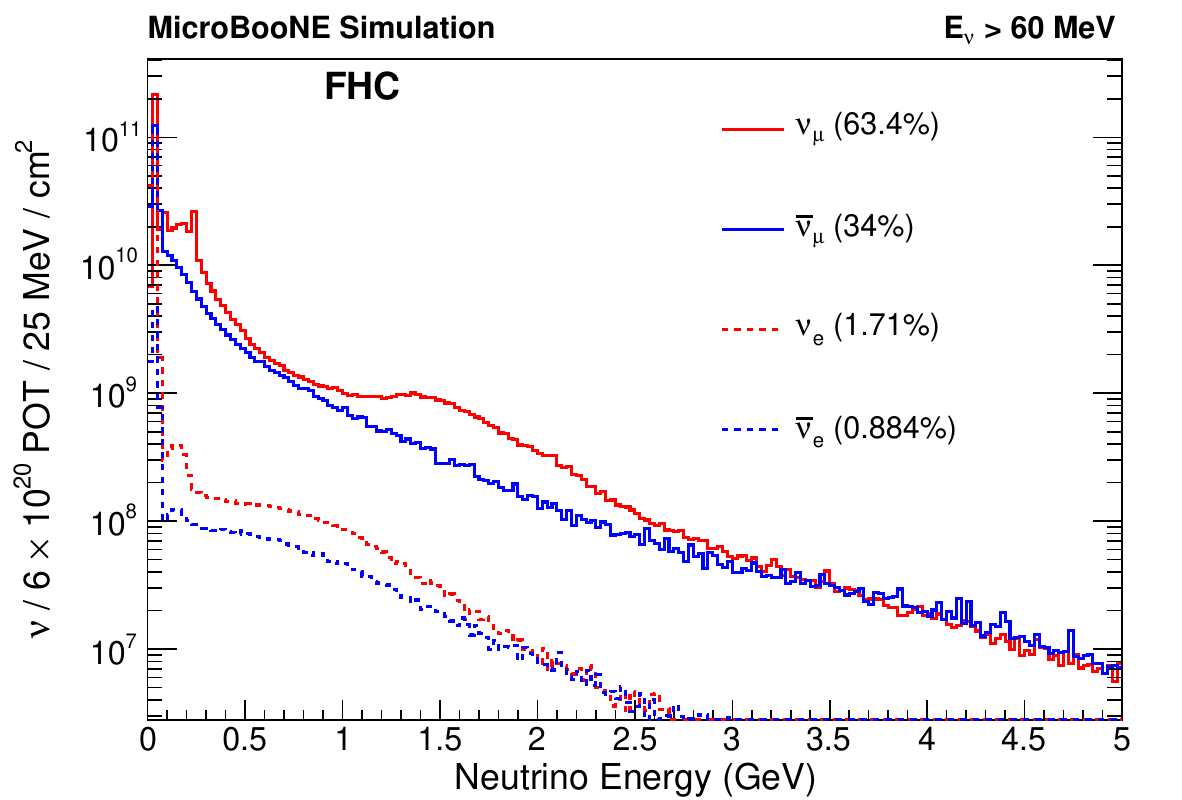}};
\draw (-2.8, 2.0) node {\textbf{(a)}};
\end{tikzpicture}
\begin{tikzpicture} 
\draw (0, 0) node[inner sep=0] {\includegraphics[width=0.495\textwidth]{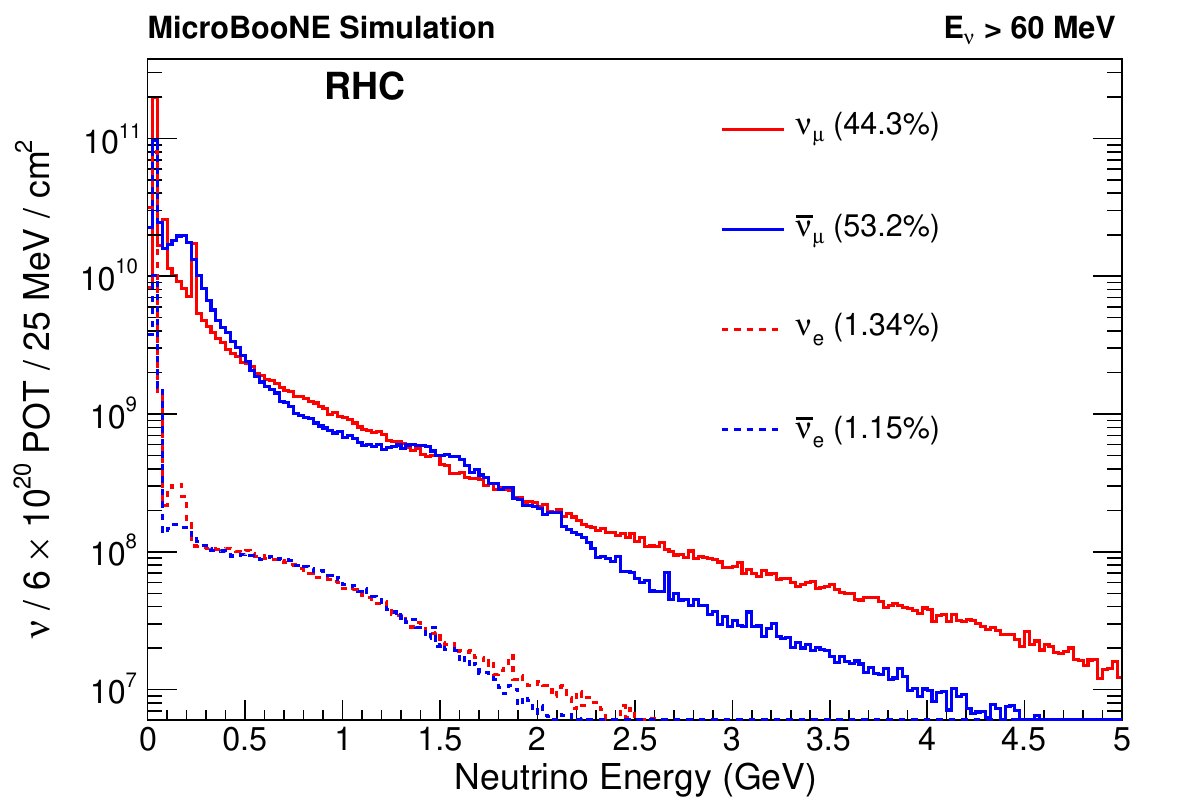}};
\draw (-2.8, 2.0) node {\textbf{(b)}};
\end{tikzpicture}
\caption{Predicted NuMI flux at MicroBooNE for (a) forward-horn-current mode and (b) reverse-horn-current mode broken down into neutrino flavor.}
\label{fig:NuMIFlux}
\end{figure}

Figure~\ref{fig:NuMIFlux} shows the predicted NuMI flux at MicroBooNE broken down into neutrino flavor for forward-horn-current (FHC) and reverse-horn-current (RHC) modes. 

\clearpage

\section{$\pi^0$-rejection BDT Variables}
\label{ShowerBDT}

The variables used in the $\pi^0$-rejection BDT are summarized in Table~\ref{tab:BDT1Variables}. The tools used in this BDT were developed as part of previous MicroBooNE $\nuewithbar$ analyses~\cite{MicroBooNE:2021wad, MicroBooNE:2022tdd, MicroBooNE:2024ryw}. The variables can be divided three categories: variables targeting the primary (most-energetic) electromagnetic shower reconstructed by \texttt{Pandora}; variables targeting a secondary electromagnetic shower reconstructed by \texttt{Pandora} if present; and variables looking for clusters of charge that could indicate shower-like activity that has been missed by the \texttt{Pandora} reconstruction. The most powerful distinguishing variables are: the dE/dx at the start of showers, since photons convert to electron-positron pairs leading to twice the ionization of an electron; the separation of the showers from the interaction vertex; and their transverse spread.

\begin{table}[htb]
    \centering
    \begin{tabular}{|l|l|}
    \hline
    {\bf Variable} & {\bf Description} \\
    \hline
    Primary Shower Vertex Distance & \makecell[l]{Distance of the primary shower from the interaction vertex. \\ Previous selection cut applied requiring distance $\leq$ 10\,cm.} \\
    Primary Shower dE/dx 1\,cm Gap  &  Primary shower start dE/dx with 1cm gap from vertex. \\
    Primary Shower dE/dx First 2\,cm  &  Primary shower start dE/dx using first 2cm only. \\
    Primary Shower Moliere Average & \makecell[l]{Primary shower Moliere average angle characterizing the spread of the shower. \\ Previous selection cut applied requiring Moliere Average angle $\leq$ 15\,\degree.} \\
    Primary Shower Energy Fraction & \makecell[l]{Fraction of energy in the primary shower relative to all showers. \\ Previous selection cut applied requiring energy fraction $>$ 0.7} \\
    Primary Shower Subclusters & Number of sub-clusters of charge the primary shower can be divided into. \\
    Primary Shower Principal Component Median & Shower principal component median~\cite{MicroBooNE:2021wad}. Metric of shower linearity. \\
    Primary Shower Cylindrical Fraction & \makecell[l]{ Fraction of the shower that is contained in a 1\,cm radius cylinder along \\ the shower direction~\cite{MicroBooNE:2021wad}.} \\
    Primary Shower Delta RMS & \makecell[l]{Shower second half root mean square~\cite{MicroBooNE:2021wad}. Metric of the spread of the \\ shower in the second half.} \\
    Primary Shower MCS Momentum & \makecell[l]{Shower Multiple Coulomb Scattering momentum~\cite{MicroBooNE:2021wad}. Metric of the \\ spread of the shower.} \\
    \hline
    Number of Showers & Number of showers starting in the fiducial volume. \\
    Secondary Shower Energy & Energy of the second shower. \\
    Secondary Shower Vertex Distance & Distance of the secondary shower from the interaction vertex. \\
    Secondary Shower Opening Angle & Angle between primary shower and secondary shower vertex. \\
    Secondary Shower Track End Distance & Distance between the secondary shower and nearest track end. \\
    Secondary Shower Generation & Secondary shower \texttt{Pandora} particle hierarchy. \\
    \hline
    Second Shower Tagger Hits & Number of reconstructed hits in the charge cluster. \\
    Second Shower Tagger Vertex Distance & Distance of the charge cluster from the interaction vertex. \\
    Second Shower Tagger Cluster Direction & Dot product of direction to the charge cluster start and cluster direction. \\
    Second Shower Tagger Angle & Angle between the charge cluster and the primary shower. \\
    \hline
    \end{tabular}
    \caption{Variables used in the $\pi^0$-rejection BDT.}
    \label{tab:BDT1Variables}
\end{table}

\begin{figure*}[htb]
\centering
\begin{tikzpicture} 
\draw (0, 0) node[inner sep=0] {\includegraphics[width=0.495\textwidth]{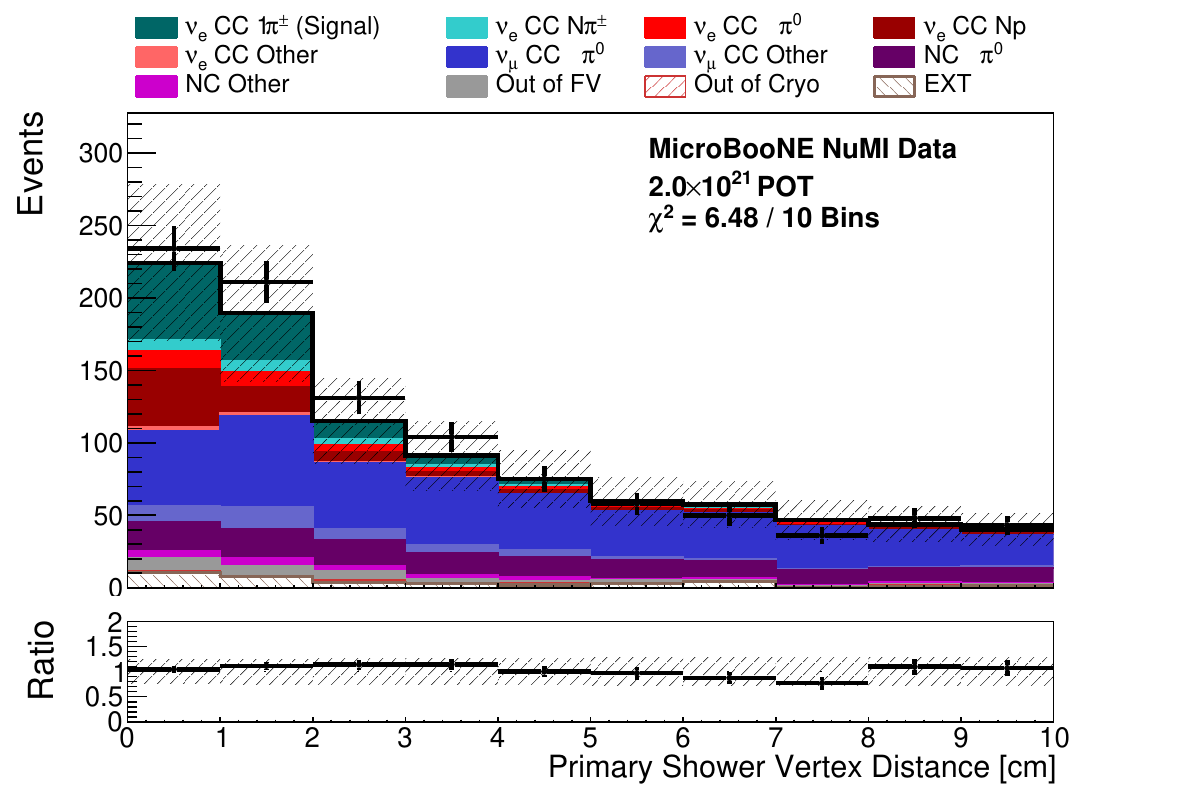}};
\draw (0.6, 1.0) node {\textbf{(a)}};
\end{tikzpicture}
\begin{tikzpicture} 
\draw (0, 0) node[inner sep=0] {\includegraphics[width=0.495\textwidth]{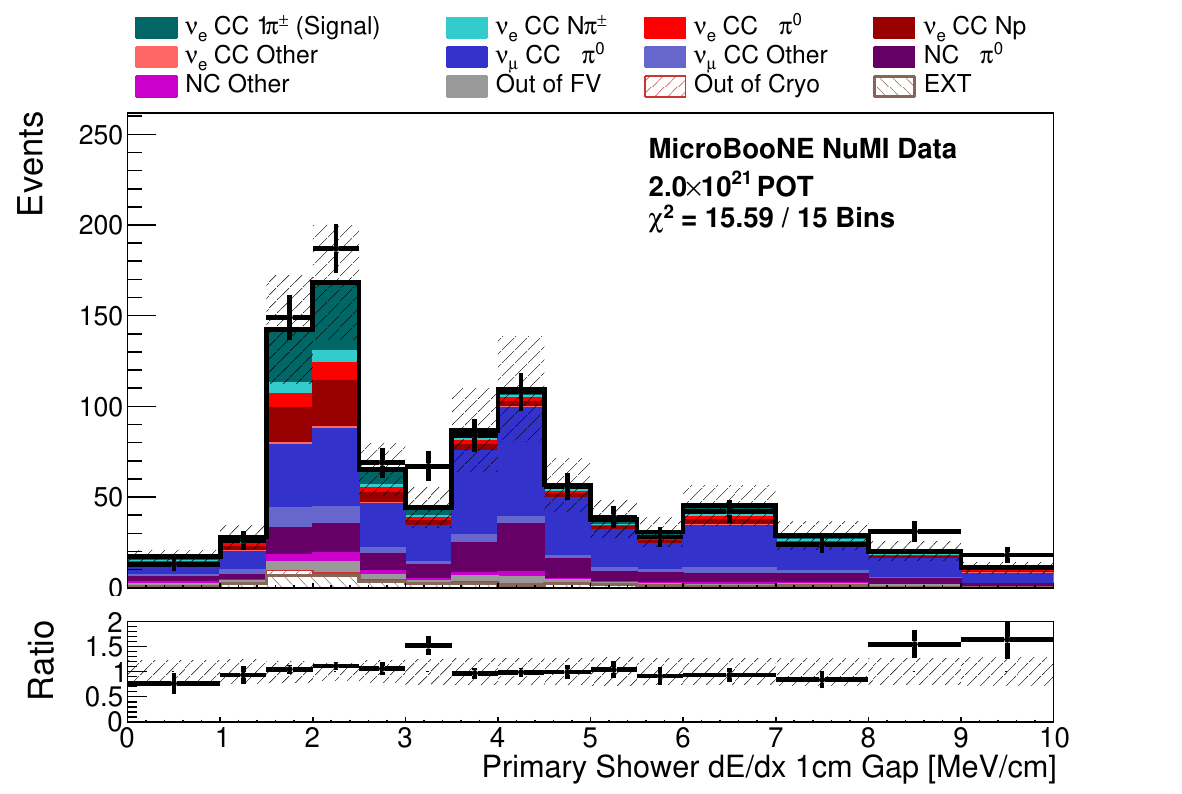}};
\draw (0.6, 1.0) node {\textbf{(b)}};
\end{tikzpicture}
\begin{tikzpicture} 
\draw (0, 0) node[inner sep=0] {\includegraphics[width=0.495\textwidth]{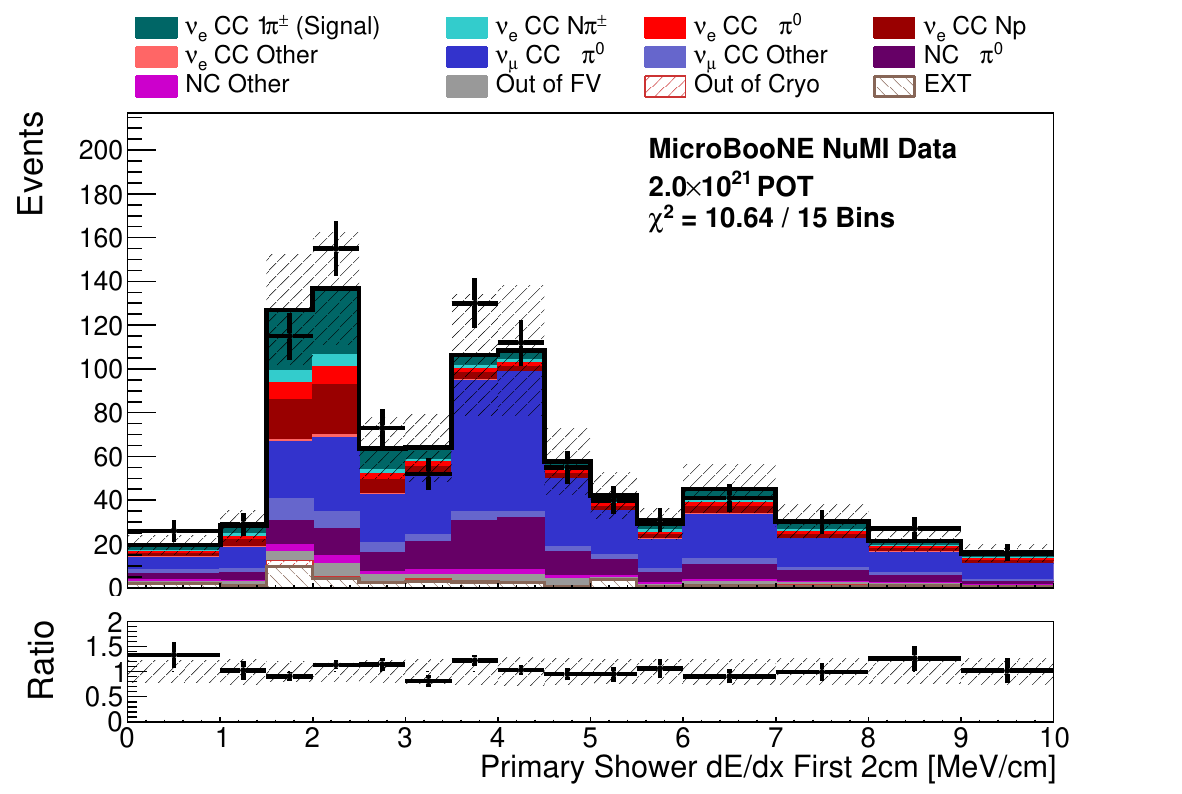}};
\draw (0.6, 1.0) node {\textbf{(c)}};
\end{tikzpicture}
\begin{tikzpicture} 
\draw (0, 0) node[inner sep=0] {\includegraphics[width=0.495\textwidth]{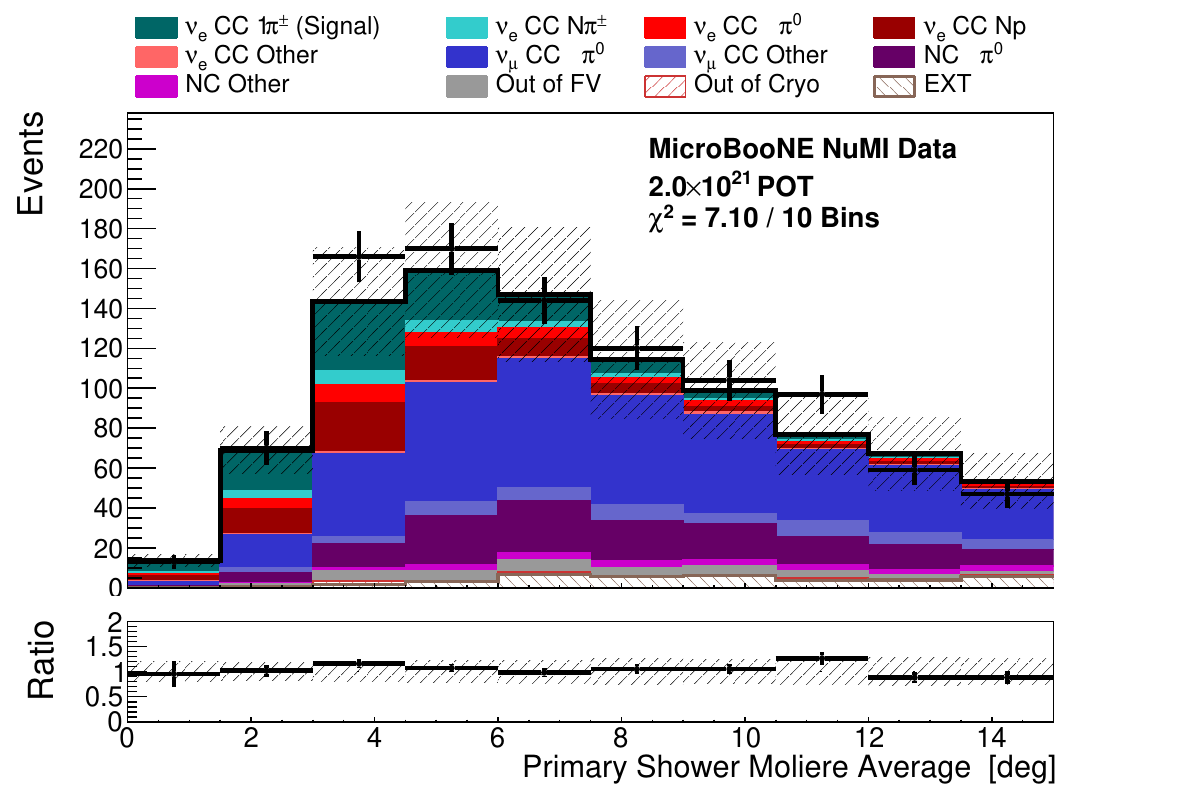}};
\draw (0.6, 1.0) node {\textbf{(d)}};
\end{tikzpicture}
\begin{tikzpicture} 
\draw (0, 0) node[inner sep=0] {\includegraphics[width=0.495\textwidth]{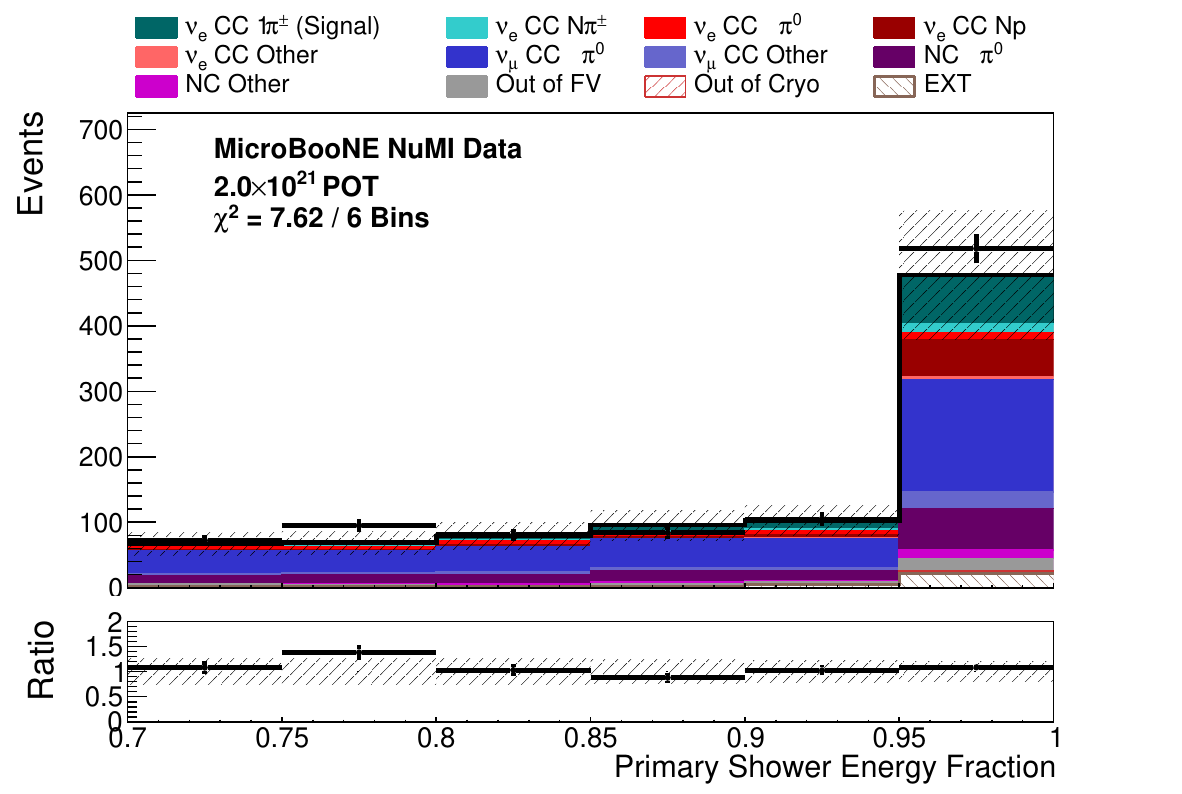}};
\draw (-2.65, 1.0) node {\textbf{(e)}};
\end{tikzpicture}
\begin{tikzpicture} 
\draw (0, 0) node[inner sep=0] {\includegraphics[width=0.495\textwidth]{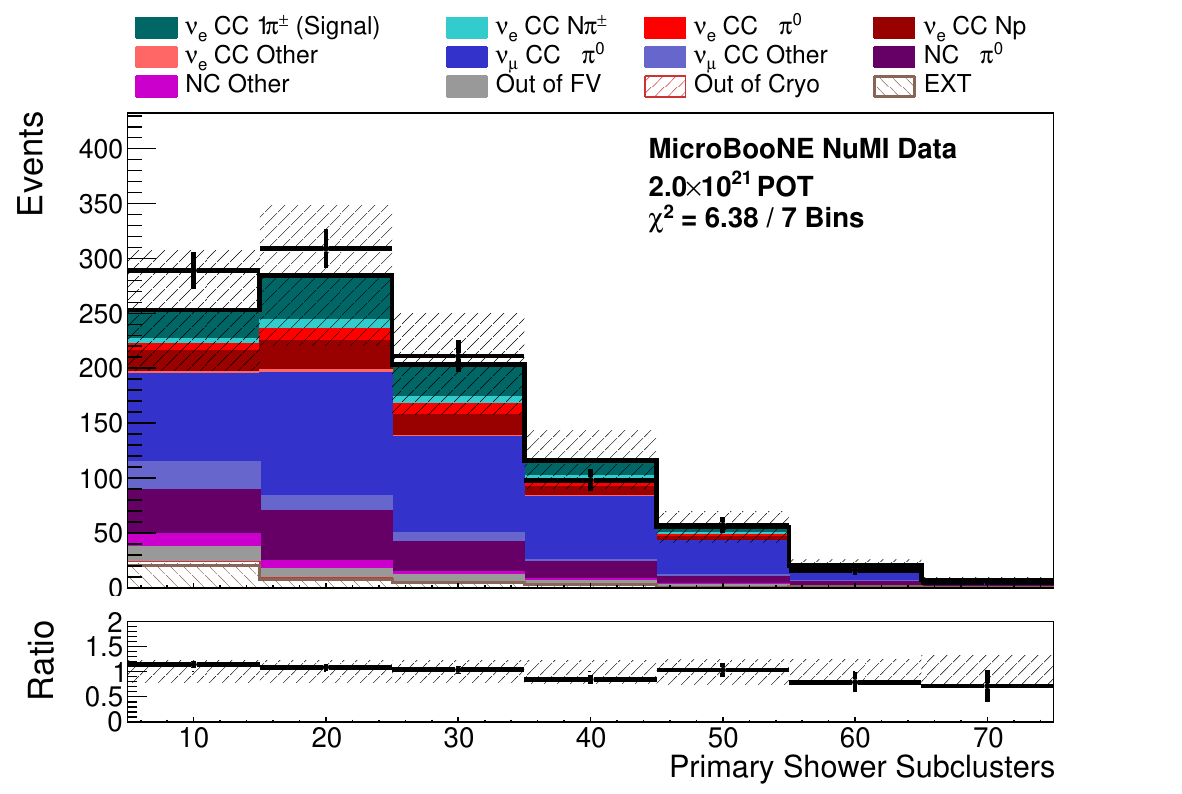}};
\draw (0.6, 1.0) node {\textbf{(f)}};
\end{tikzpicture}
\caption{Event distributions for the $\pi^0$-rejection BDT input variables (a) Primary Shower Vertex Distance, (b) Primary Shower dE/dx 1\,cm Gap, (c) Primary Shower dE/dx First 2\,cm, (d) Primary Shower Moliere Average, (e) Primary Shower Energy Fraction and (f) Primary Shower Subclusters. The shaded band shows the systematic and statistical uncertainty on the MC prediction and the black points show the data with statistical uncertainties.}
\label{fig:ShowerBDTVars_1}
\end{figure*}

\begin{figure*}[htb]
\centering
\begin{tikzpicture} 
\draw (0, 0) node[inner sep=0] {\includegraphics[width=0.495\textwidth]{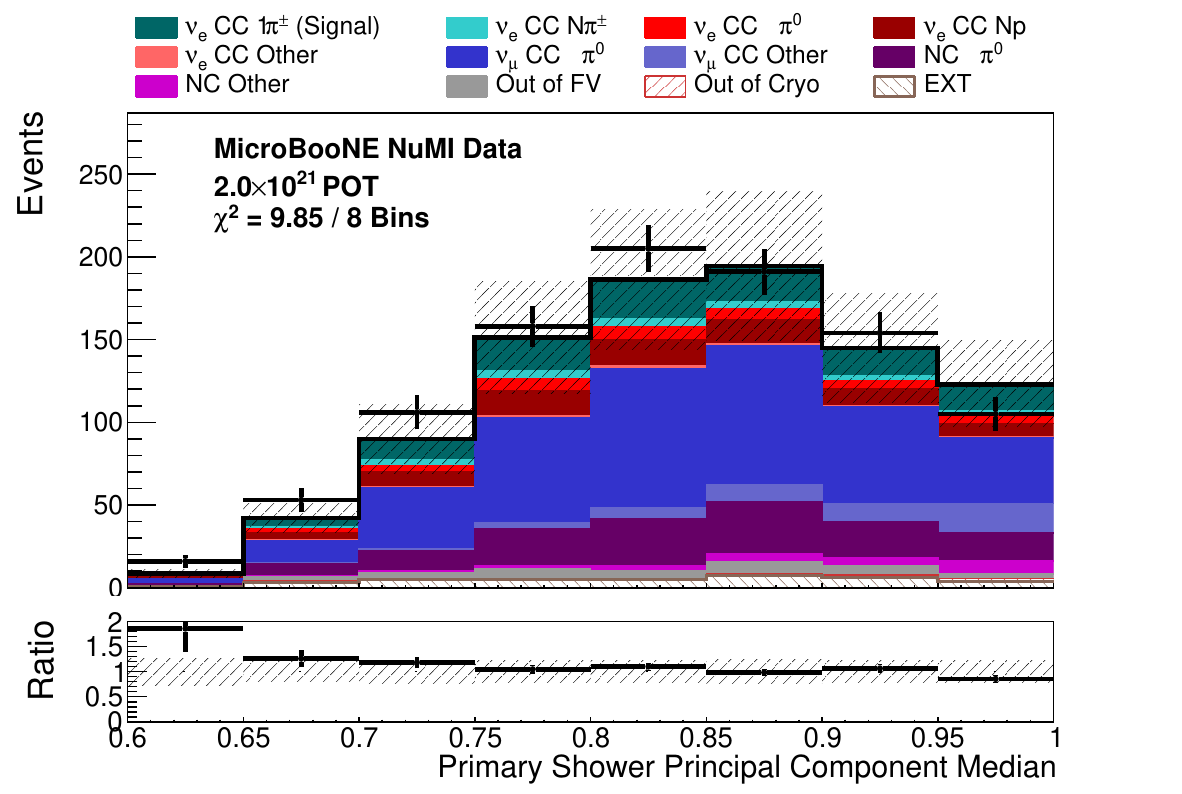}};
\draw (-2.65, 1.0) node {\textbf{(a)}};
\end{tikzpicture}
\begin{tikzpicture} 
\draw (0, 0) node[inner sep=0] {\includegraphics[width=0.495\textwidth]{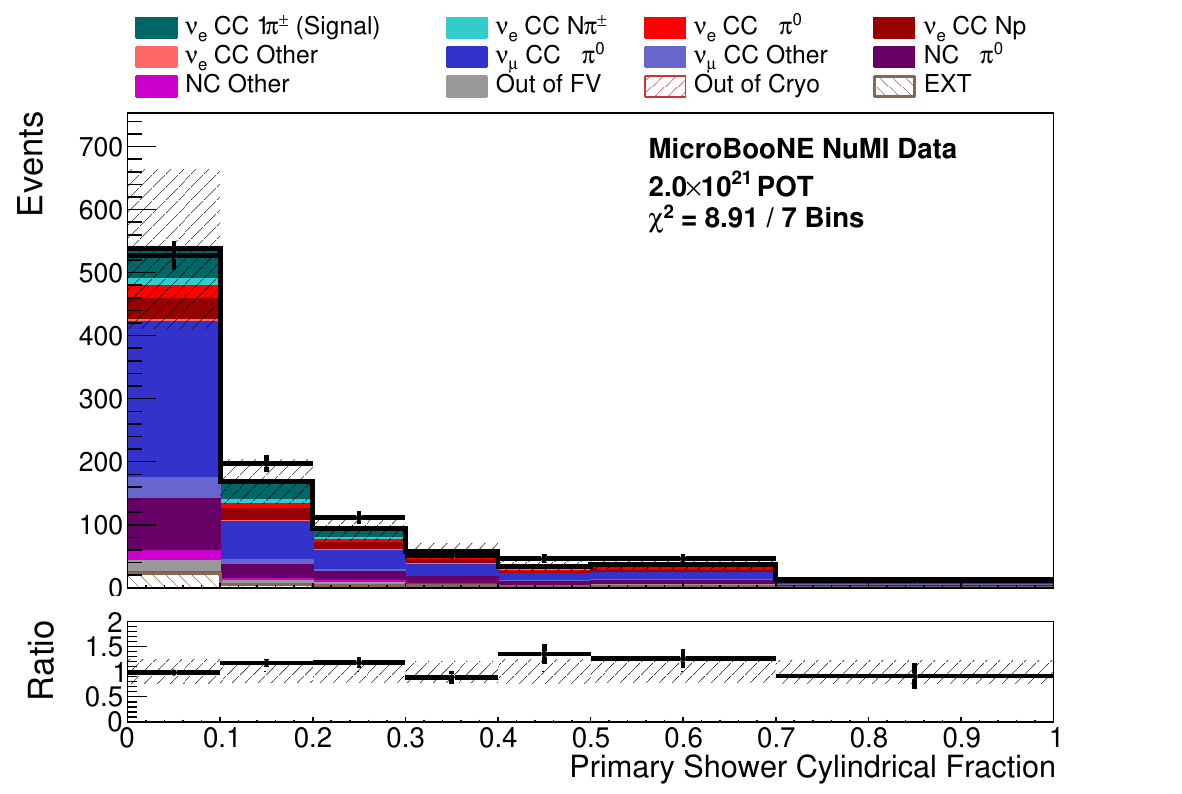}};
\draw (0.6, 1.0) node {\textbf{(b)}};
\end{tikzpicture}
\begin{tikzpicture} 
\draw (0, 0) node[inner sep=0] {\includegraphics[width=0.495\textwidth]{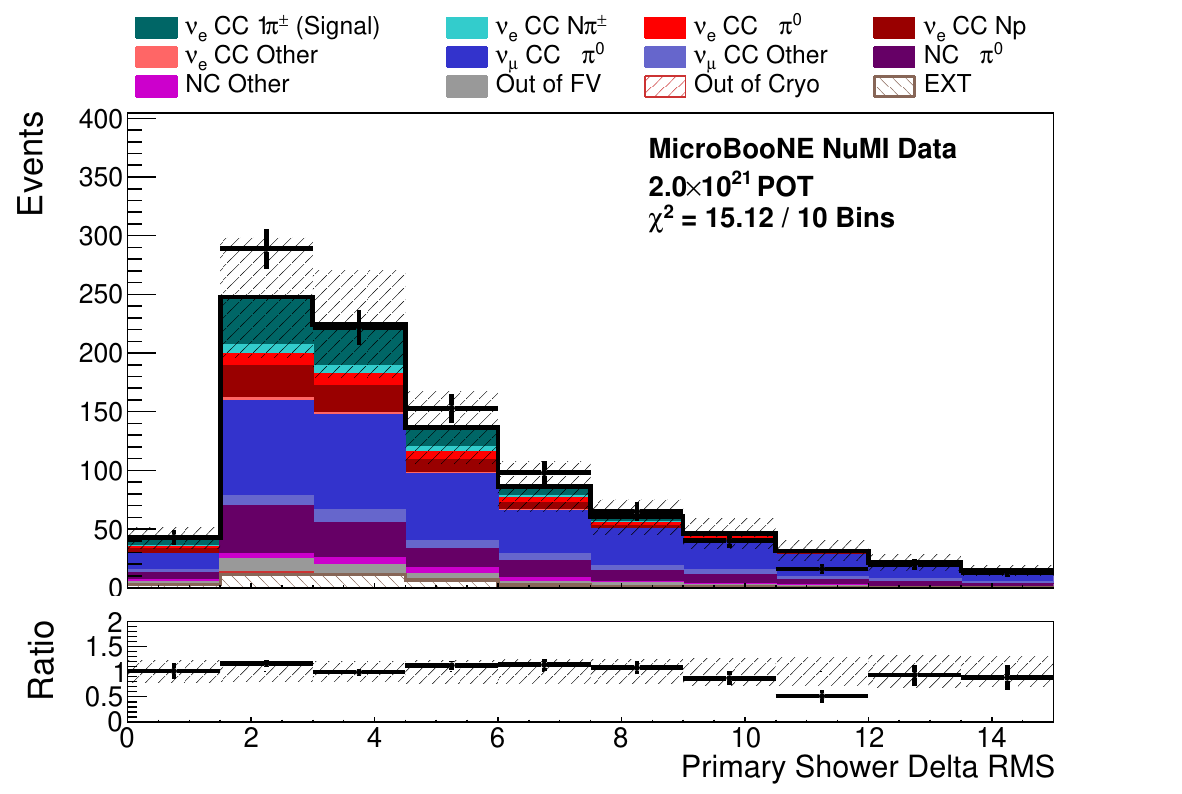}};
\draw (0.6, 1.0) node {\textbf{(c)}};
\end{tikzpicture}
\begin{tikzpicture} 
\draw (0, 0) node[inner sep=0] {\includegraphics[width=0.495\textwidth]{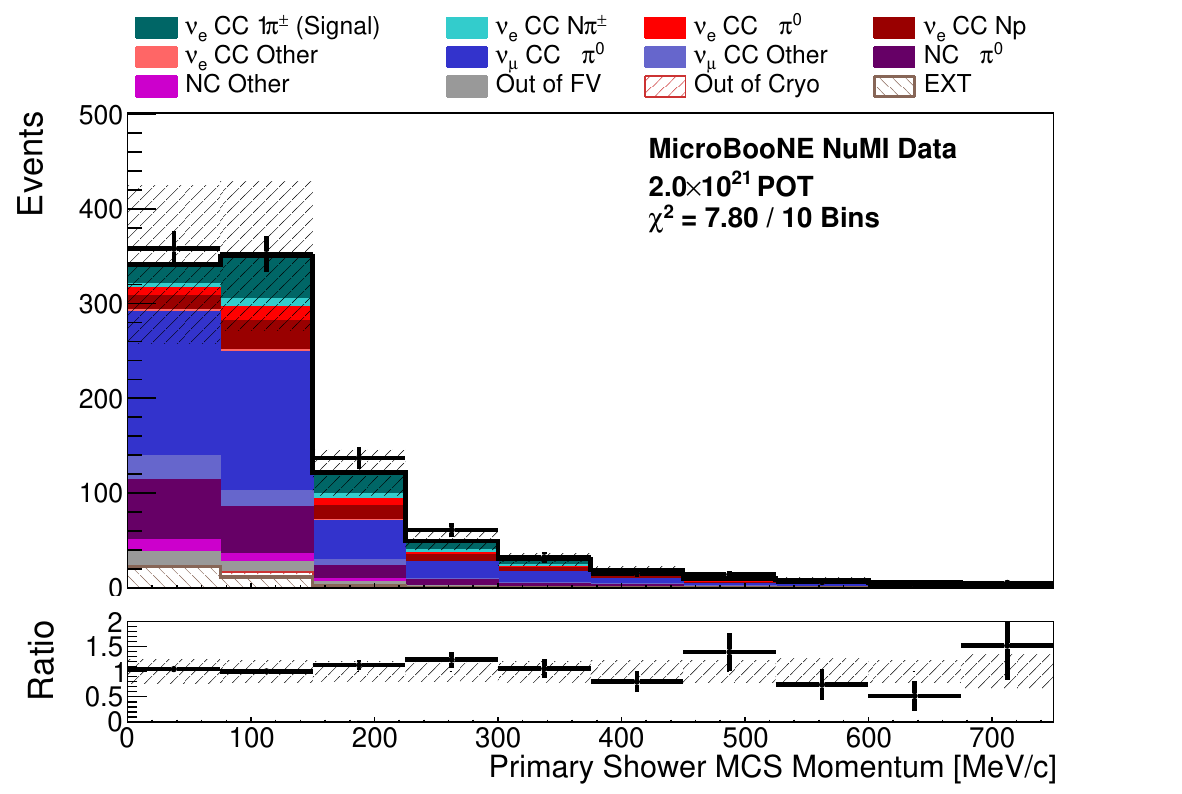}};
\draw (0.6, 1.0) node {\textbf{(d)}};
\end{tikzpicture}
\begin{tikzpicture} 
\draw (0, 0) node[inner sep=0] {\includegraphics[width=0.495\textwidth]{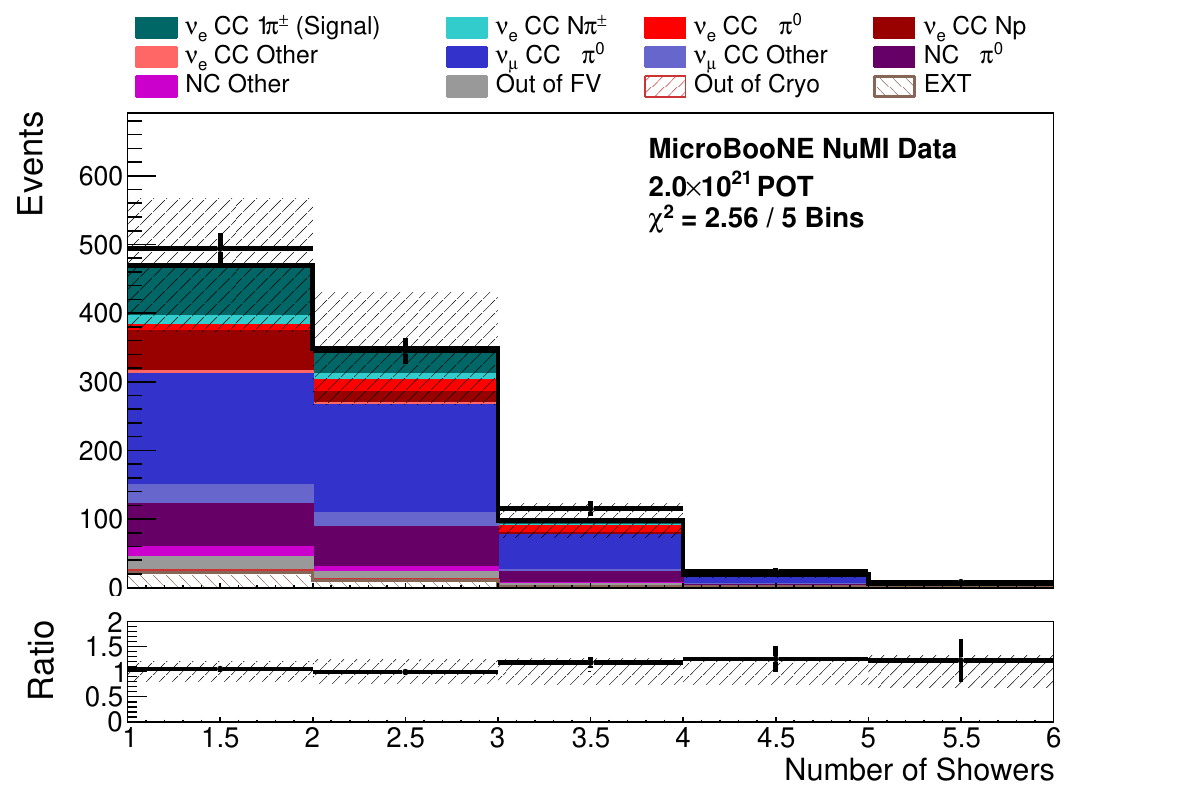}};
\draw (0.6, 1.0) node {\textbf{(e)}};
\end{tikzpicture}
\begin{tikzpicture} 
\draw (0, 0) node[inner sep=0] {\includegraphics[width=0.495\textwidth]{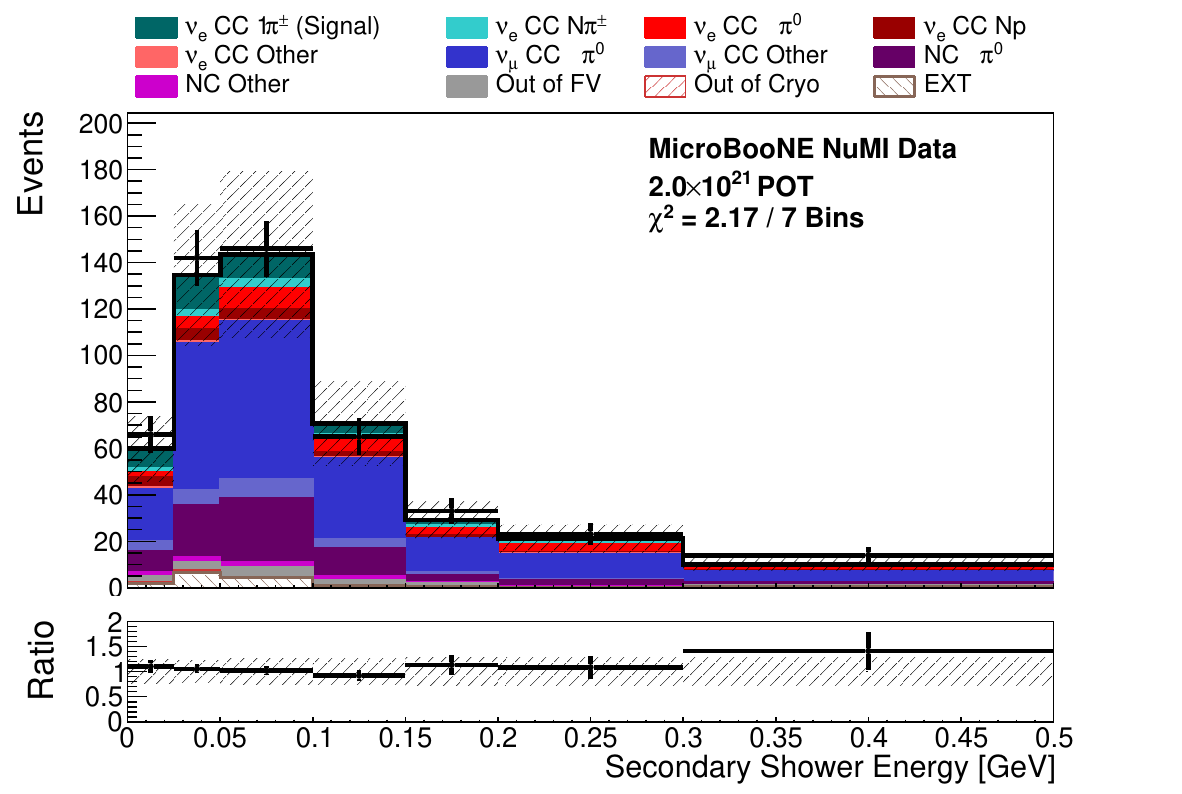}};
\draw (0.6, 1.0) node {\textbf{(f)}};
\end{tikzpicture}
\caption{Event distributions for the $\pi^0$-rejection BDT input variables (a) Primary Shower Principal Component Median, (b) Primary Shower Cylindrical Fraction, (c) Primary Shower Delta RMS, (d) Primary Shower MCS Momentum, (e) Number of Showers and (f) Secondary Shower Energy . The shaded band shows the systematic and statistical uncertainty on the MC prediction and the black points show the data with statistical uncertainties.}
\label{fig:ShowerBDTVars_2}
\end{figure*}

\begin{figure*}[htb]
\centering
\begin{tikzpicture} 
\draw (0, 0) node[inner sep=0] {\includegraphics[width=0.495\textwidth]{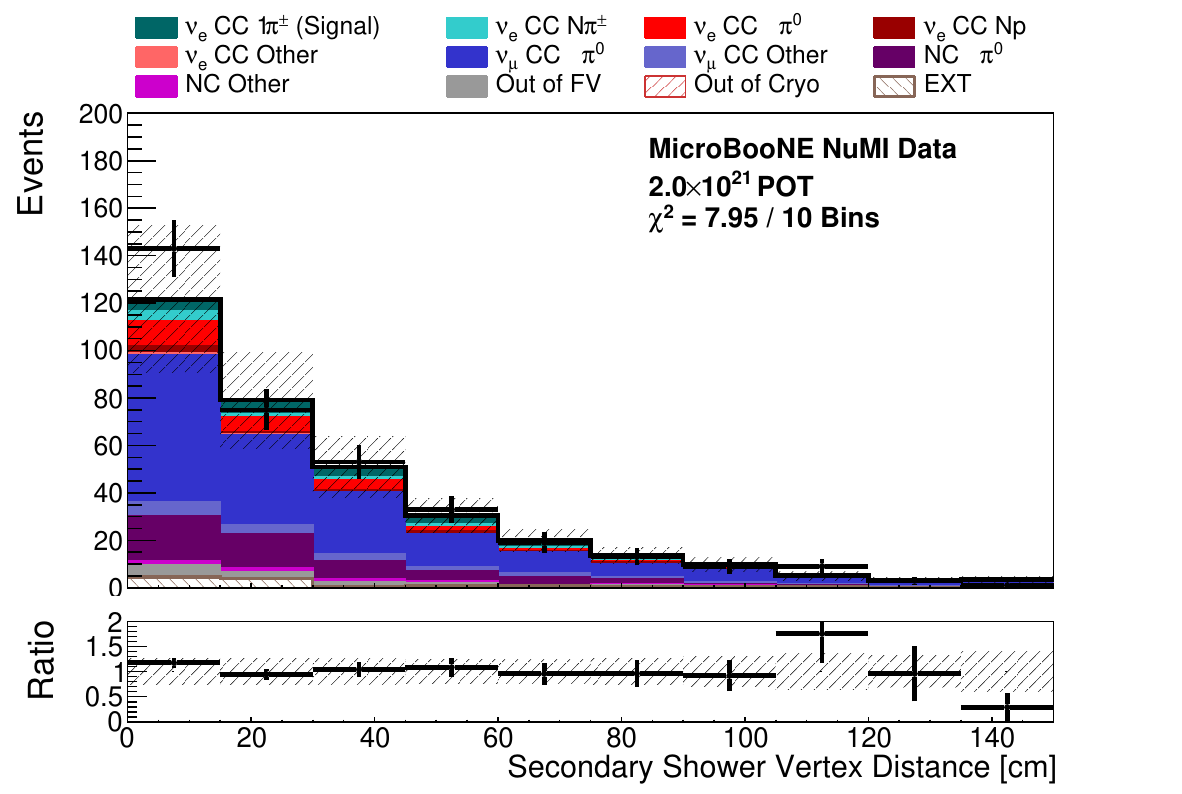}};
\draw (0.6, 1.0) node {\textbf{(a)}};
\end{tikzpicture}
\begin{tikzpicture} 
\draw (0, 0) node[inner sep=0] {\includegraphics[width=0.495\textwidth]{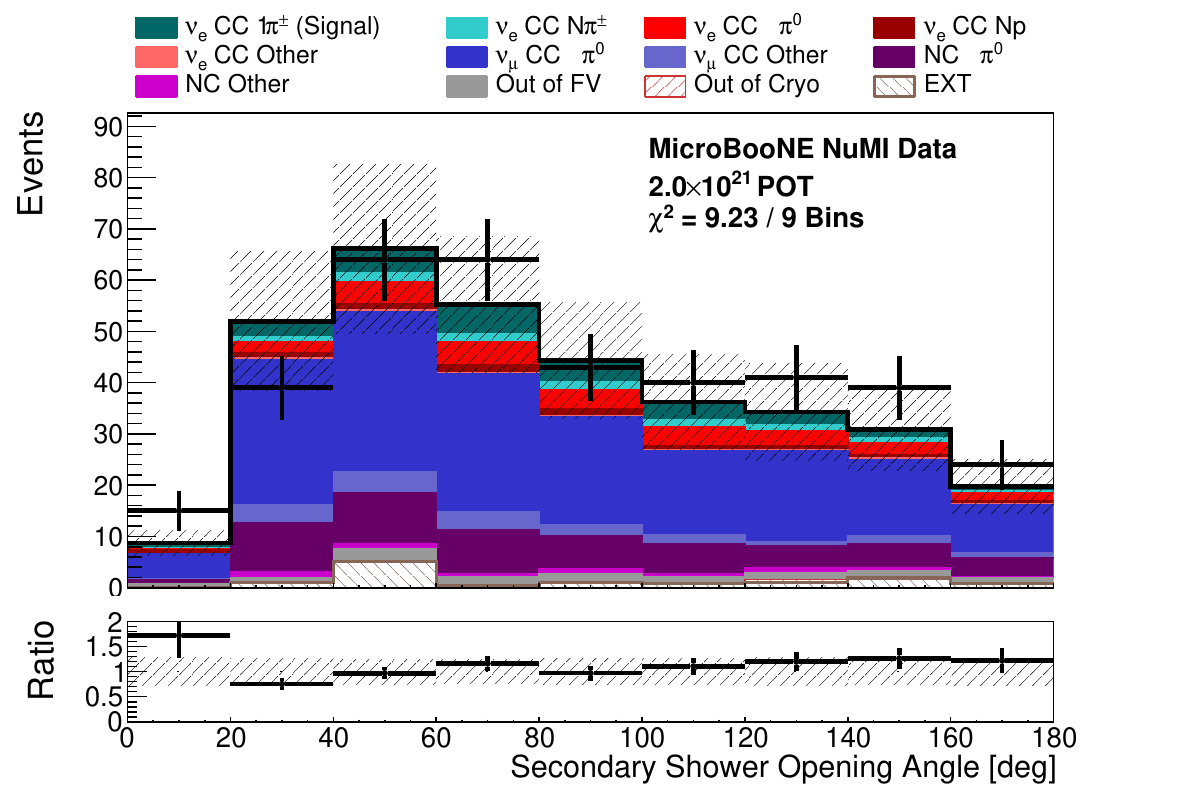}};
\draw (0.6, 1.0) node {\textbf{(b)}};
\end{tikzpicture}
\begin{tikzpicture} 
\draw (0, 0) node[inner sep=0] {\includegraphics[width=0.495\textwidth]{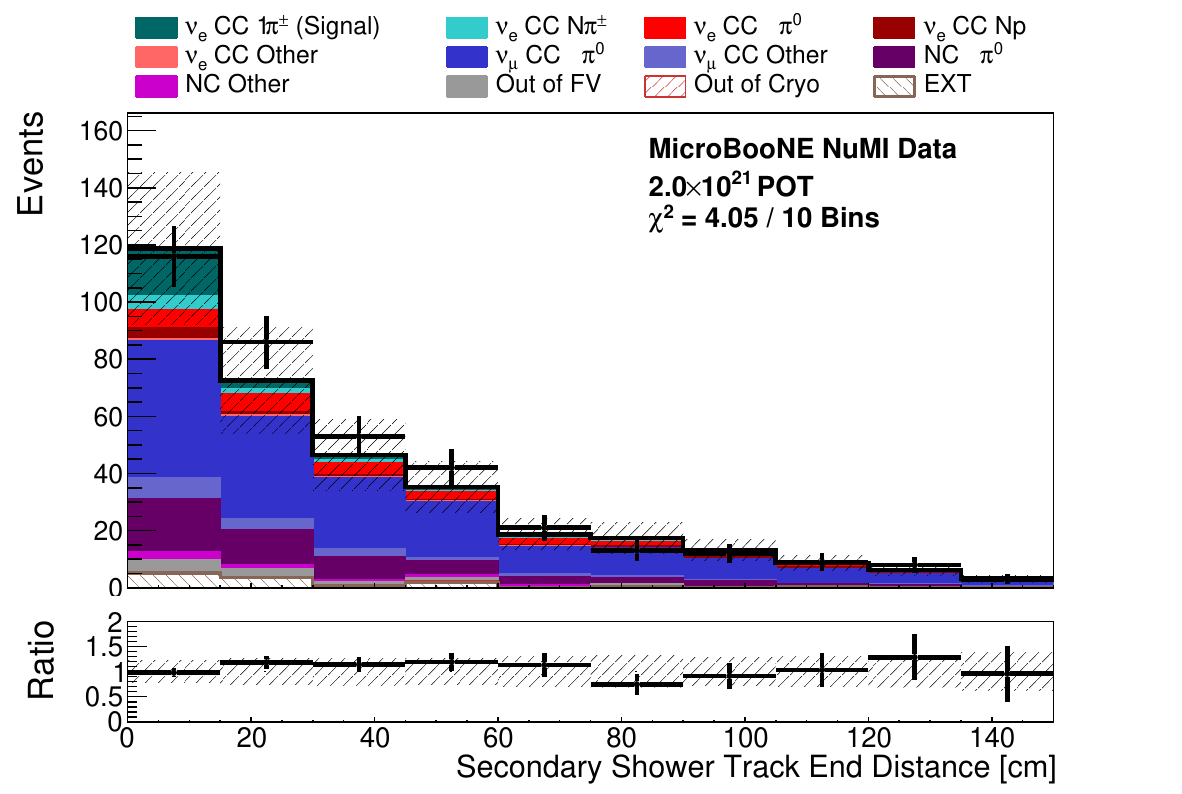}};
\draw (0.6, 1.0) node {\textbf{(c)}};
\end{tikzpicture}
\begin{tikzpicture} 
\draw (0, 0) node[inner sep=0] {\includegraphics[width=0.495\textwidth]{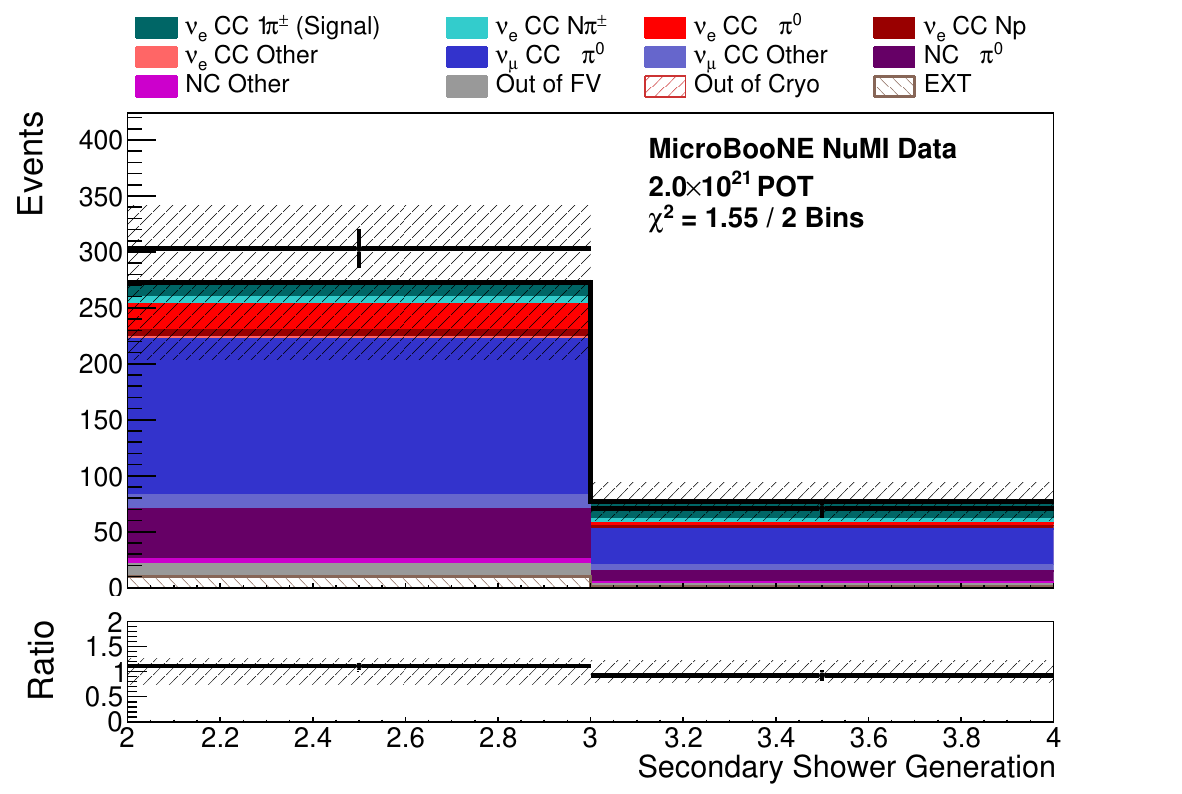}};
\draw (0.6, 1.0) node {\textbf{(d)}};
\end{tikzpicture}
\begin{tikzpicture} 
\draw (0, 0) node[inner sep=0] {\includegraphics[width=0.495\textwidth]{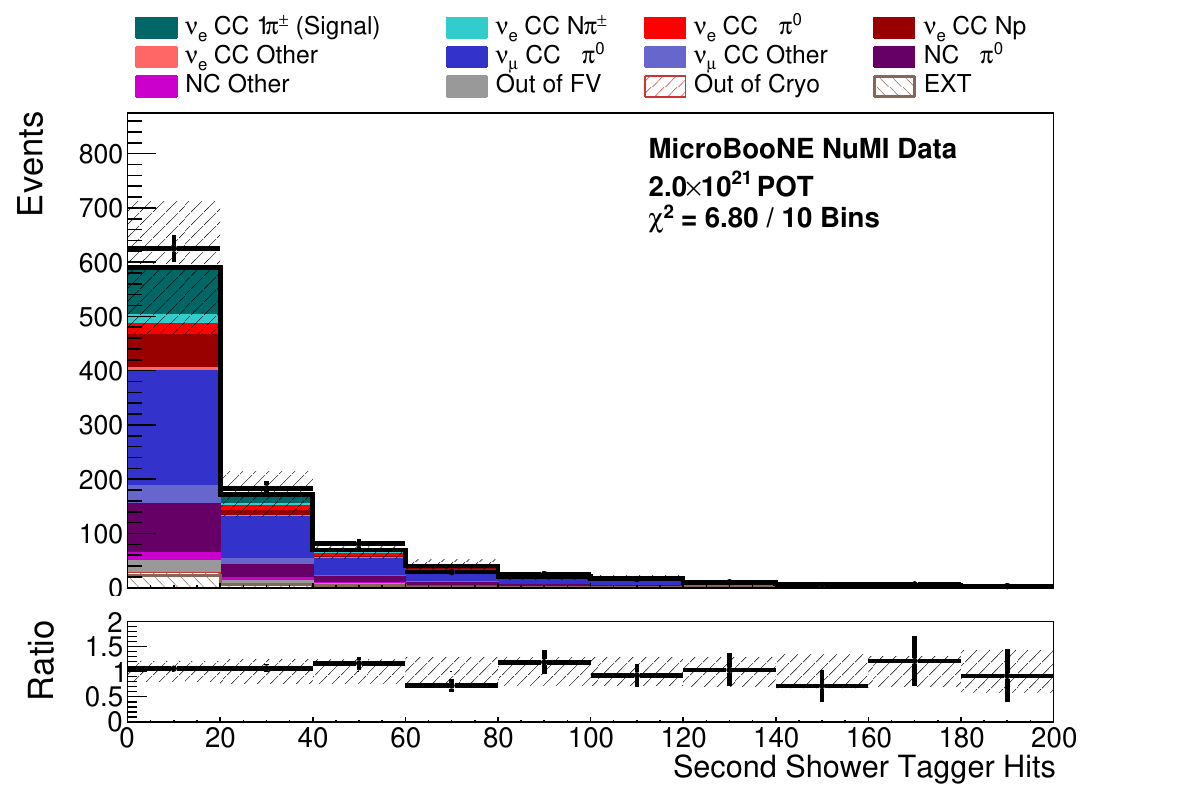}};
\draw (0.6, 1.0) node {\textbf{(e)}};
\end{tikzpicture}
\begin{tikzpicture} 
\draw (0, 0) node[inner sep=0] {\includegraphics[width=0.495\textwidth]{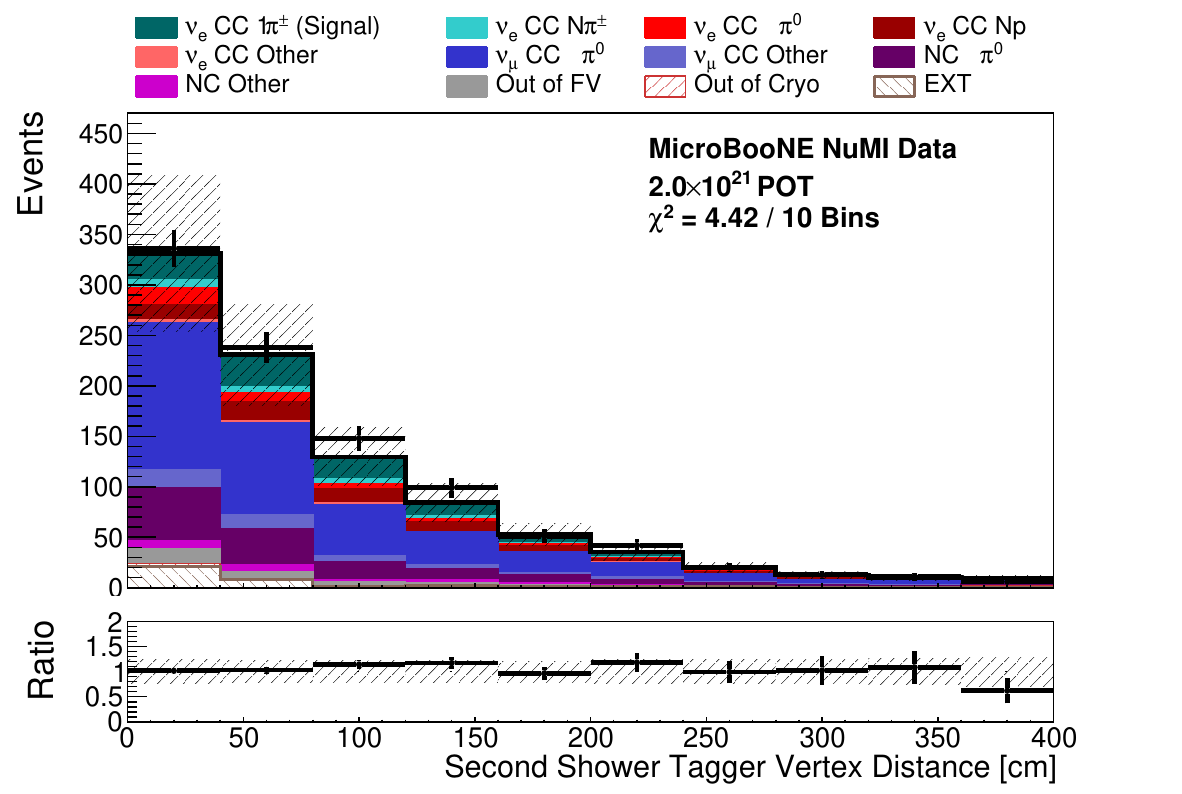}};
\draw (0.6, 1.0) node {\textbf{(f)}};
\end{tikzpicture}
\caption{Event distributions for the $\pi^0$-rejection BDT input variables (a) Secondary Shower Vertex Distance, (b) Secondary Shower Opening Angle, (c) Secondary Shower Track End Distance, (d) Secondary Shower Generation, (e) Second Shower Tagger Hits and (f) Second Shower Tagger Vertex Distance. The shaded band shows the systematic and statistical uncertainty on the MC prediction and the black points show the data with statistical uncertainties.}
\label{fig:ShowerBDTVars_3}
\end{figure*}

\begin{figure*}[htb]
\centering
\begin{tikzpicture} 
\draw (0, 0) node[inner sep=0] {\includegraphics[width=0.495\textwidth]{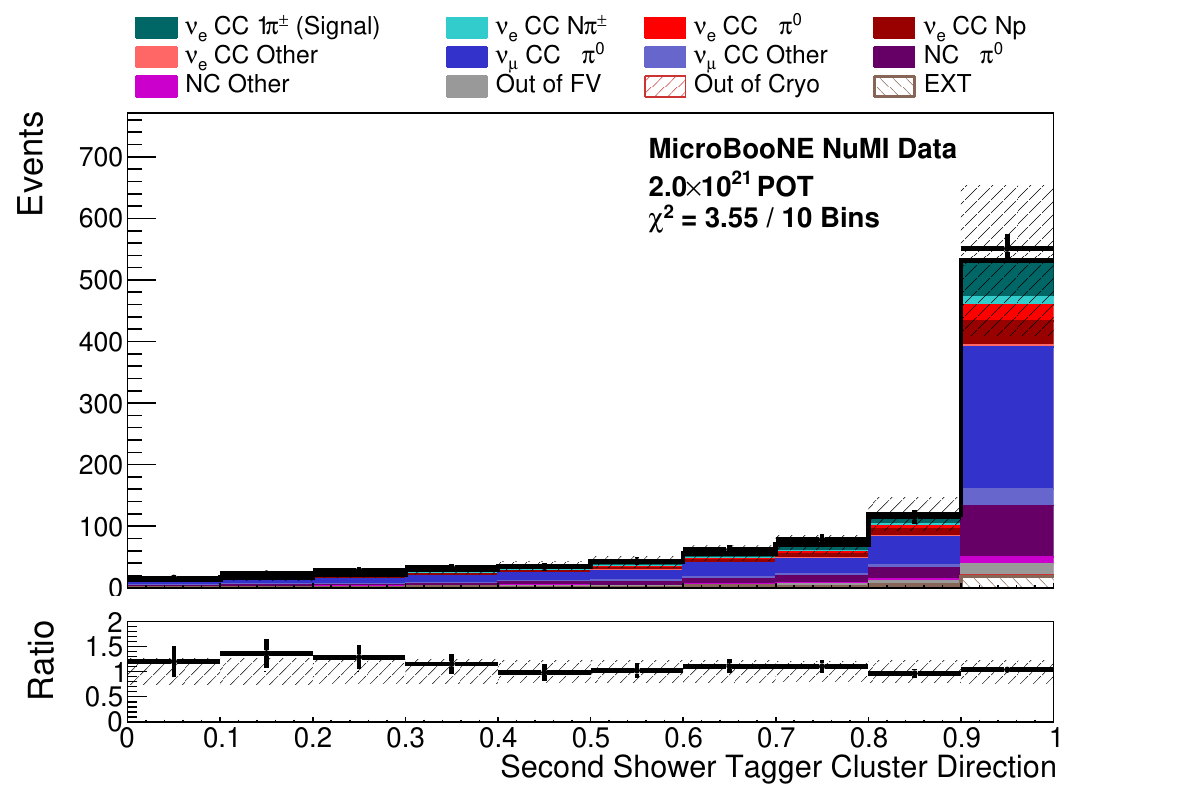}};
\draw (0.6, 1.0) node {\textbf{(a)}};
\end{tikzpicture}
\begin{tikzpicture} 
\draw (0, 0) node[inner sep=0] {\includegraphics[width=0.495\textwidth]{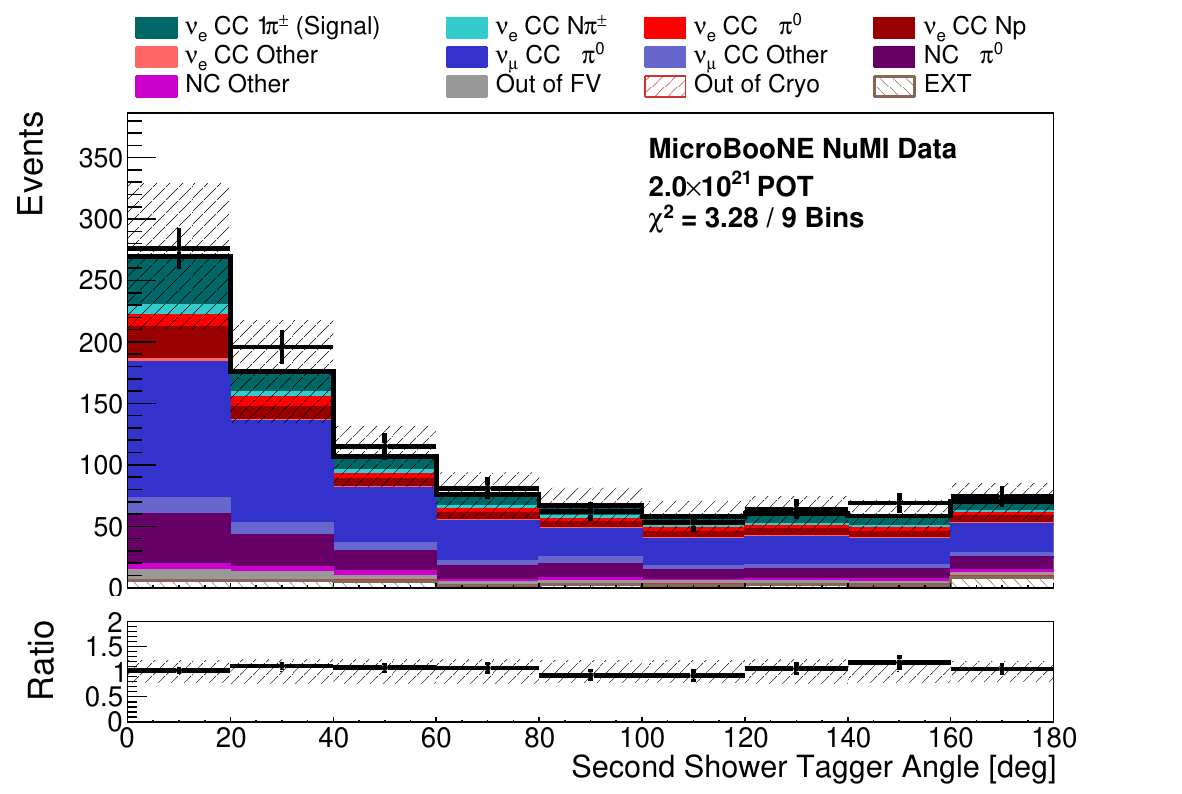}};
\draw (0.6, 1.0) node {\textbf{(b)}};
\end{tikzpicture}
\caption{Event distributions for the $\pi^0$-rejection BDT input variables (a) Second Shower Tagger Cluster Direction and (b) Second Shower Tagger Angle. The shaded band shows the systematic and statistical uncertainty on the MC prediction and the black points show the data with statistical uncertainties.}
\label{fig:ShowerBDTVars_4}
\end{figure*}

Figures~\ref{fig:ShowerBDTVars_1}, \ref{fig:ShowerBDTVars_2}, \ref{fig:ShowerBDTVars_3} and \ref{fig:ShowerBDTVars_4} show the event distributions for the $\pi^0$-rejection BDT input variables. The plots are shown after the pre-selection and loose background rejection cuts have been applied. Across all variables good data-MC agreement is seen within statistical and systematic uncertainties. 

\clearpage

\section{$\nuewithbar$ CC N$p$-rejection BDT Variables}
\label{TrackBDT}

The variables used in the $\nuewithbar$ CC N$p$-rejection BDT are summarized in Table~\ref{tab:BDT2Variables}. The majority of protons are rejected during the loose rejection stage by cutting on the Track dE/dx Profile Proton-Muon Hypothesis Score~\cite{MicroBooNE:2021ddy}. The remaining protons are those that do not have a clear reconstructed Bragg peak. The BDT aims to distinguish these from pion tracks by looking at the dE/dx profile under various hypotheses as well as looking for evidence activity near the ends of tracks that could indicate pion re-interactions or decays. The most powerful distinguishing variables are the dE/dx compared with a minimally ionizing particle hypothesis; the average dE/dx over the first third of the track; and the number of daughter particles of the track identified by \texttt{Pandora}.

\begin{table}[htb]
    \centering
    \begin{tabular}{|l|l|}
    \hline
    {\bf Variable} & {\bf Description} \\
    \hline
    \texttt{Pandora} Track Score & \makecell[l]{\texttt{Pandora} Track-Shower score. \\ Previous selection cut applied requiring score $>$ 0.5.} \\
    Track dE/dx Profile Proton-Muon Hypothesis Score &  \makecell[l]{Particle identification score with proton versus muon Bragg peak \\ hypothesis. Previous selection cut applied at score $>$ 0.}  \\
    Track dE/dx Profile MIP Hypothesis Score  &  \makecell[l]{Particle identification score with minimally ionizing particle (MIP) \\ dE/dx hypothesis.} \\
    Track dE/dx Profile Pion Hypothesis Score  &  Particle identification score with pion Bragg peak hypothesis. \\
    Track Start dE/dx  &  Average dE/dx over the first third of the track. \\
    Track End Number Daughter Particles  &  Number of daughter particles of the track reconstructed by \texttt{Pandora}. \\
    Track End Number Spacepoints & \makecell[l]{Number of reconstruct 3D hits (space-points) near the track end. \\ Metric for activity from decays or re-interactions missed by \texttt{Pandora}.} \\
    \hline
    \end{tabular}
    \caption{Variables used in the $\nuewithbar$ CC N$p$-rejection BDT.}
    \label{tab:BDT2Variables}
\end{table}

\begin{figure*}[htb]
\centering
\begin{tikzpicture} 
\draw (0, 0) node[inner sep=0] {\includegraphics[width=0.495\textwidth]{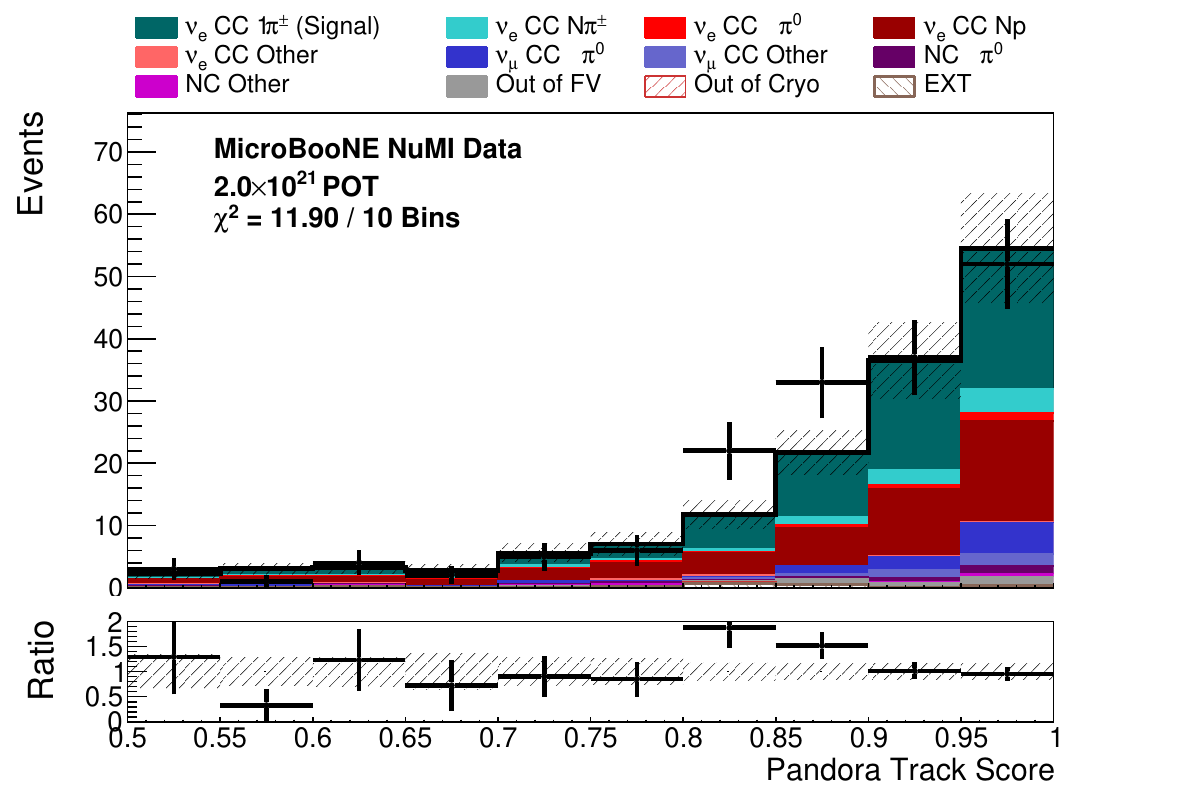}};
\draw (-2.65, 1.0) node {\textbf{(a)}};
\end{tikzpicture}
\begin{tikzpicture} 
\draw (0, 0) node[inner sep=0] {\includegraphics[width=0.495\textwidth]{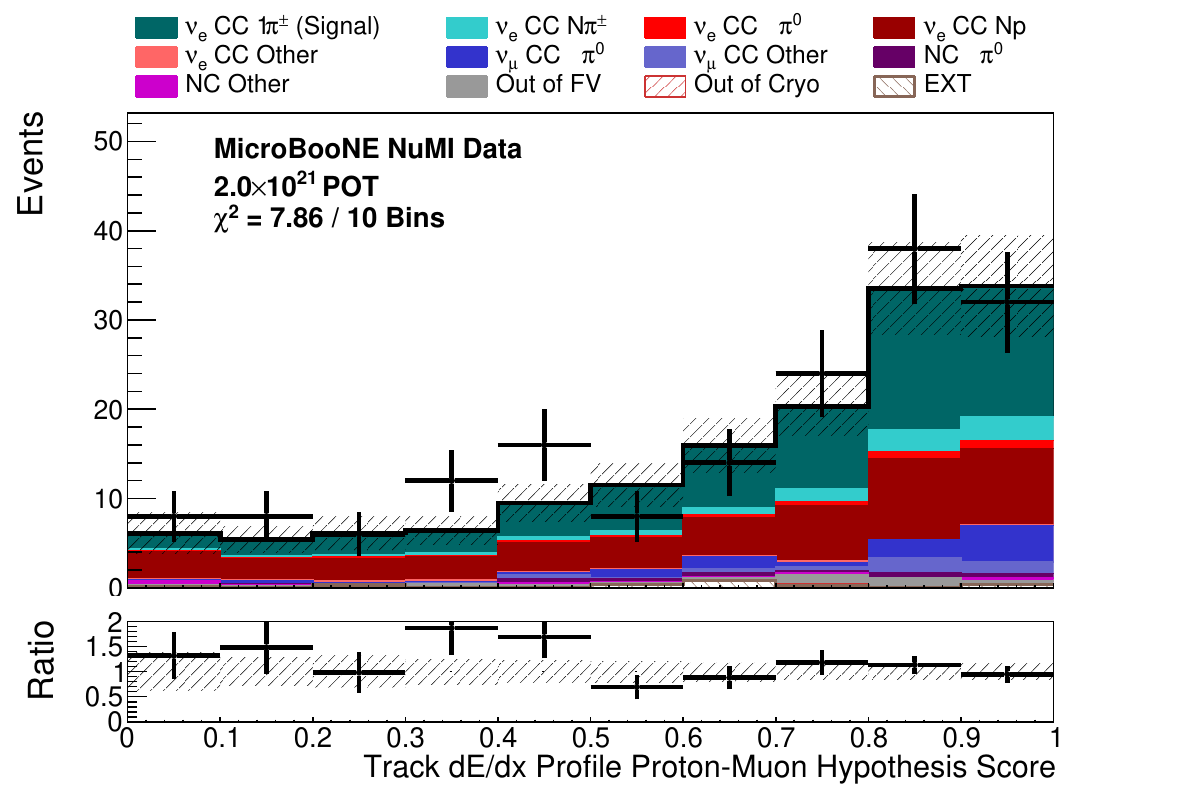}};
\draw (-2.65, 1.0) node {\textbf{(b)}};
\end{tikzpicture}
\caption{Event distributions for the $\nuewithbar$ CC N$p$-rejection BDT input variables (a) \texttt{Pandora} Track Score and (b) Track dE/dx Profile Proton-Muon Hypothesis Score. The shaded band shows the systematic and statistical uncertainty on the MC prediction and the black points show the data with statistical uncertainties.}
\label{fig:TrackBDTVars_1}
\end{figure*}

\begin{figure*}[htb]
\centering
\begin{tikzpicture} 
\draw (0, 0) node[inner sep=0] {\includegraphics[width=0.495\textwidth]{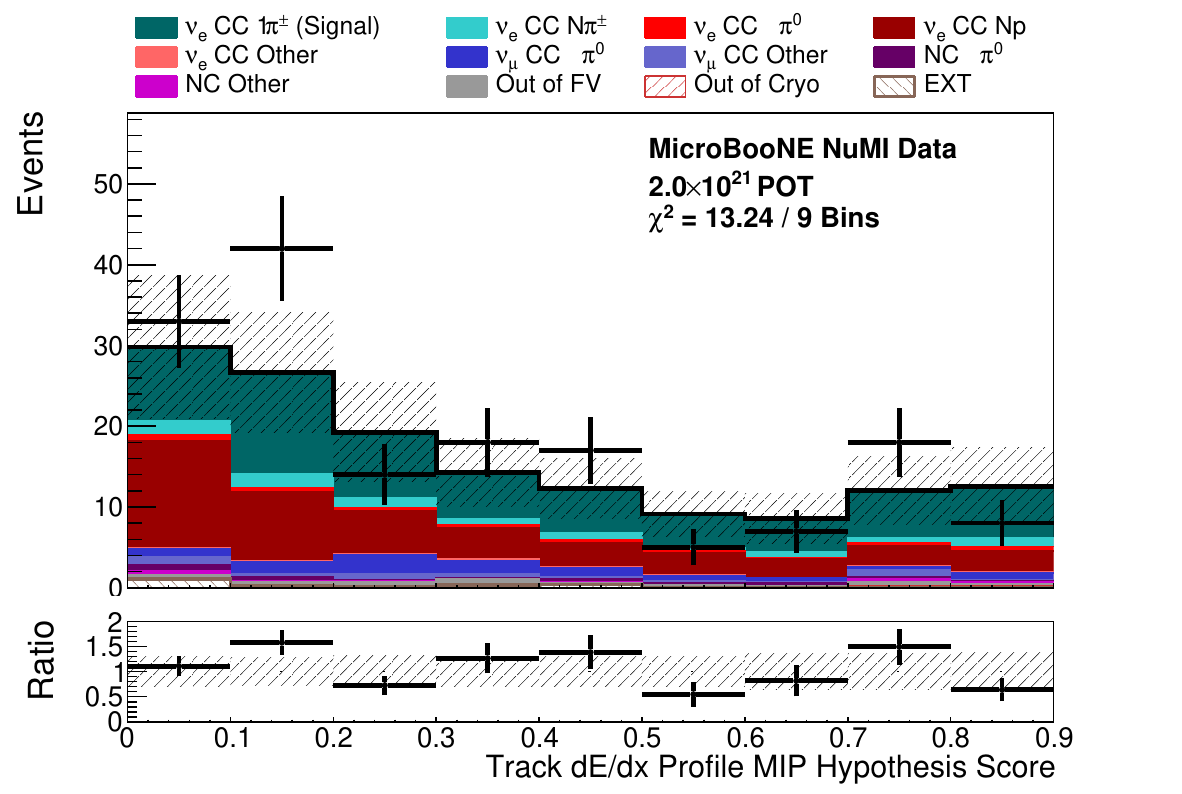}};
\draw (0.6, 1.0) node {\textbf{(a)}};
\end{tikzpicture}
\begin{tikzpicture} 
\draw (0, 0) node[inner sep=0] {\includegraphics[width=0.495\textwidth]{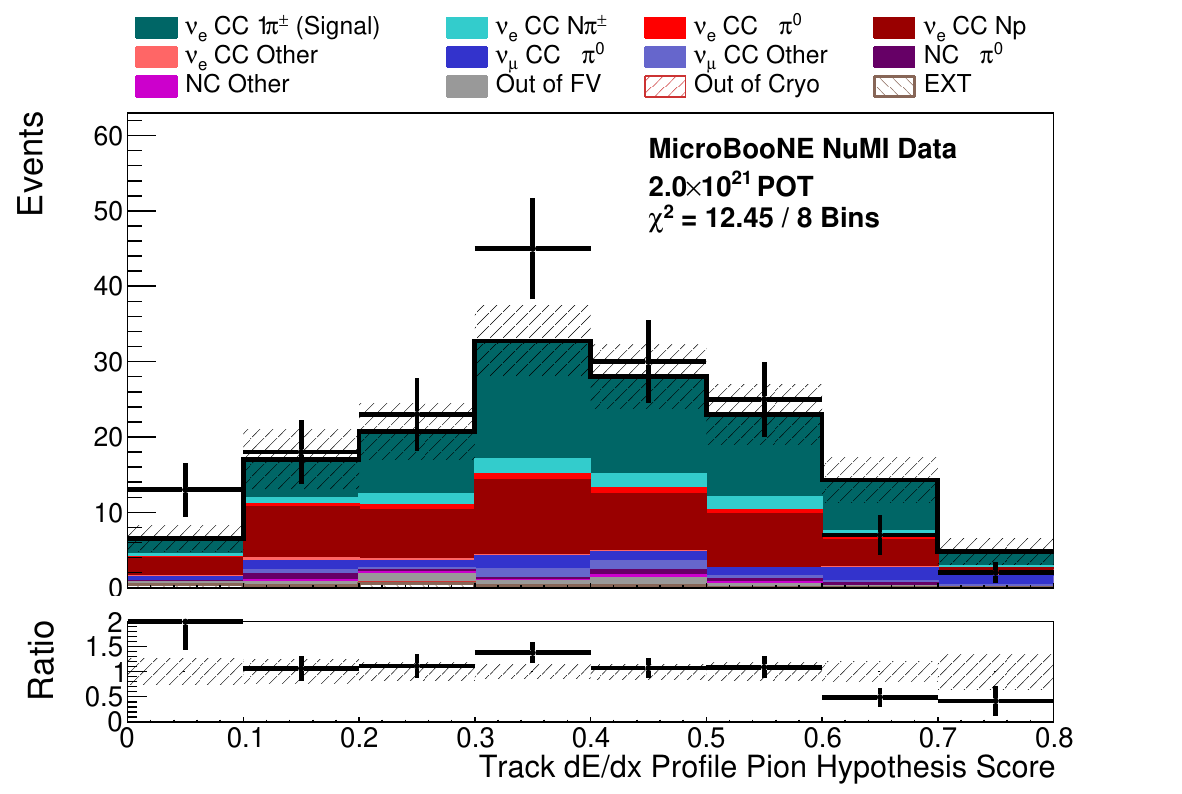}};
\draw (0.6, 1.0) node {\textbf{(b)}};
\end{tikzpicture}
\begin{tikzpicture} 
\draw (0, 0) node[inner sep=0] {\includegraphics[width=0.495\textwidth]{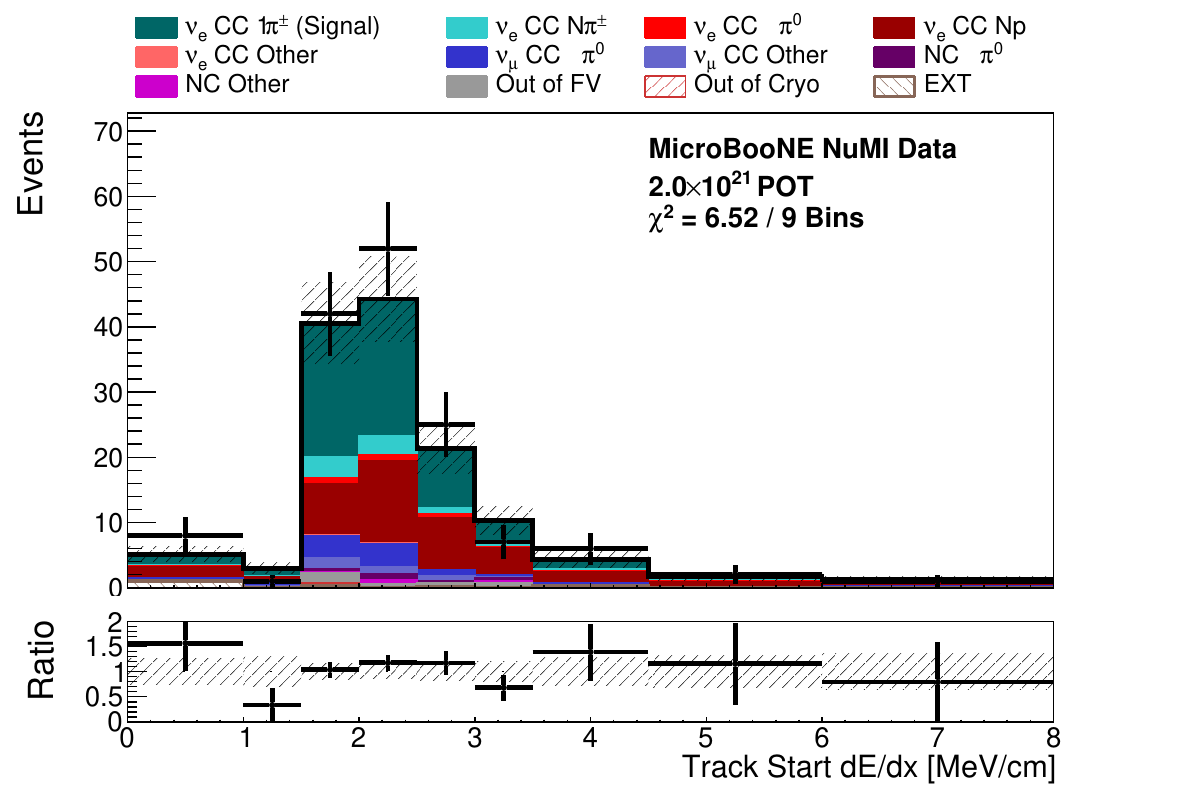}};
\draw (0.6, 1.0) node {\textbf{(c)}};
\end{tikzpicture}
\begin{tikzpicture} 
\draw (0, 0) node[inner sep=0] {\includegraphics[width=0.495\textwidth]{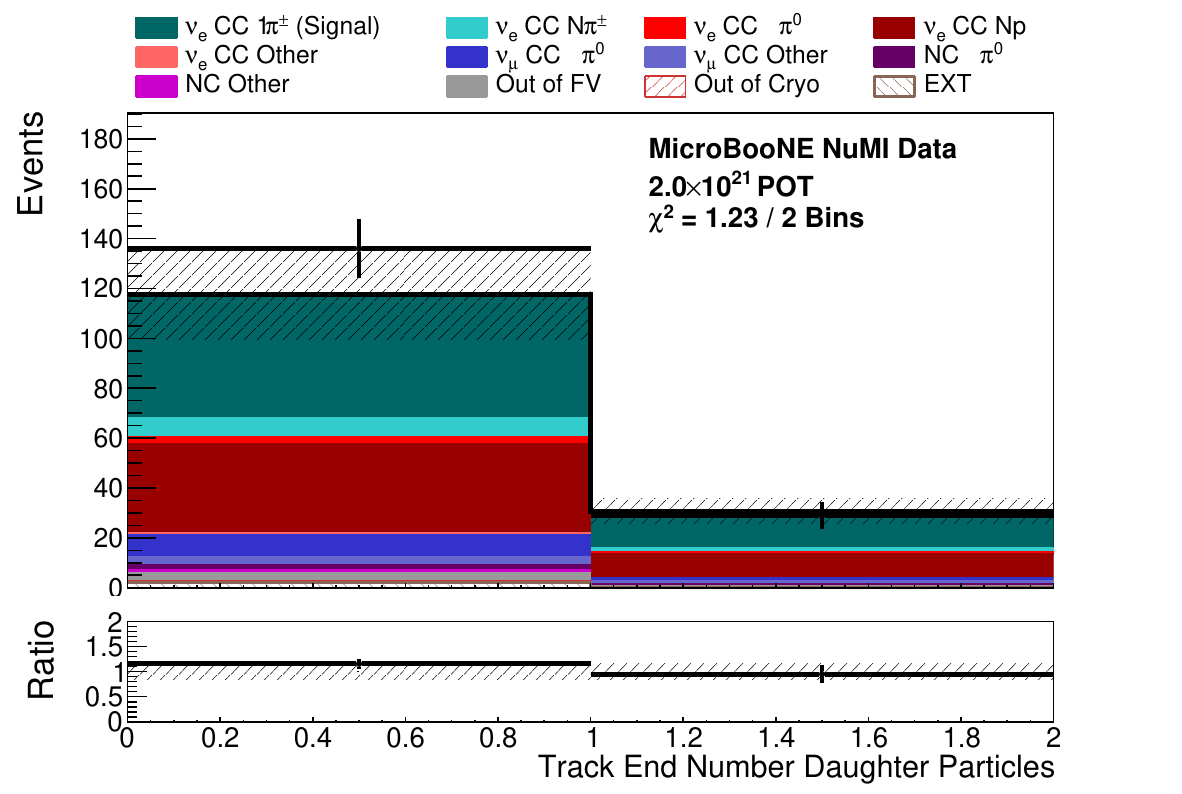}};
\draw (0.6, 1.0) node {\textbf{(d)}};
\end{tikzpicture}
\begin{tikzpicture} 
\draw (0, 0) node[inner sep=0] {\includegraphics[width=0.495\textwidth]{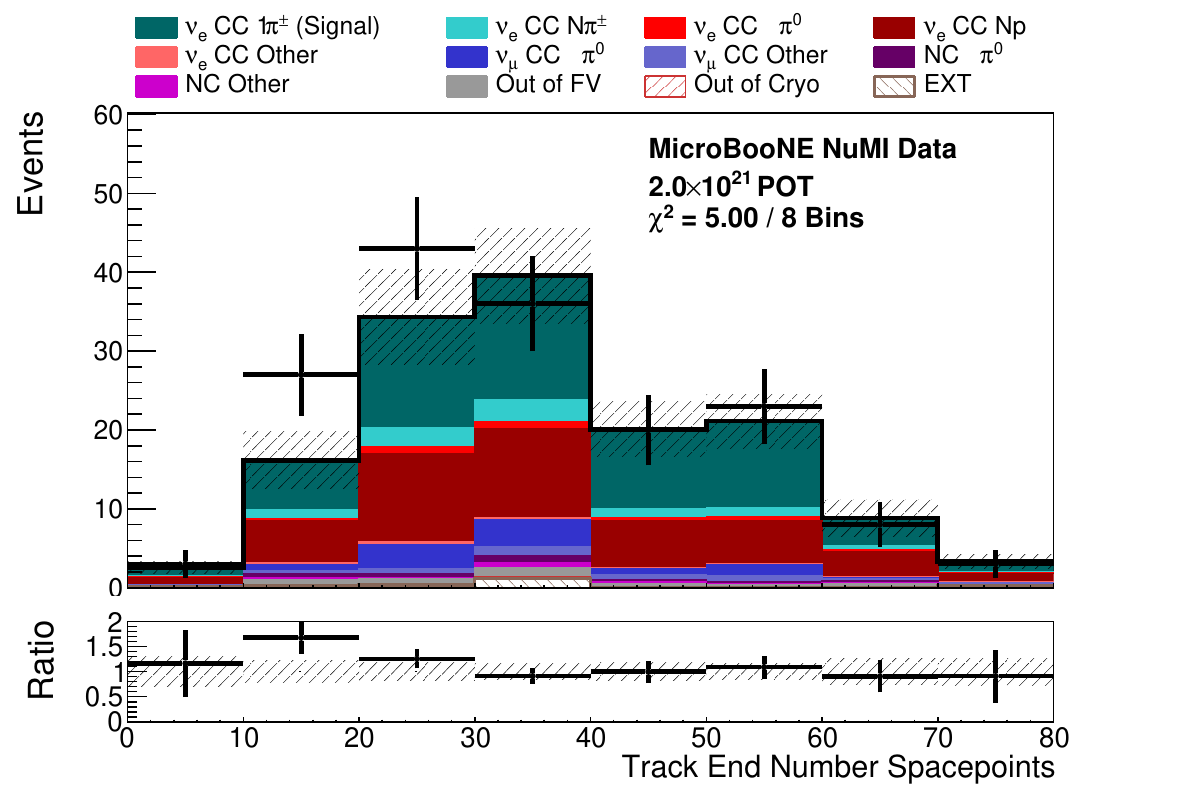}};
\draw (0.6, 1.0) node {\textbf{(e)}};
\end{tikzpicture}
\caption{Event distributions for the $\nuewithbar$ CC N$p$-rejection BDT input variables (a) Track dE/dx Profile MIP Hypothesis Score, (b) Track dE/dx Profile Pion Hypothesis Score, (c) Track Start dE/dx, (d) Track End Number Daughter Particles and (e) Track End Number Spacepoints. The shaded band shows the systematic and statistical uncertainty on the MC prediction and the black points show the data with statistical uncertainties.}
\label{fig:TrackBDTVars_2}
\end{figure*}

Figures~\ref{fig:TrackBDTVars_1} and \ref{fig:TrackBDTVars_2} show the event distributions for the $\nuewithbar$ CC N$p$-rejection BDT input variables. The plots are shown after the pre-selection, loose background rejection and $\pi^0$-rejection BDT cuts have been applied. Across all variables good data-MC agreement is seen within statistical and systematic uncertainties. 

\clearpage

\section{Reverse-horn-current BDT score distributions}

\begin{figure}[htb]
\centering
\begin{tikzpicture} 
\draw (0, 0) node[inner sep=0] {\includegraphics[width=0.495\textwidth]{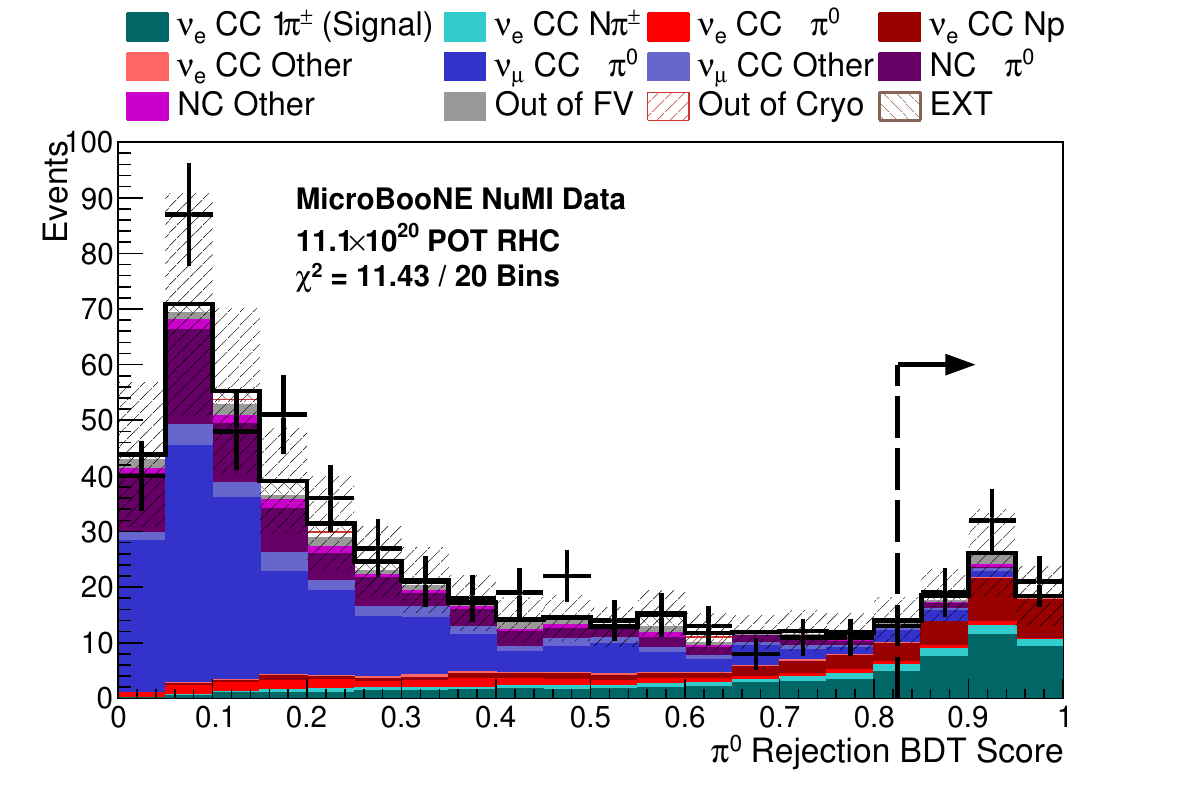}};
\draw (2.5, 1.5) node {\textbf{(a)}};
\end{tikzpicture}
\begin{tikzpicture} 
\draw (0, 0) node[inner sep=0] {\includegraphics[width=0.495\textwidth]{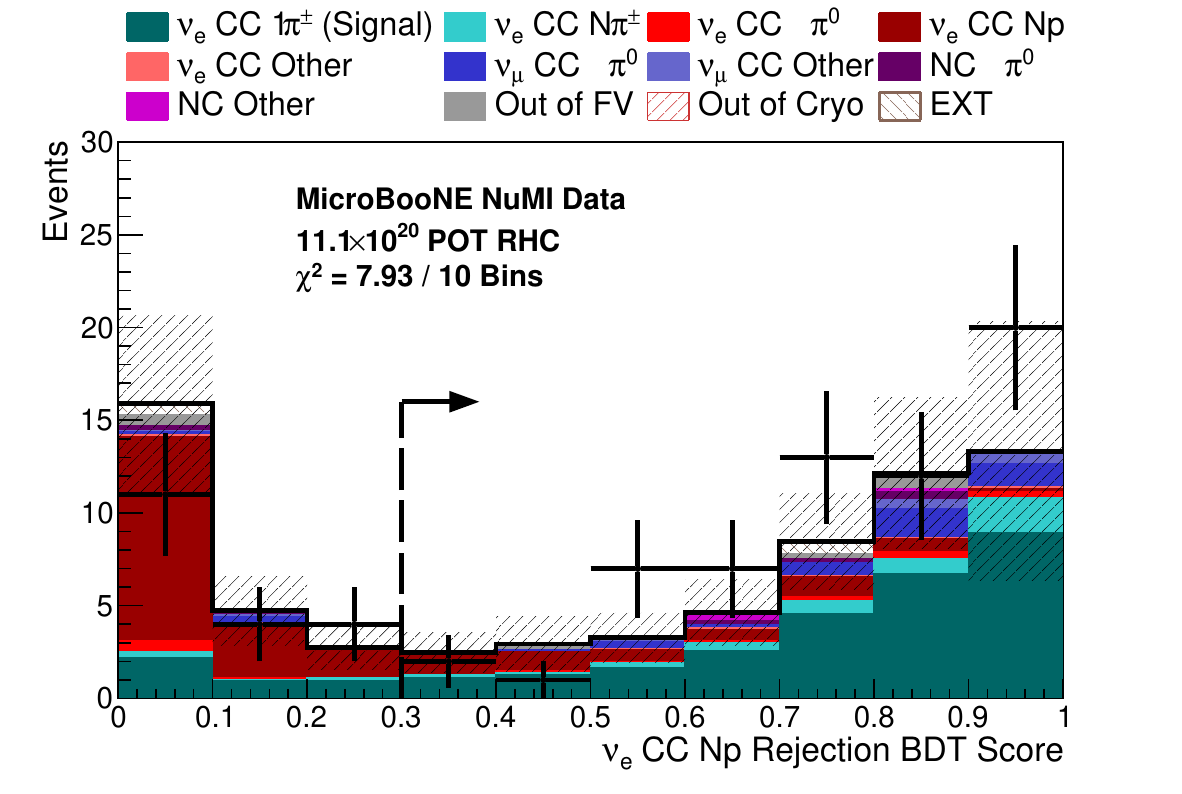}};
\draw (2.5, 1.5) node {\textbf{(b)}};
\end{tikzpicture}
\caption{Distribution of scores for (a) the $\pi^0$-rejection BDT and (b) the $\nuewithbar$ CC N$p$-rejection BDT compared with RHC mode data. The shaded band shows the systematic and statistical uncertainty on the MC prediction and the black points show the data with statistical uncertainties. The dashed lines show the cuts applied, where events to the right are selected.}
\label{fig:RHCBDTsWithData}
\end{figure}

Figure~\ref{fig:RHCBDTsWithData} shows the distribution of scores for the $\pi^0$-rejection BDT and the $\nuewithbar$ CC N$p$-rejection BDT compared with data for reverse-horn-current (RHC) mode. The $\nuewithbar$ CC N$p$-rejection BDT is shown after the cut on the $\pi^0$-rejection BDT has been applied. Both BDTs achieve good signal-background separation for their target topologies. In addition, excellent data--MC agreement within statistical and systematic uncertainties is seen across the full BDT score distributions.  

\clearpage

\section{Selection performance}

Tables~\ref{tab:fhc_cut_table} and \ref{tab:rhc_cut_table} show the efficiency and purity at each stage of the selection for FHC and RHC modes, respectively. The pre-selection requires that a neutrino candidate is identified by the \texttt{Pandora} reconstruction with a vertex within the fiducial volume and at least one shower and one track. In addition, the majority of energy depositions reconstructed in the event are required to be within the fiducial volume and associated with \texttt{Pandora} track or shower objects. The next two stages focus on the quality of the primary (most-energetic) reconstructed shower and the reconstructed tracks. The candidate particles are required to be identified as daughters of the neutrino candidate by \texttt{Pandora}, be reconstructed on all three wire planes, be contained within the detector, be identified as unambiguous showers or tracks by \texttt{Pandora}, and are checked for signs of common reconstruction failures. In addition, they are required to have energies and angles that are consistent with the signal definition. 

The next three selection stages focus on rejection of obvious background events. This then allows the BDT rejection to focus on the more challenging to reject event topologies. Cosmic-ray-induced interactions that haven't been removed by the previous selection stages are rejected through \texttt{Pandora} topological classification. Events containing neutral pions are rejected through loose cuts on the Primary Shower Vertex Distance, Primary Shower Moliere Average angle and the Primary Shower Energy Fraction. As previously mentioned, each of these variables are also provided to the subsequent $\pi^0$ rejection BDT to allow tighter cuts to be applied. Next, $\nuewithbar$ interactions containing protons but not pions are rejected by cutting on the Track dE/dx Profile Proton-Muon Hypothesis Score~\cite{MicroBooNE:2021ddy}.

Finally, the two BDTs described previously are applied. The first uses the reconstructed electromagnetic shower information to reject interactions containing neutral pions. The second uses reconstructed track information to reject $\nuewithbar$ CC N$p$ interactions.

\begin{table}[htb]
    \centering
    \begin{tabular}{|c|c|c|}
    \hline
    \multicolumn{3}{|c|}{\textbf{Forward Horn Current}} \\
    \hline
    Stage & Efficiency & Purity \\ \hline
    Pre-selection & 62.1\%  & 0.7\%   \\
    Candidate Shower Identification & 45.6\%  & 2.5\% \\
    Candidate Track Identification & 25.5\% & 3.7\% \\ 
    Cosmic Rejection & 22.0\% & 4.4\% \\
    Loose $\pi^0$ Rejection & 16.7\%  & 9.9\%  \\
    Loose $\nuewithbar$ CC N$p$ Rejection & 12.4\% & 12.8\%  \\
    BDT $\pi^0$ Rejection & 7.1\%  & 42.1\%  \\
    BDT $\nuewithbar$ CC N$p$ Rejection & 6.2\%  & 57.4\% \\
    \hline
    \end{tabular}
    \caption{Selection performance for forward horn current mode.}
    \label{tab:fhc_cut_table}
\end{table}

\begin{table}[htb]
    \centering
    \begin{tabular}{|c|c|c|}
    \hline
    \multicolumn{3}{|c|}{\textbf{Reverse Horn Current}} \\
    \hline
    Stage & Efficiency & Purity \\ \hline
    Pre-selection & 63.2\%  & 0.7\%   \\
    Candidate Shower Identification & 46.7\%  & 2.5\% \\
    Candidate Track Identification & 25.3\% & 3.7\% \\ 
    Cosmic Rejection & 21.8\% & 4.4\% \\
    Loose $\pi^0$ Rejection & 16.6\%  & 9.8\%  \\
    Loose $\nuewithbar$ CC N$p$ Rejection & 12.6\% & 12.5\%  \\
    BDT $\pi^0$ Rejection & 6.5\%  & 44.2\%  \\
    BDT $\nuewithbar$ CC N$p$ Rejection & 5.8\% & 58.5\% \\
    \hline
    \end{tabular}
    \caption{Selection performance for reverse horn current mode.}
    \label{tab:rhc_cut_table}
\end{table}

\clearpage

\section{Neutrino direction approximation}

\begin{figure}[htb]
\centering
\begin{tikzpicture} 
\draw (0, 0) node[inner sep=0] {\includegraphics[width=0.495\textwidth]{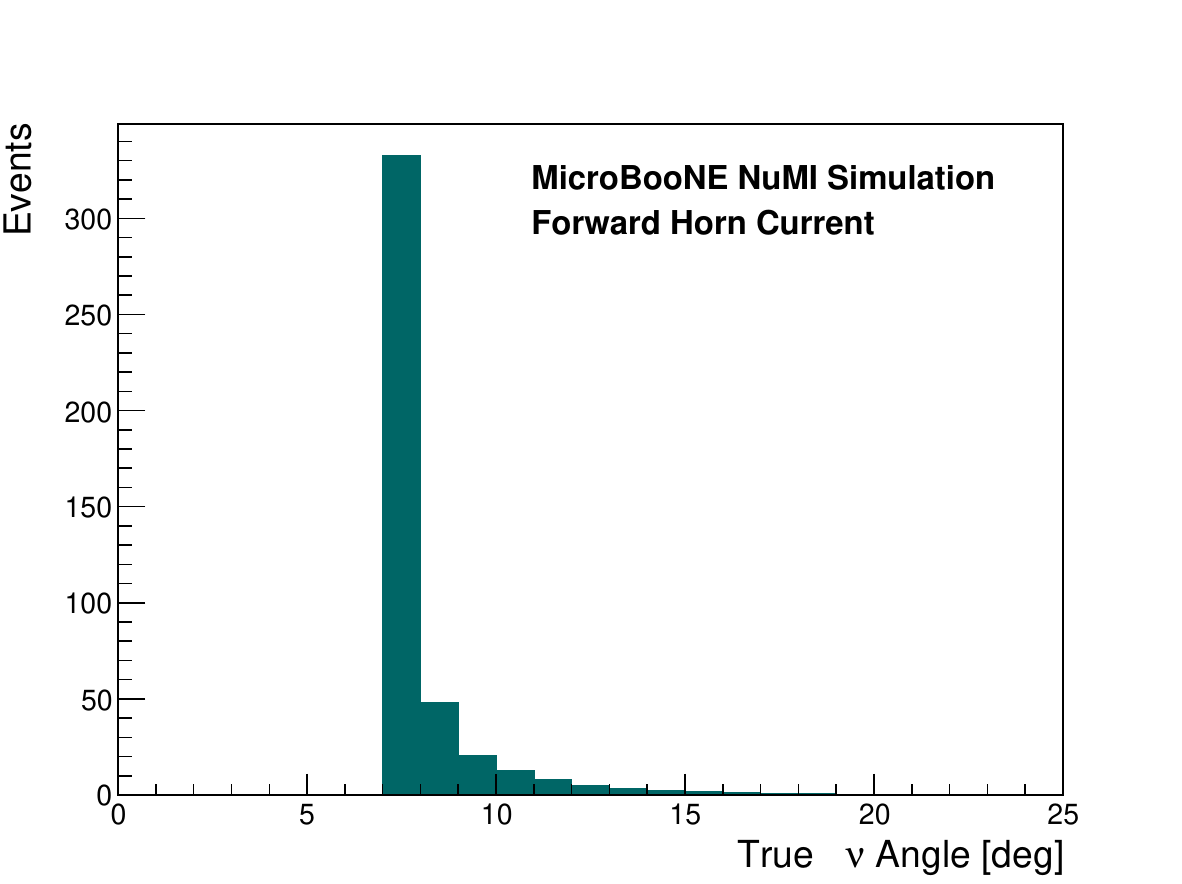}};
\draw (-2.8, 2.0) node {\textbf{(a)}};
\end{tikzpicture}
\begin{tikzpicture} 
\draw (0, 0) node[inner sep=0] {\includegraphics[width=0.495\textwidth]{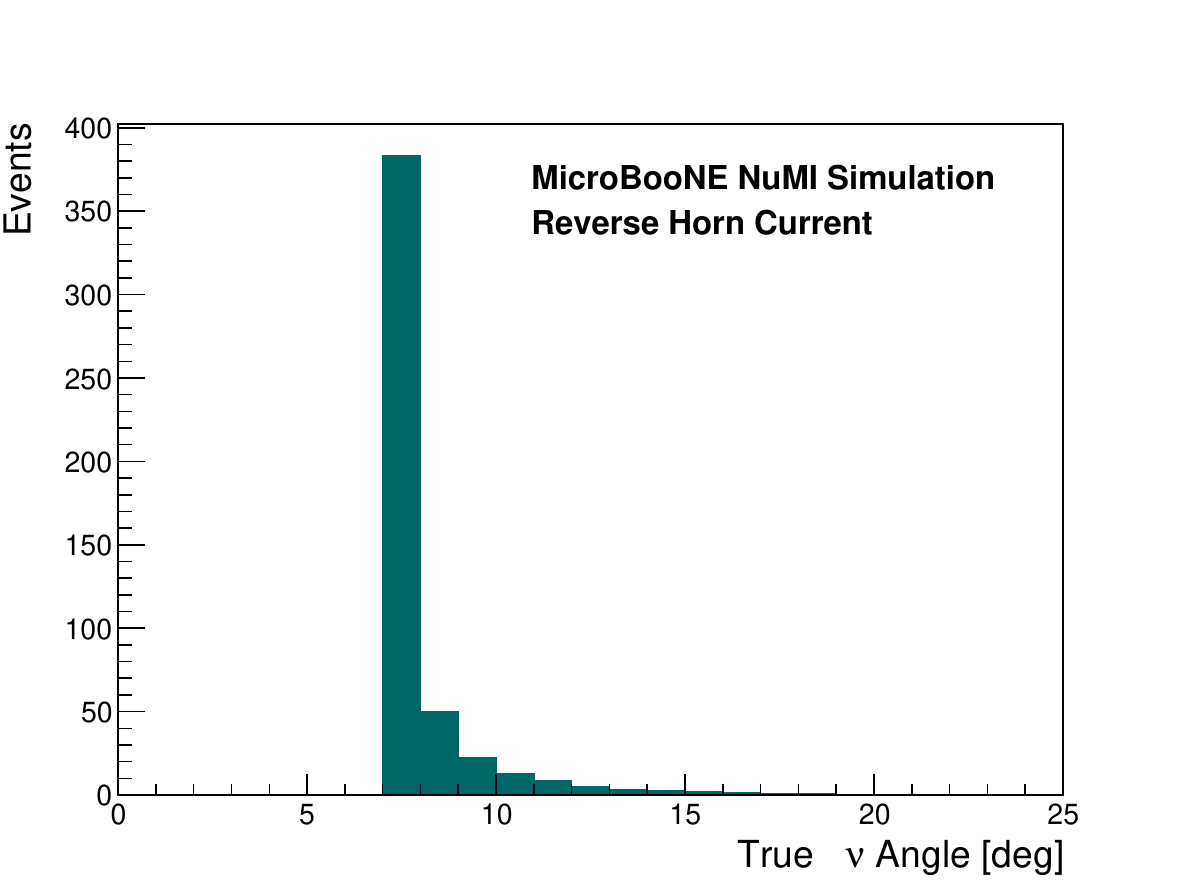}};
\draw (-2.8, 2.0) node {\textbf{(b)}};
\end{tikzpicture}
\caption{Angular distribution of signal $\nuewithbar \mathrm{CC}\,1\pi^\pm$ interactions for (a) FHC mode and (b) RHC mode.}
\label{fig:neutrinoDirection}
\end{figure}

As a result of the NuMI beam being off-axis to the MicroBooNE detector, neutrinos can arrive at MicroBooNE with a range of directions depending on where along the beam line they are produced. Figure~\ref{fig:neutrinoDirection} shows the angular distribution of signal $\nuewithbar \mathrm{CC}\,1\pi^\pm$ interactions for forward horn current (FHC) mode and RHC mode. The majority are produced close to the NuMI target approximately $8\degree$ off-axis. However, a subset are produced at larger off-axis angles as a result of meson decays further along the NuMI decay pipe. 

During reconstruction the exact neutrino production position and, hence, direction is not known. Instead, the neutrino direction is approximated by assuming that the neutrino is produced at the NuMI target location approximately $8\degree$ off-axis from MicroBooNE. This provides an effective approximation for reconstructing angles $\theta$ relative to the neutrino direction. For signal $\nuewithbar \mathrm{CC}\,1\pi^\pm$ interactions, 85\% have true neutrino directions within $1\degree$ of the approximated neutrino direction and 95\% within $3\degree$ of the approximated neutrino direction.

\begin{figure}
\centering
\begin{tikzpicture} 
\draw (0, 0) node[inner sep=0] {\includegraphics[width=0.495\textwidth]{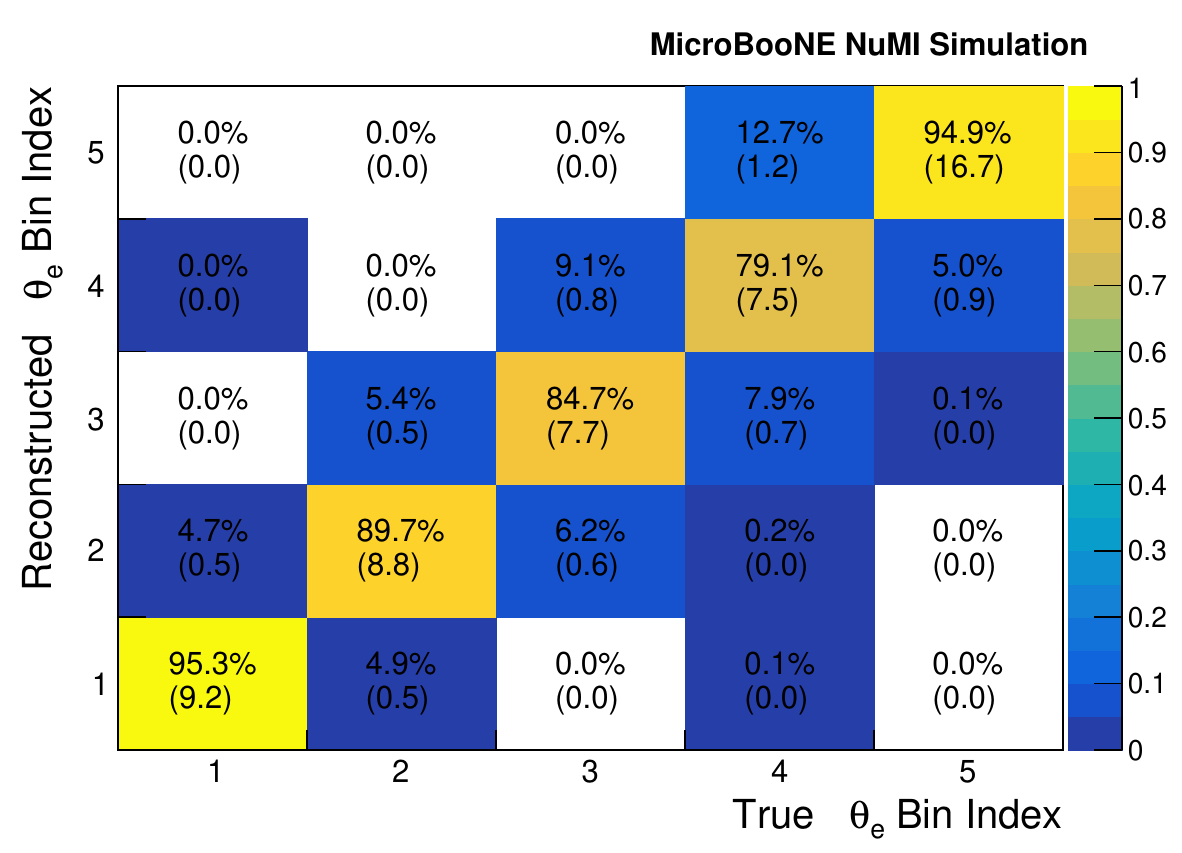}};
\draw (-2.8, 2.8) node {\textbf{(a)}};
\end{tikzpicture}
\begin{tikzpicture} 
\draw (0, 0) node[inner sep=0] {\includegraphics[width=0.495\textwidth]{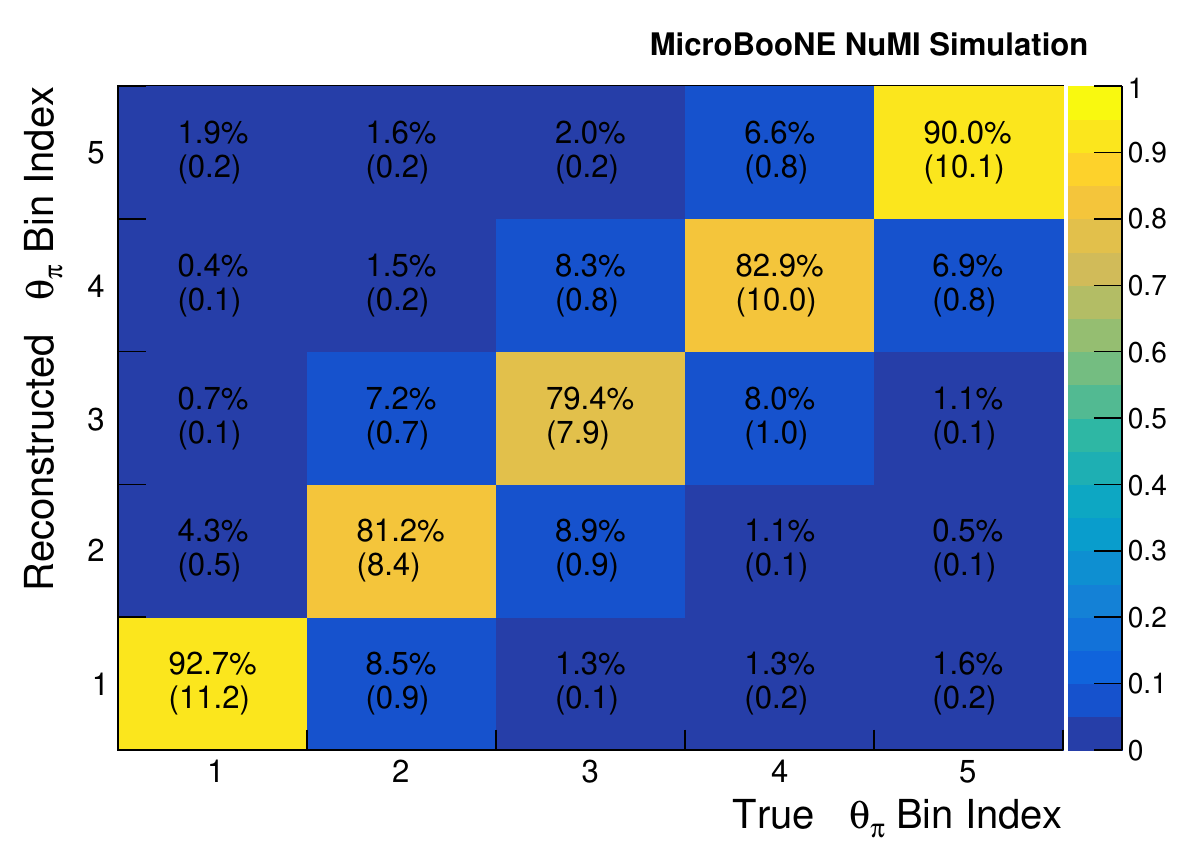}};
\draw (-2.8, 2.8) node {\textbf{(b)}};
\end{tikzpicture}
\caption{Smearing matrices for (a) the electron angle, $\theta_e$, and (b) the pion angle, $\theta_\pi$, for the binning schemes considered. The number of predicted signal events in each bin is shown in parentheses.}
\label{fig:betaResponseMatrices}
\end{figure}

For true $\theta$ angles, the true neutrino direction is used rather than the approximation. The smearing resulting from the approximation is then accounted for during unfolding allowing the extracted cross-sections to be reported in terms of true $\theta$ angles that are directly comparable with predictions of scattering angles from event generators. Figure~\ref{fig:betaResponseMatrices} shows the smearing matrices for the electron angle, $\theta_e$, and pion angle, $\theta_\pi$, for the binning schemes considered. These encode the impact of the neutrino direction approximation along with other detector response and reconstruction effects on these variables. Greater than 75\% of reconstructed angles fall within the correct truth bin for all bins considered indicating this approximation is performing well. The bin edges are tabulated in the data release at the end of the Supplemental Material.

\clearpage

\section{Selected Events}

\begin{figure*} [htb]
\centering
\begin{tikzpicture} 
\draw (0, 0) node[inner sep=0] {\includegraphics[width=0.495\textwidth]{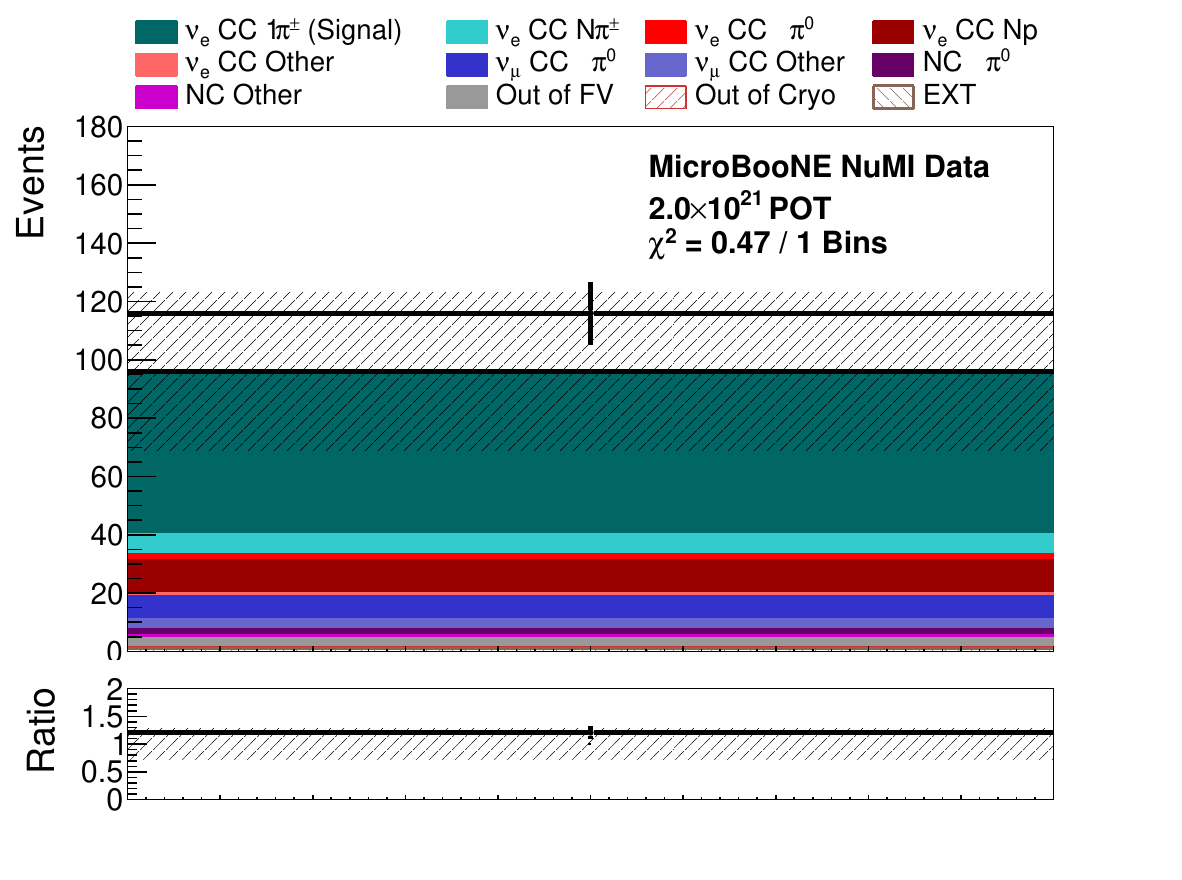}};
\end{tikzpicture}
\caption{Total selected events. The shaded band shows the systematic and statistical uncertainty on the MC prediction and the black points show the data with statistical uncertainties.}
\label{fig:Total}
\end{figure*}

\begin{figure*}[htb]
\centering
\begin{tikzpicture} 
\draw (0, 0) node[inner sep=0] {\includegraphics[width=0.495\textwidth]{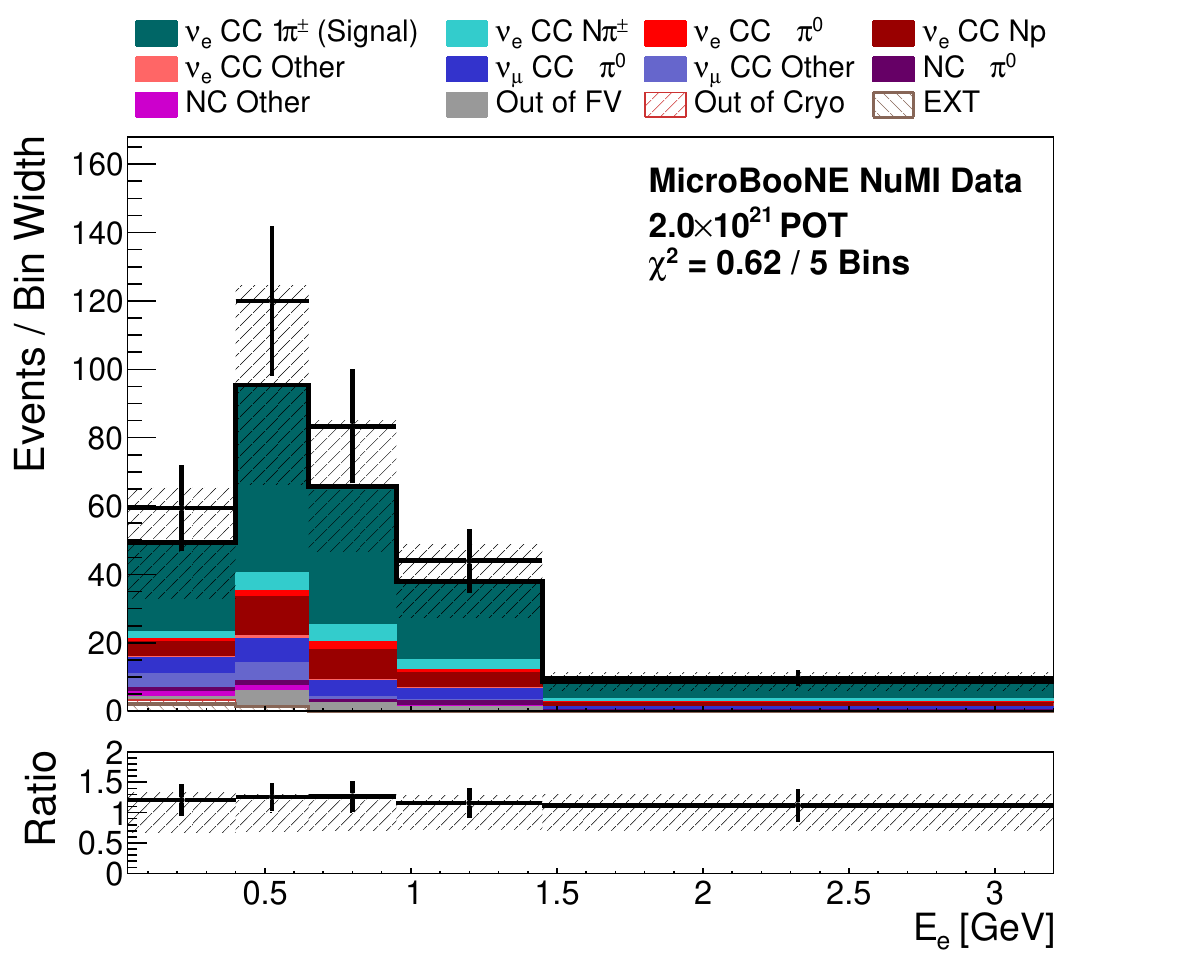}};
\draw (0.6, 1.0) node {\textbf{(a)}};
\end{tikzpicture}
\begin{tikzpicture} 
\draw (0, 0) node[inner sep=0] {\includegraphics[width=0.495\textwidth]{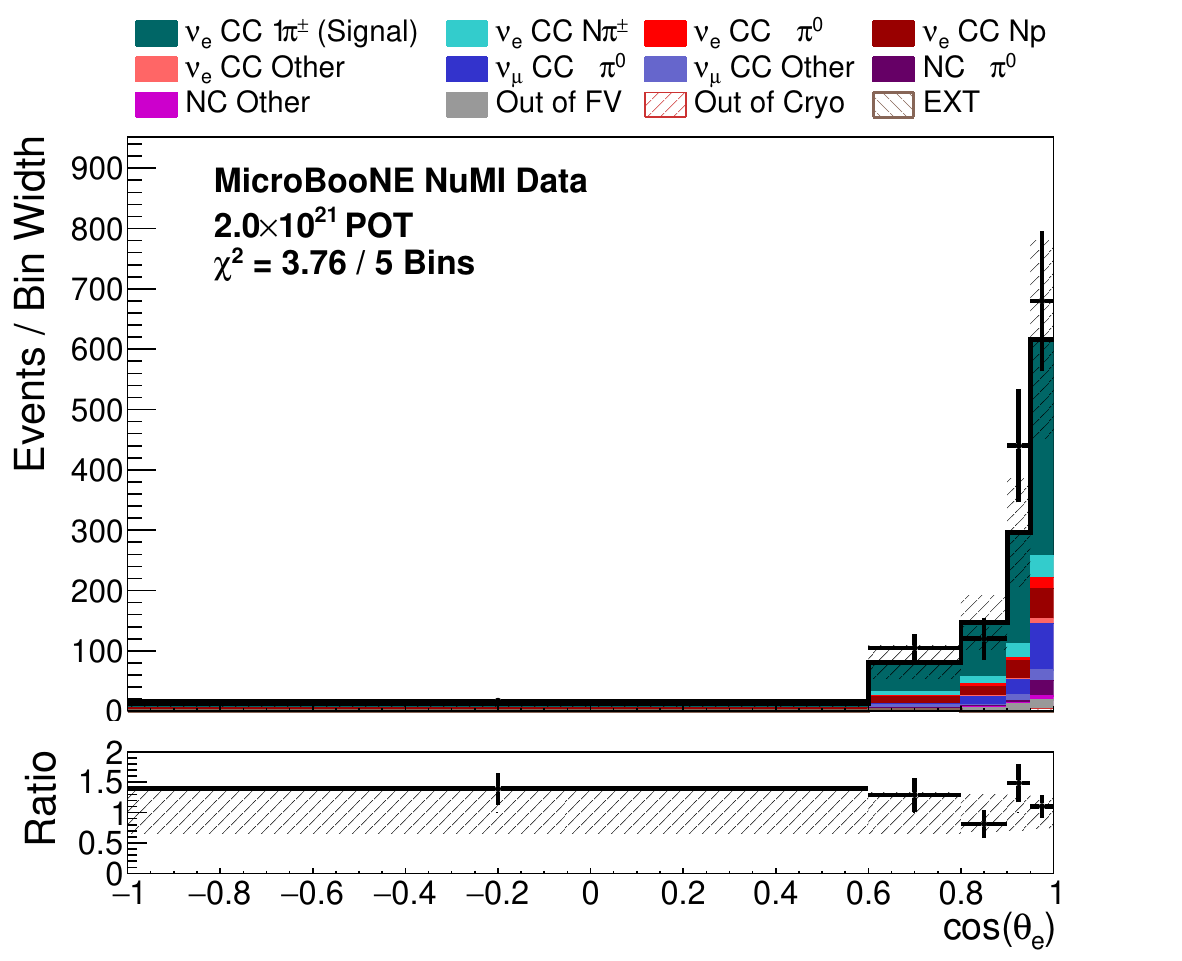}};
\draw (-2.65, 1.0) node {\textbf{(b)}};
\end{tikzpicture}
\begin{tikzpicture} 
\draw (0, 0) node[inner sep=0] {\includegraphics[width=0.495\textwidth]{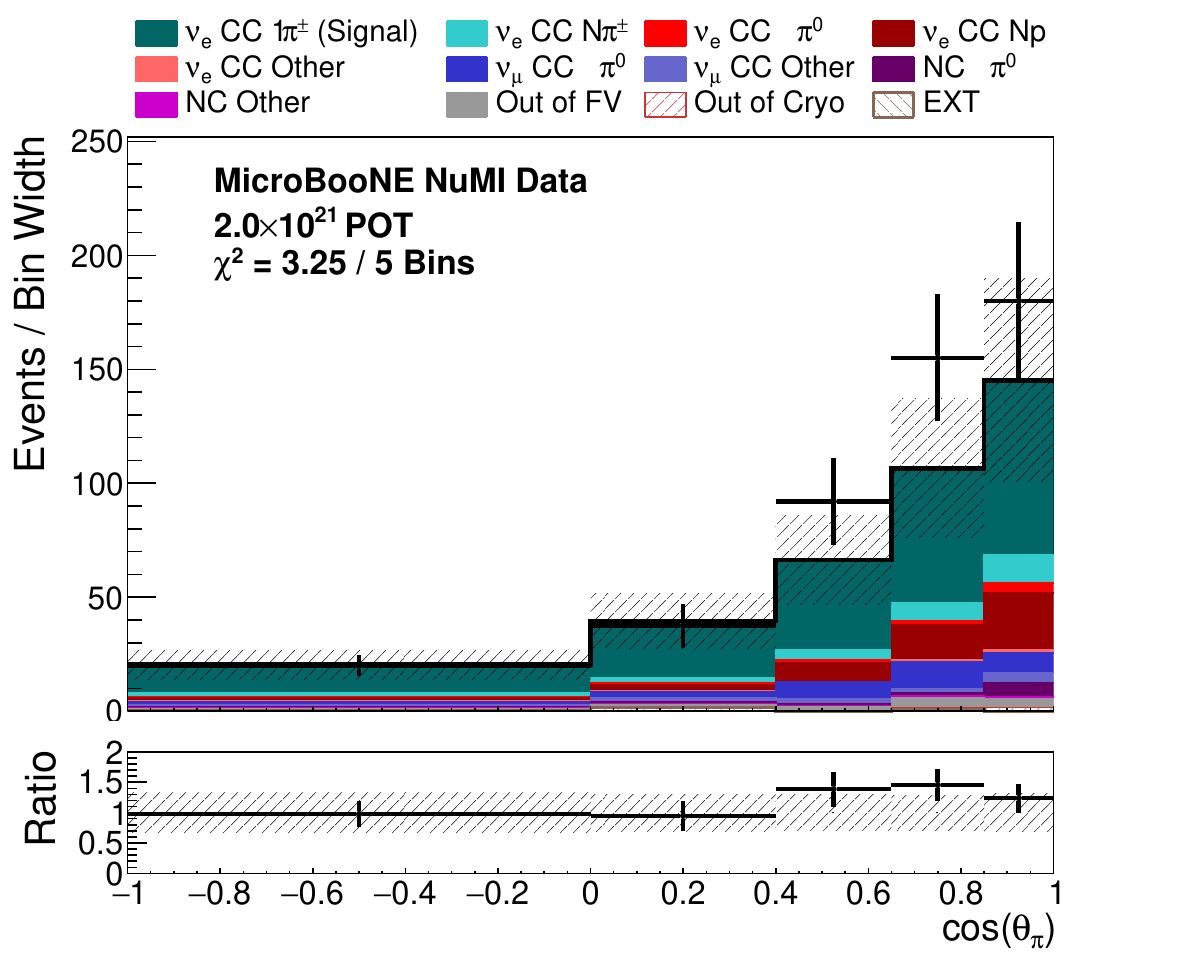}};
\draw (-2.65, 1.0) node {\textbf{(c)}};
\end{tikzpicture}
\begin{tikzpicture} 
\draw (0, 0) node[inner sep=0] {\includegraphics[width=0.495\textwidth]{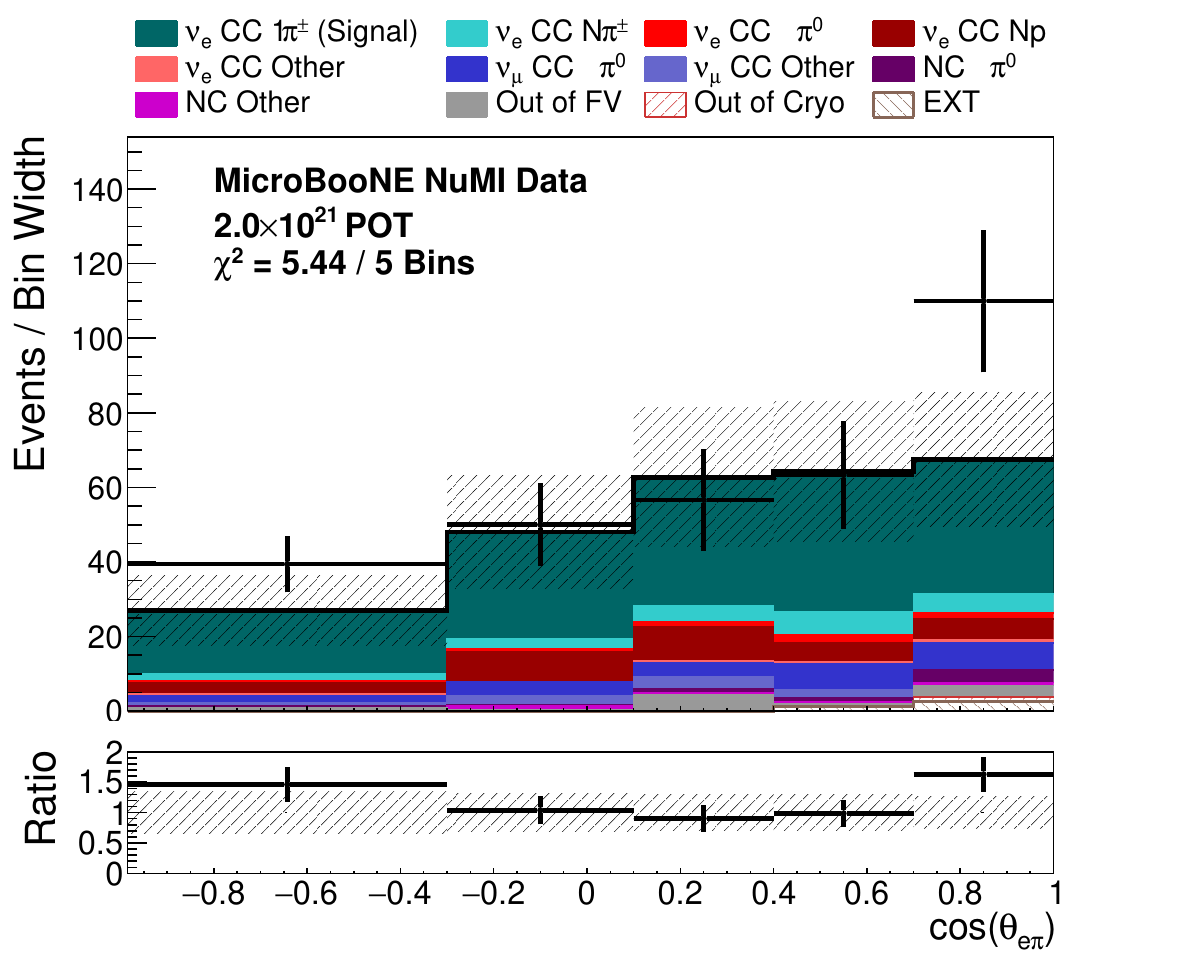}};
\draw (-2.65, 1.0) node {\textbf{(d)}};
\end{tikzpicture}
\caption{Selected event distributions for (a) electron energy, (b) electron angle, (c) pion angle and (d) electron-pion opening angle. The shaded band shows the systematic and statistical uncertainty on the MC prediction and the black points show the data with statistical uncertainties.}
\label{fig:SelectedEvents}
\end{figure*}

Figure~\ref{fig:Total} shows the total selected events and Fig.~\ref{fig:SelectedEvents} shows the selected event distributions for the electron energy, $E_e$, electron angle, $\theta_e$, pion angle, $\theta_\pi$, and electron-pion opening angle, $\theta_{e\pi}$. Each distribution is divided by bin width. Good data-MC agreement is seen across all distributions within statistical and systematic uncertainties. 

\clearpage

\section{Sidebands}

\begin{figure*} [htb]
\centering
\begin{tikzpicture} 
\draw (0, 0) node[inner sep=0] {\includegraphics[width=0.495\textwidth]{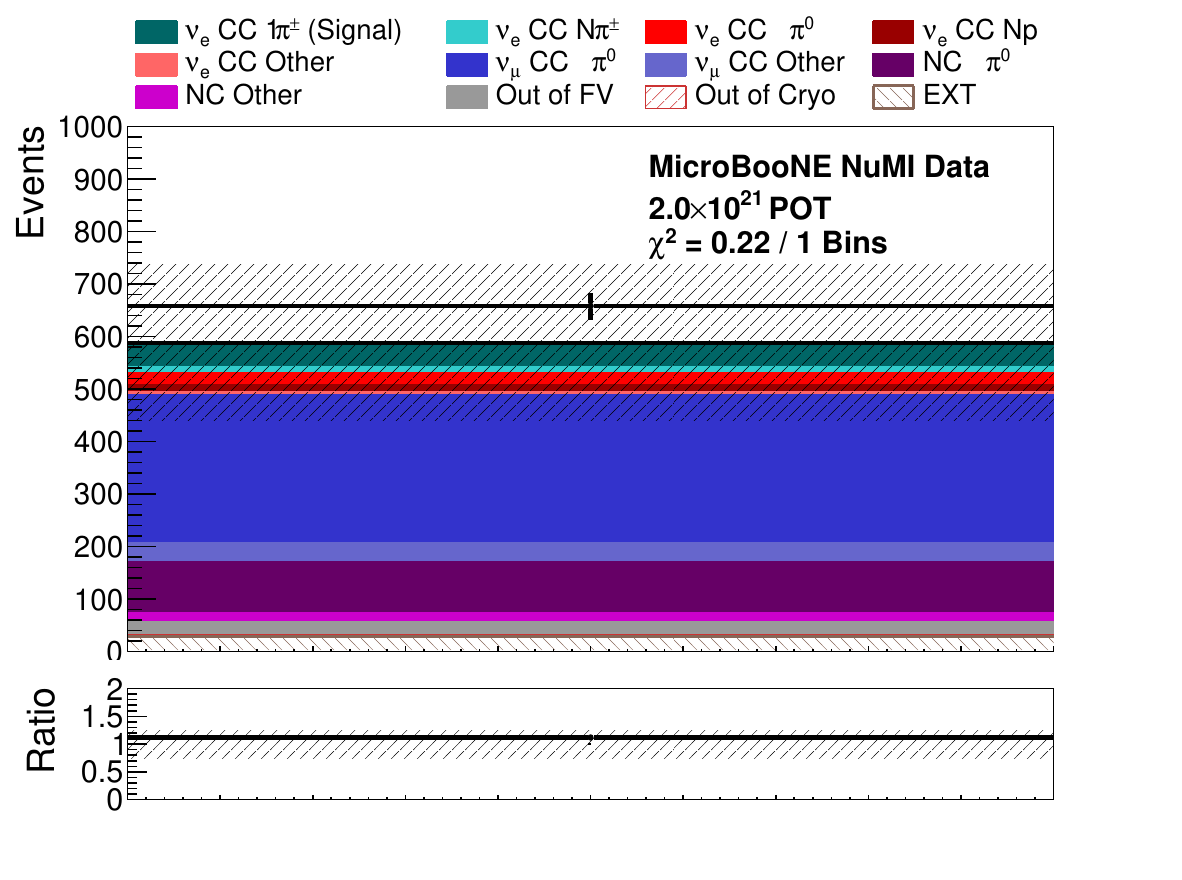}};
\end{tikzpicture}
\caption{Total selected event rate for the $\pi^0$-rich sideband. The shaded band shows the systematic and statistical uncertainty on the MC prediction and the black points show the data with statistical uncertainties.}
\label{fig:Sideband1Total}
\end{figure*}

\begin{figure*}[htb]
\centering
\begin{tikzpicture} 
\draw (0, 0) node[inner sep=0] {\includegraphics[width=0.495\textwidth]{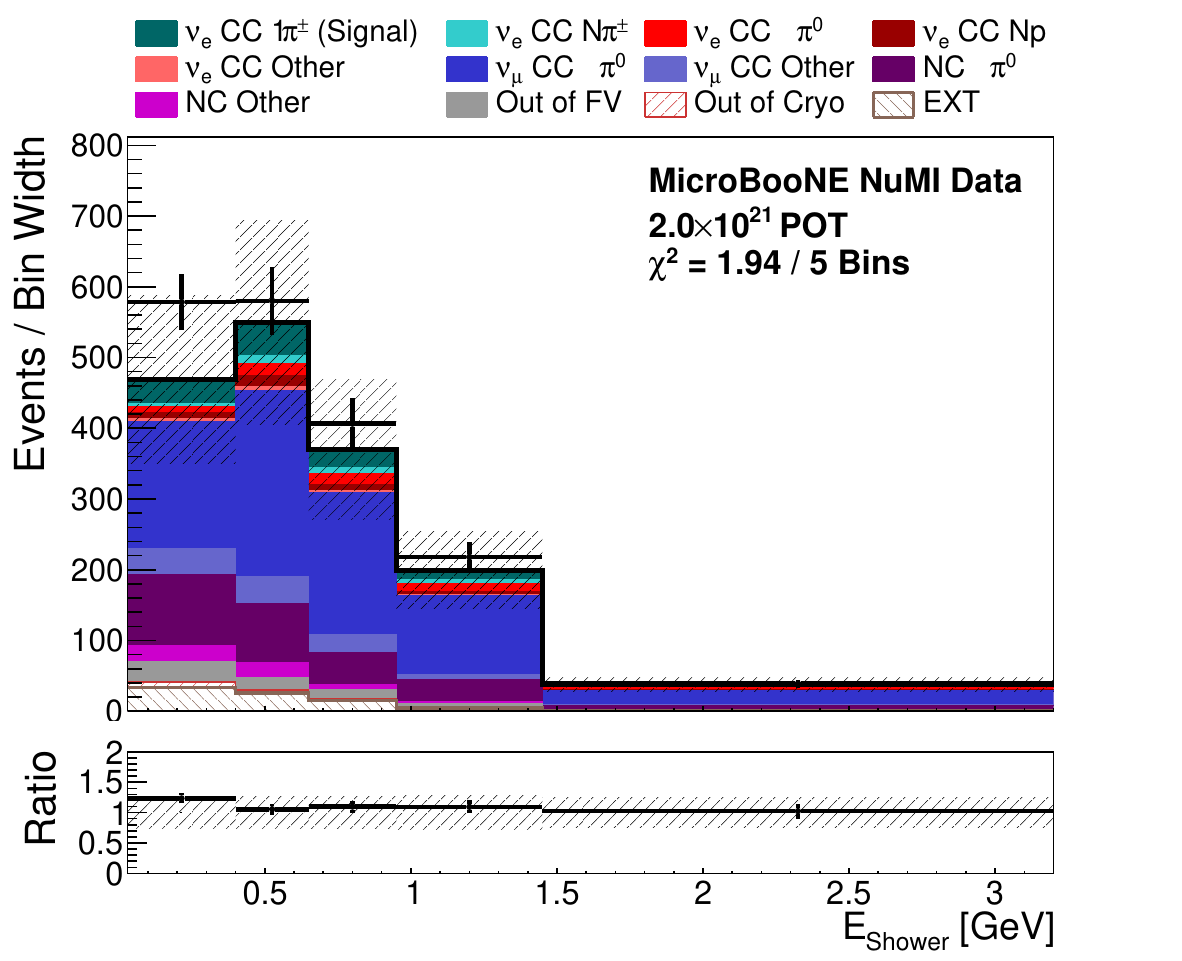}};
\draw (0.6, 1.0) node {\textbf{(a)}};
\end{tikzpicture}
\begin{tikzpicture} 
\draw (0, 0) node[inner sep=0] {\includegraphics[width=0.495\textwidth]{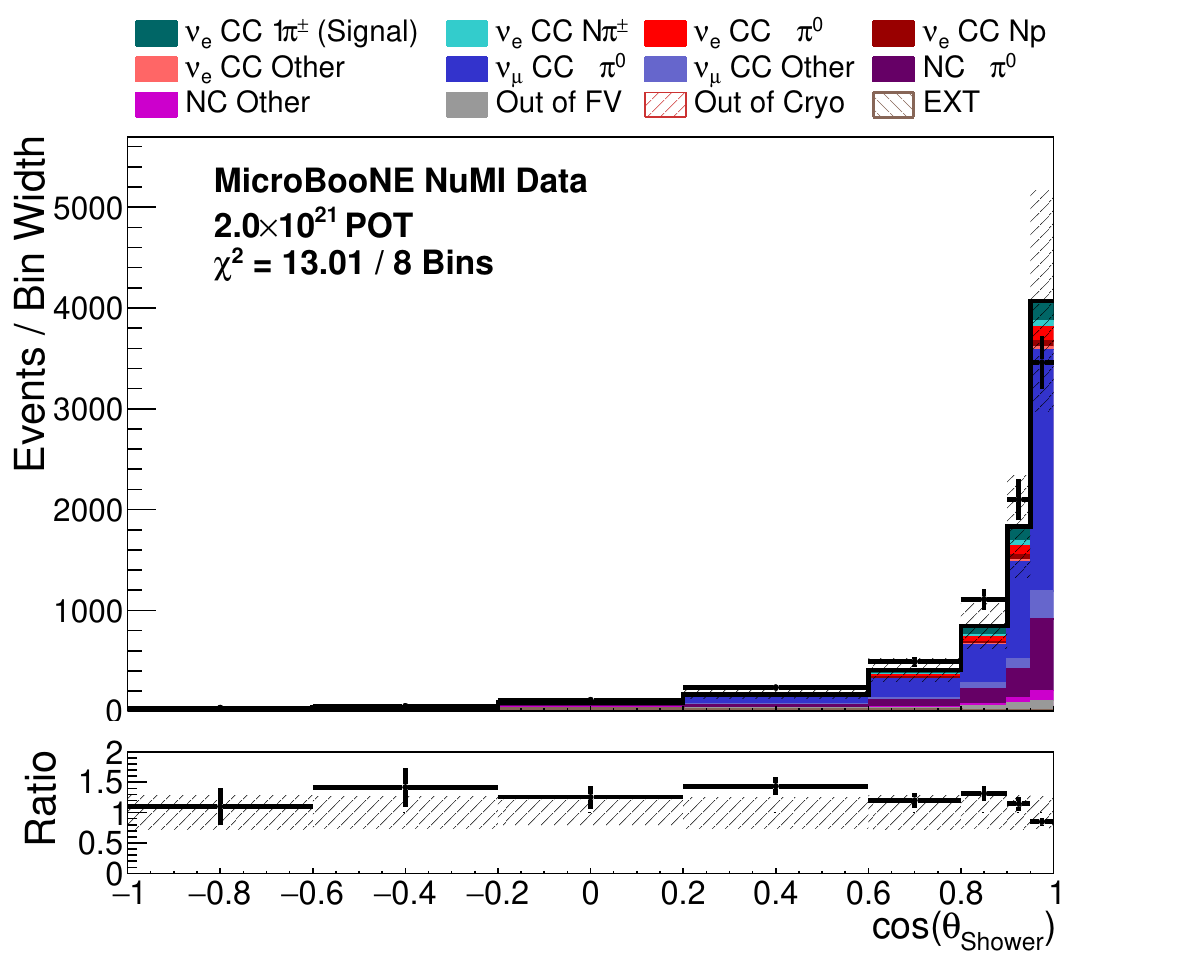}};
\draw (-2.65, 1.0) node {\textbf{(b)}};
\end{tikzpicture}
\begin{tikzpicture} 
\draw (0, 0) node[inner sep=0] {\includegraphics[width=0.495\textwidth]{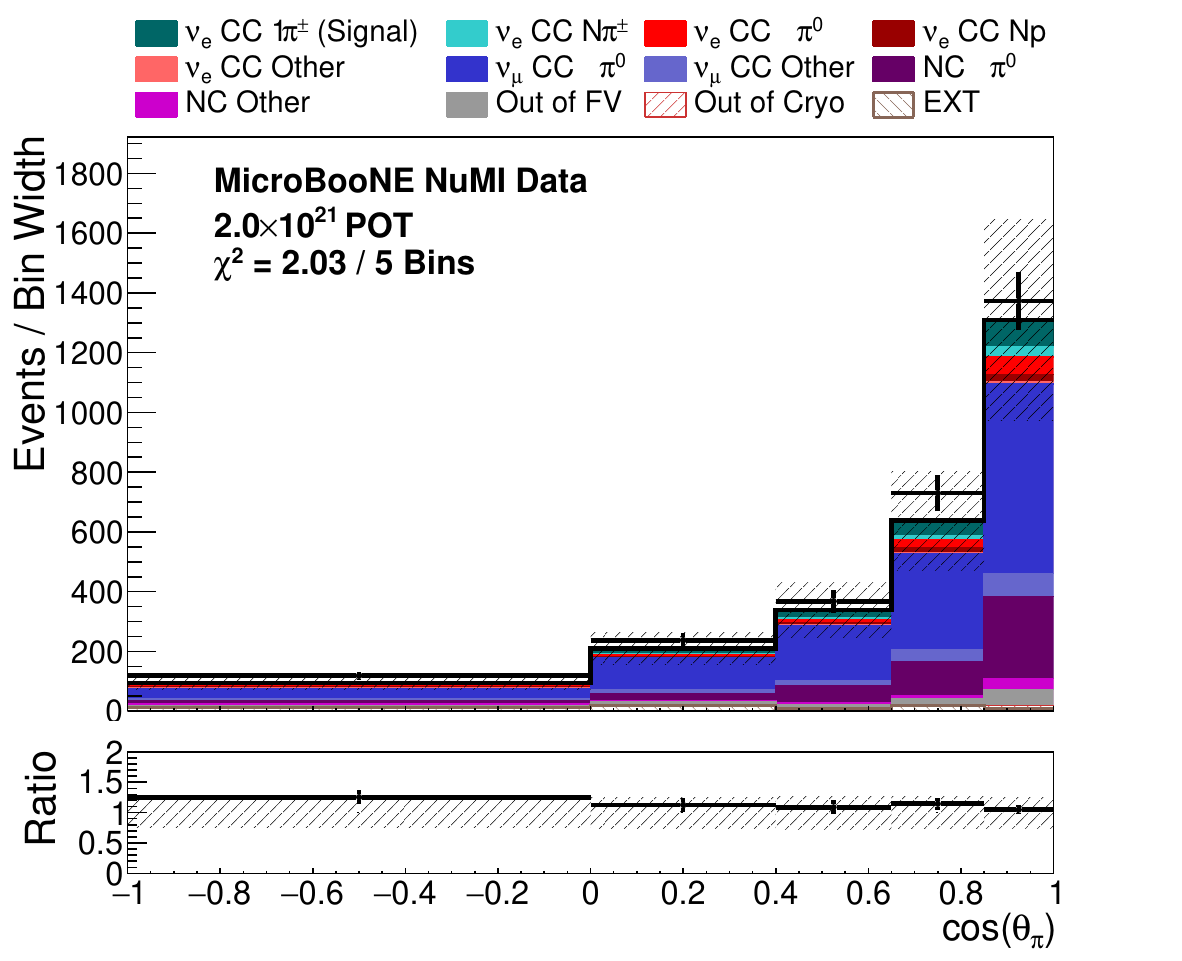}};
\draw (-2.65, 1.0) node {\textbf{(c)}};
\end{tikzpicture}
\begin{tikzpicture} 
\draw (0, 0) node[inner sep=0] {\includegraphics[width=0.495\textwidth]{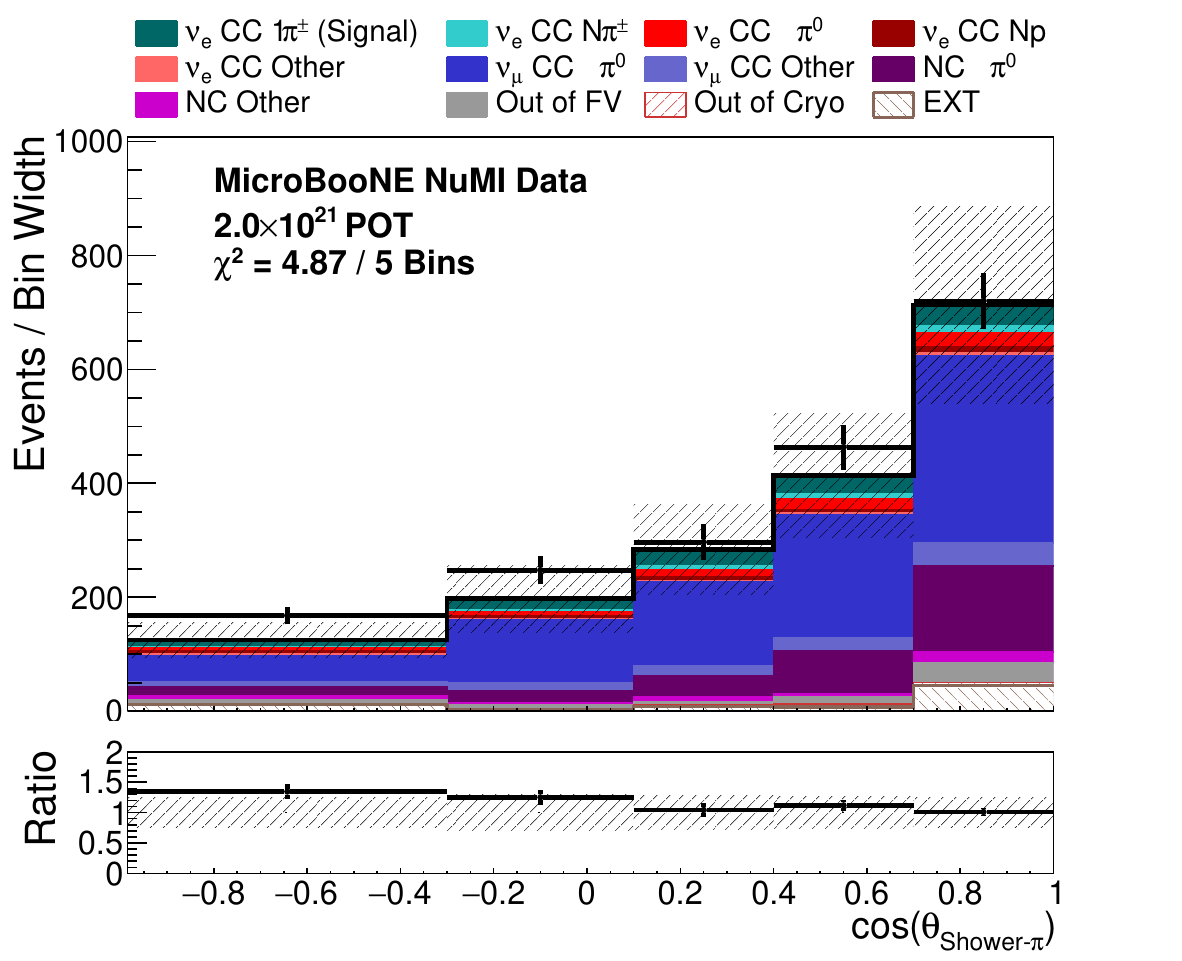}};
\draw (-2.65, 1.0) node {\textbf{(d)}};
\end{tikzpicture}
\caption{Selected event distributions for the $\pi^0$-rich sideband in variables (a) primary shower energy, (b) primary shower angle, (c) pion angle and (d) primary shower-pion opening angle. The shaded band shows the systematic and statistical uncertainty on the MC prediction and the black points show the data with statistical uncertainties.}
\label{fig:Sideband1SelectedEvents}
\end{figure*}

Figure~\ref{fig:Sideband1Total} shows the total selected events for the $\pi^0$-rich sideband and Fig.~\ref{fig:Sideband1SelectedEvents} shows the selected event distributions in variables primary shower energy, $E_{\mathrm{Shower}}$, primary shower angle, $\theta_{\mathrm{Shower}}$, pion angle, $\theta_\pi$, and primary shower-pion opening angle, $\theta_{\mathrm{Shower}-\pi}$. Each distribution is divided by bin width. Data-MC agreement within $2\sigma$ is seen across all distributions within statistical and systematic uncertainties indicating the $\pi^0$ background modeling is sufficient to proceed with cross-section extraction. 

\clearpage

\begin{figure*} [htb]
\centering
\begin{tikzpicture} 
\draw (0, 0) node[inner sep=0] {\includegraphics[width=0.495\textwidth]{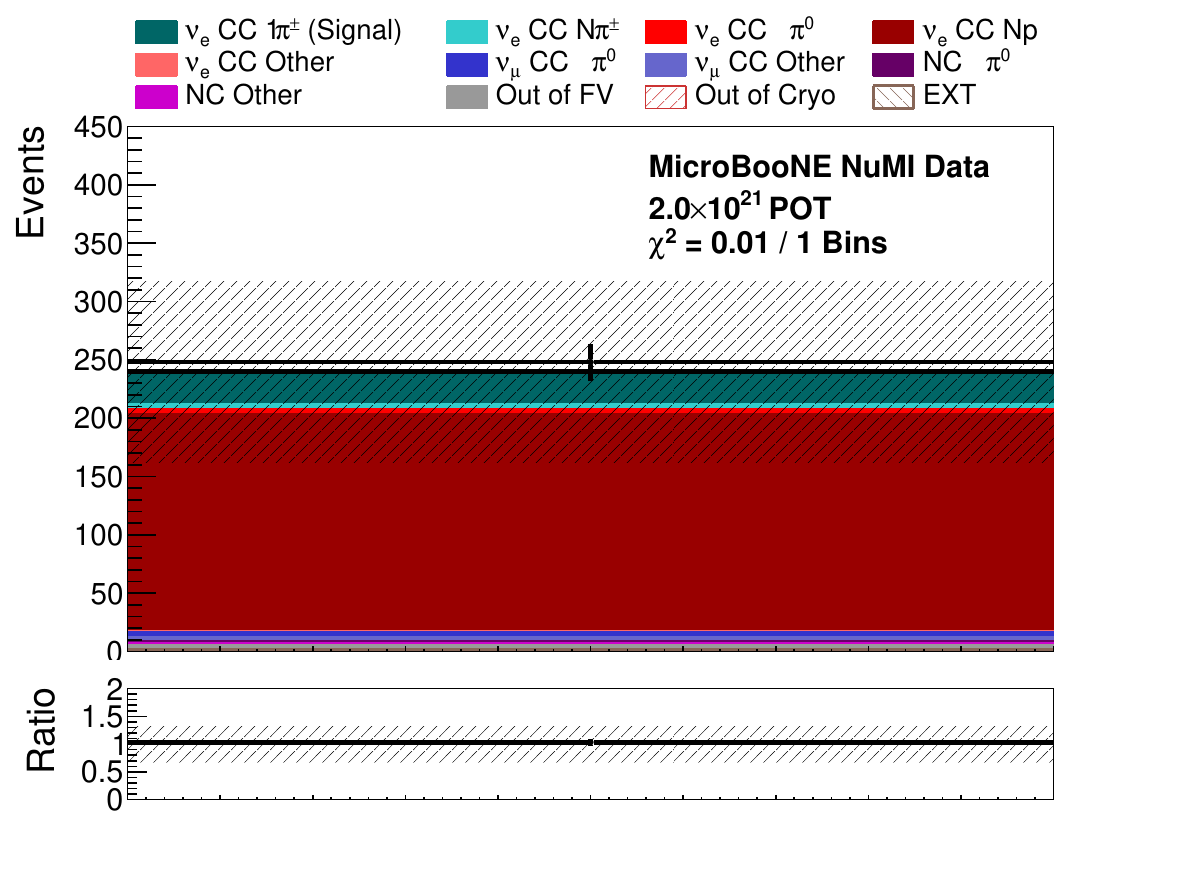}};
\end{tikzpicture}
\caption{Total selected event rate for the $\nuewithbar$ CC N$p$-rich sideband. The shaded band shows the systematic and statistical uncertainty on the MC prediction and the black points show the data with statistical uncertainties.}
\label{fig:Sideband2Total}
\end{figure*}

\begin{figure*}[htb]
\centering
\begin{tikzpicture} 
\draw (0, 0) node[inner sep=0] {\includegraphics[width=0.495\textwidth]{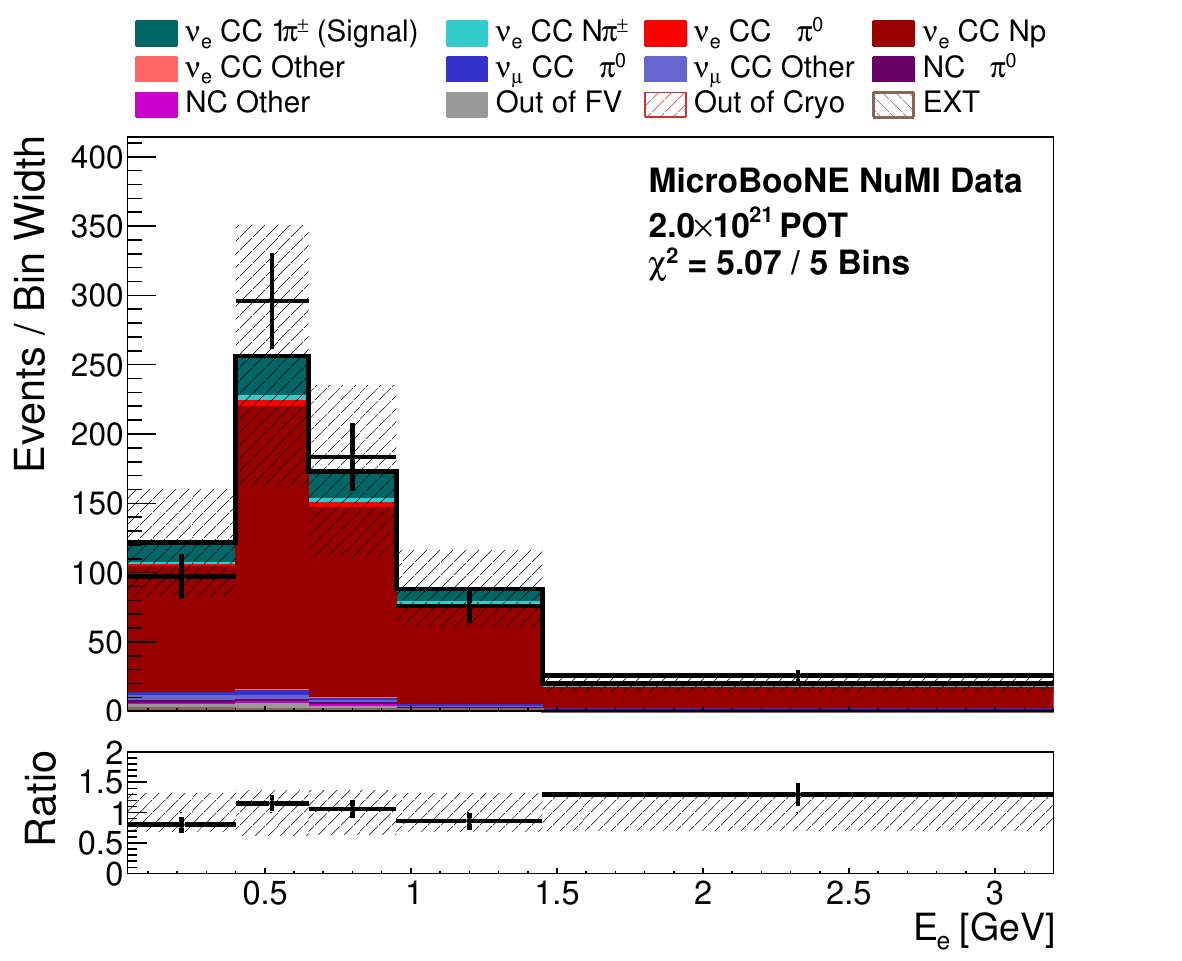}};
\draw (0.6, 1.0) node {\textbf{(a)}};
\end{tikzpicture}
\begin{tikzpicture} 
\draw (0, 0) node[inner sep=0] {\includegraphics[width=0.495\textwidth]{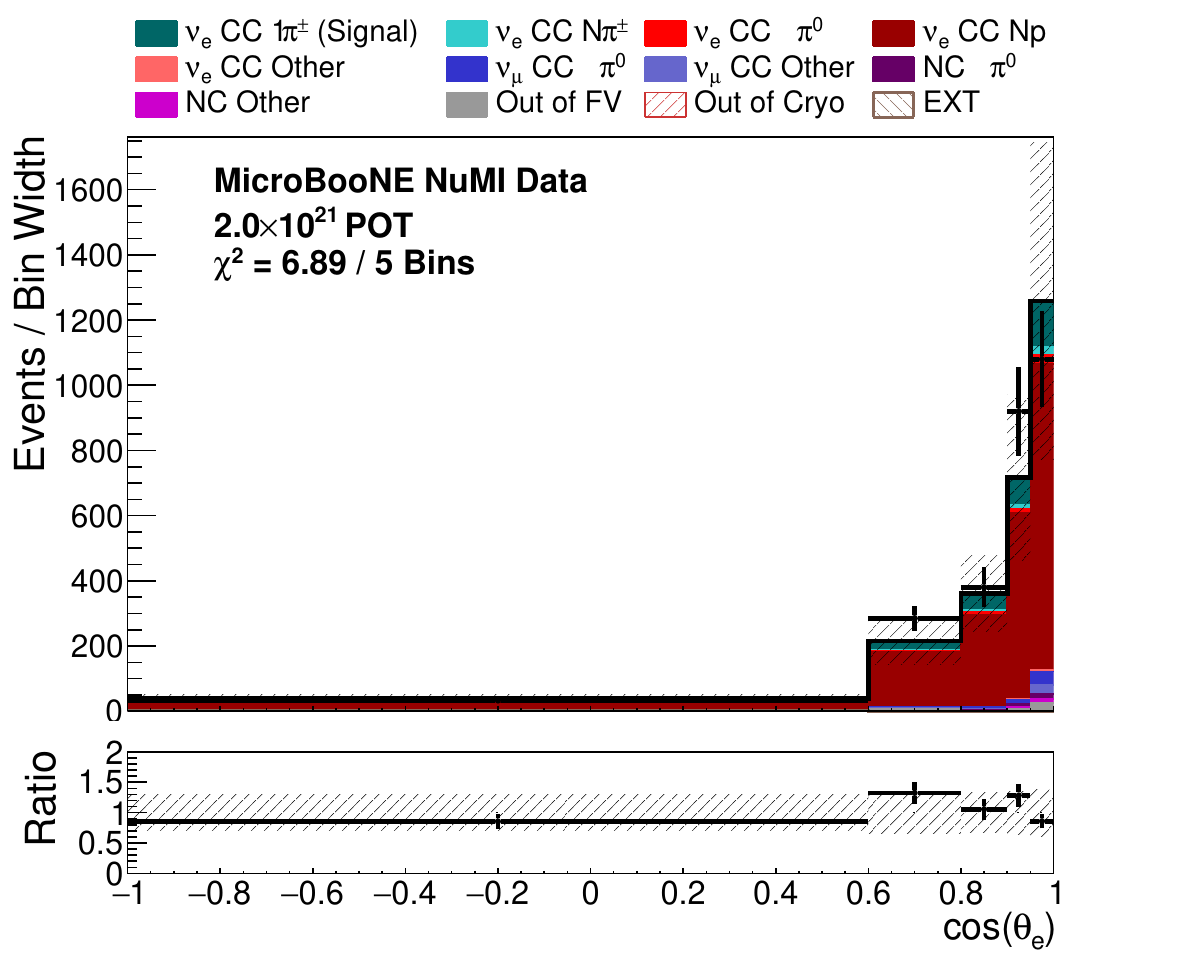}};
\draw (-2.65, 1.0) node {\textbf{(b)}};
\end{tikzpicture}
\begin{tikzpicture} 
\draw (0, 0) node[inner sep=0] {\includegraphics[width=0.495\textwidth]{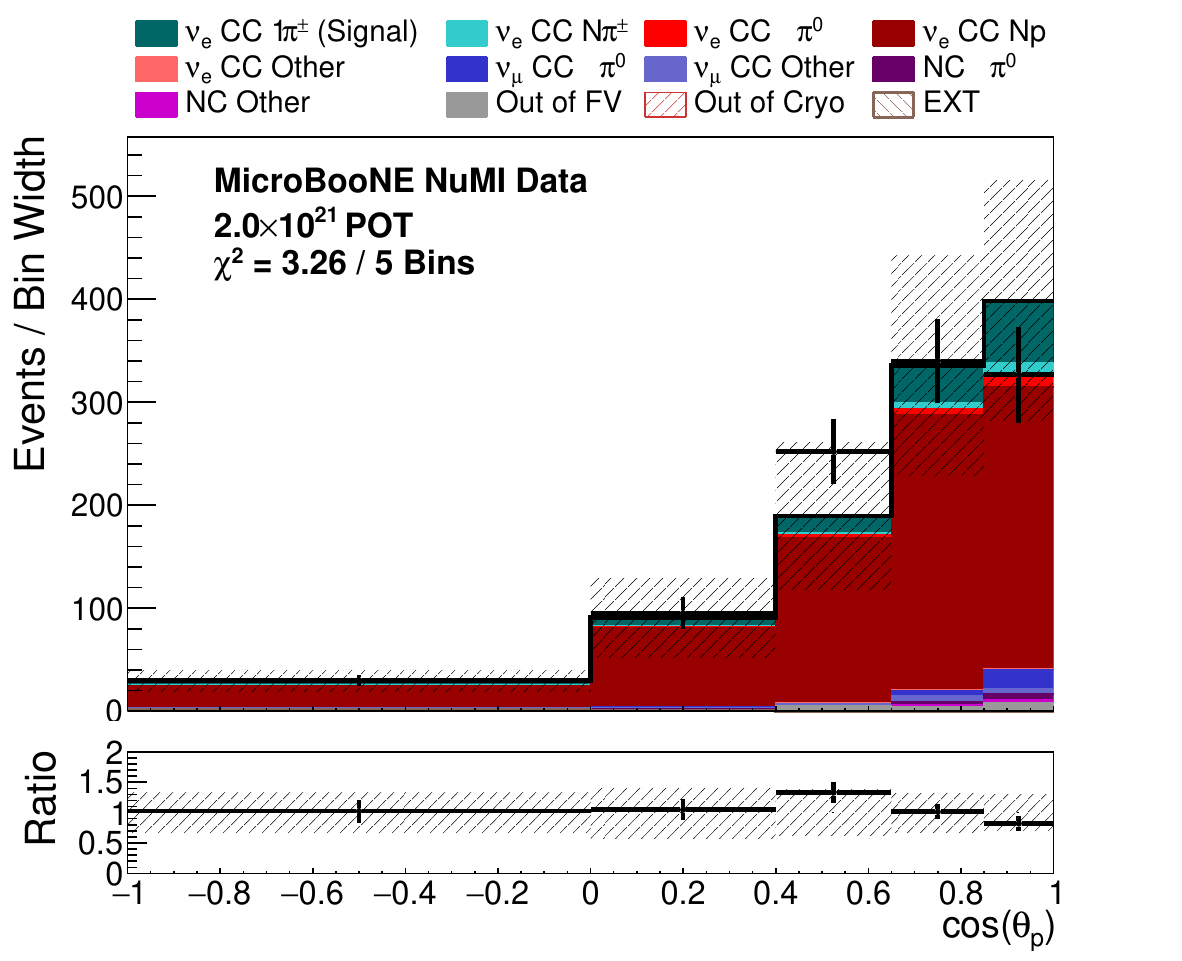}};
\draw (-2.65, 1.0) node {\textbf{(c)}};
\end{tikzpicture}
\begin{tikzpicture} 
\draw (0, 0) node[inner sep=0] {\includegraphics[width=0.495\textwidth]{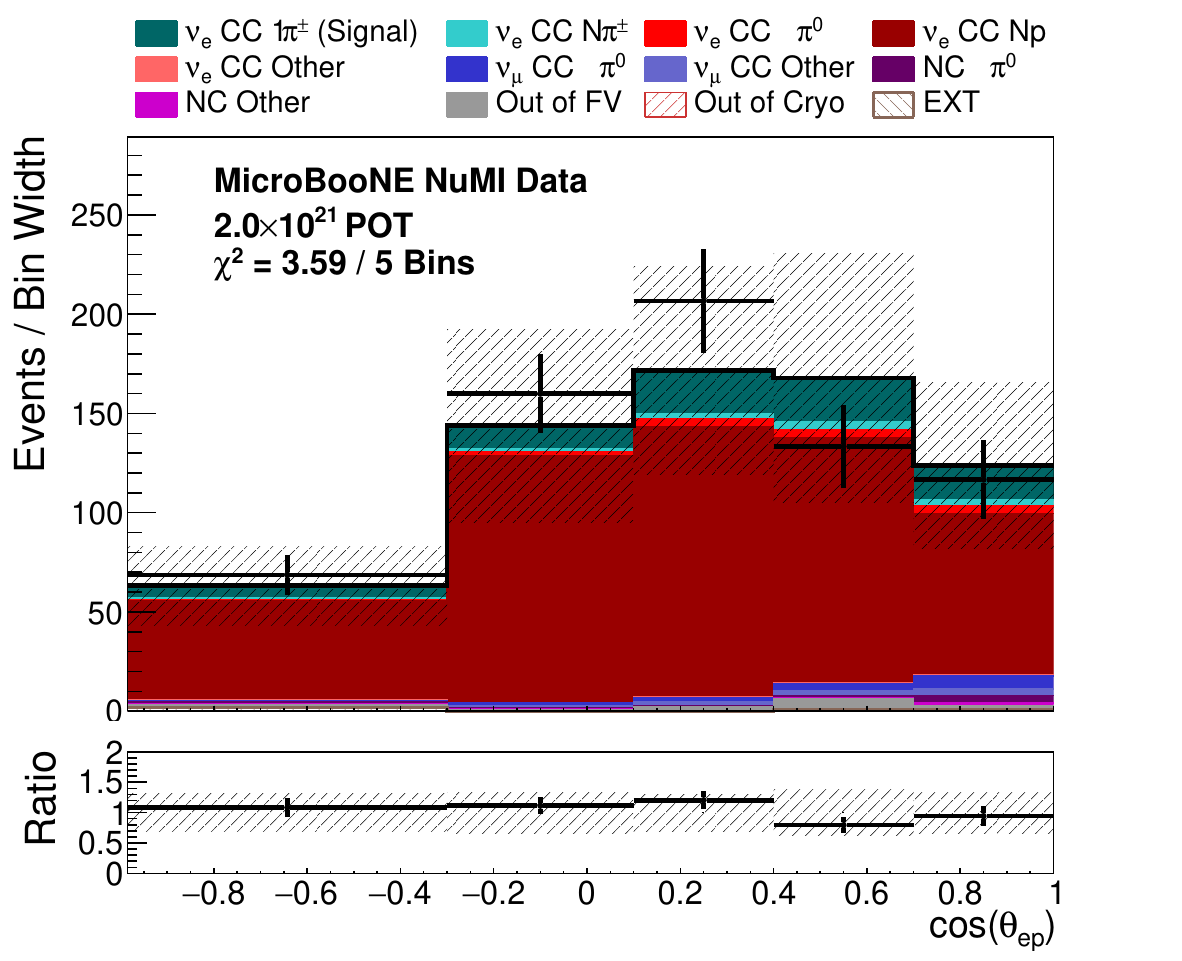}};
\draw (-2.65, 1.0) node {\textbf{(d)}};
\end{tikzpicture}
\caption{Selected event distributions for the $\nuewithbar$ CC N$p$-rich sideband in variables (a) electron energy, (b) electron angle, (c) proton angle and (d) electron-proton opening angle. The shaded band shows the systematic and statistical uncertainty on the MC prediction and the black points show the data with statistical uncertainties.}
\label{fig:Sideband2SelectedEvents}
\end{figure*}

Figure~\ref{fig:Sideband2Total} shows the total selected events for the $\nuewithbar$ CC N$p$-rich sideband and Fig.~\ref{fig:Sideband2SelectedEvents} shows the selected event distributions in variables electron energy, $E_e$, electron angle, $\theta_e$, proton angle, $\theta_p$, and electron-proton opening angle, $\theta_{ep}$.  Each distribution is divided by bin width. Data-MC agreement within $2\sigma$ is seen across all distributions within statistical and systematic uncertainties indicating the $\nuewithbar$ CC N$p$ background modeling is sufficient to proceed with cross-section extraction.

\clearpage

\section{Covariance and correlation matrices}

\begin{figure}[htb]
\centering
\includegraphics[width=0.995\textwidth]{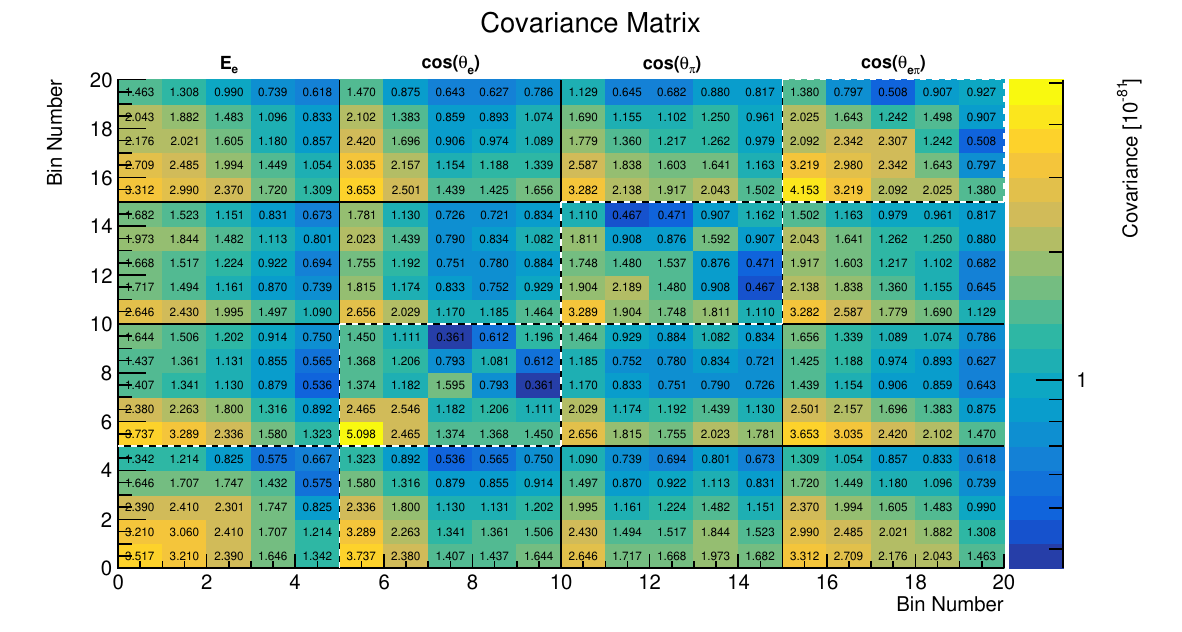}
\caption{Covariance matrix of the unfolded differential cross sections.}
\label{fig:CovarianceMatrix}
\end{figure}

\begin{figure}[htb]
\centering
\includegraphics[width=0.995\textwidth]{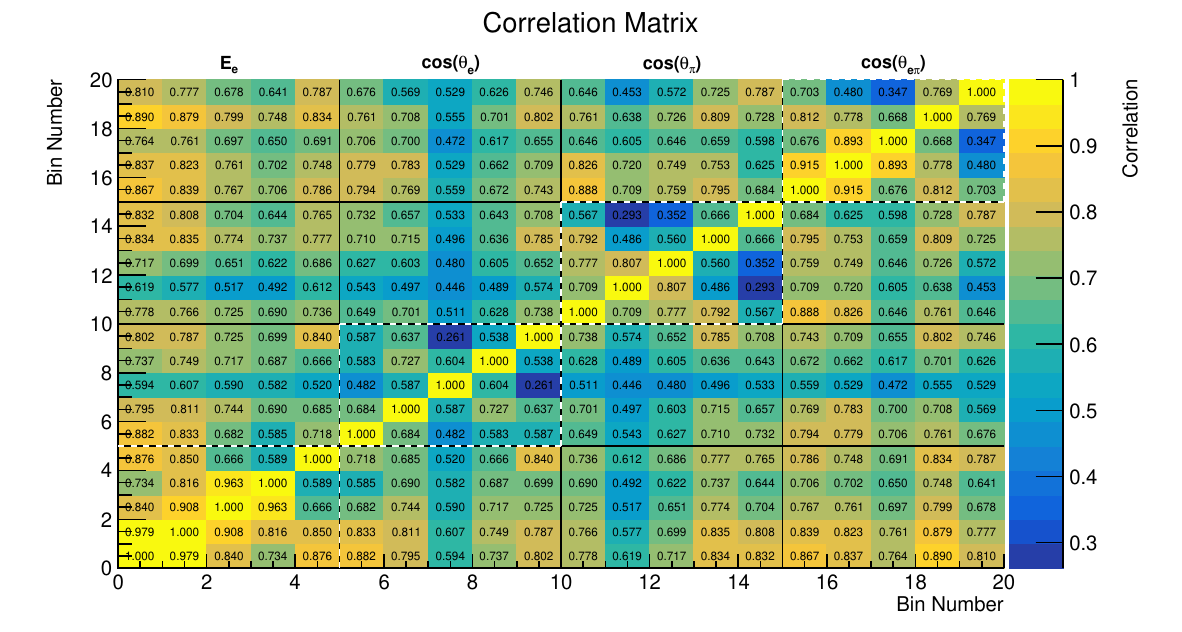}
\caption{Correlation matrix of the unfolded differential cross sections. }
\label{fig:CorrelationMatrix}
\end{figure}

Figures~\ref{fig:CovarianceMatrix} and \ref{fig:CorrelationMatrix} show the total covariance and correlation matrices of the unfolded differential cross sections, respectively. These are evaluated using a block-wise unfolding technique allowing the correlations between variables to be reported~\cite{Gardiner:2024gdy}. The blocks are highlighted by the dashed lines and correspond to the electron energy (bins 1-5), the electron angle (bins 6-10), the pion angle (bins 11-15) and the electron-pion opening angle (bins 16-20).

\clearpage

\section{Regularization Matrix}

\begin{figure}[htb]
\centering
\includegraphics[width=0.995\textwidth]{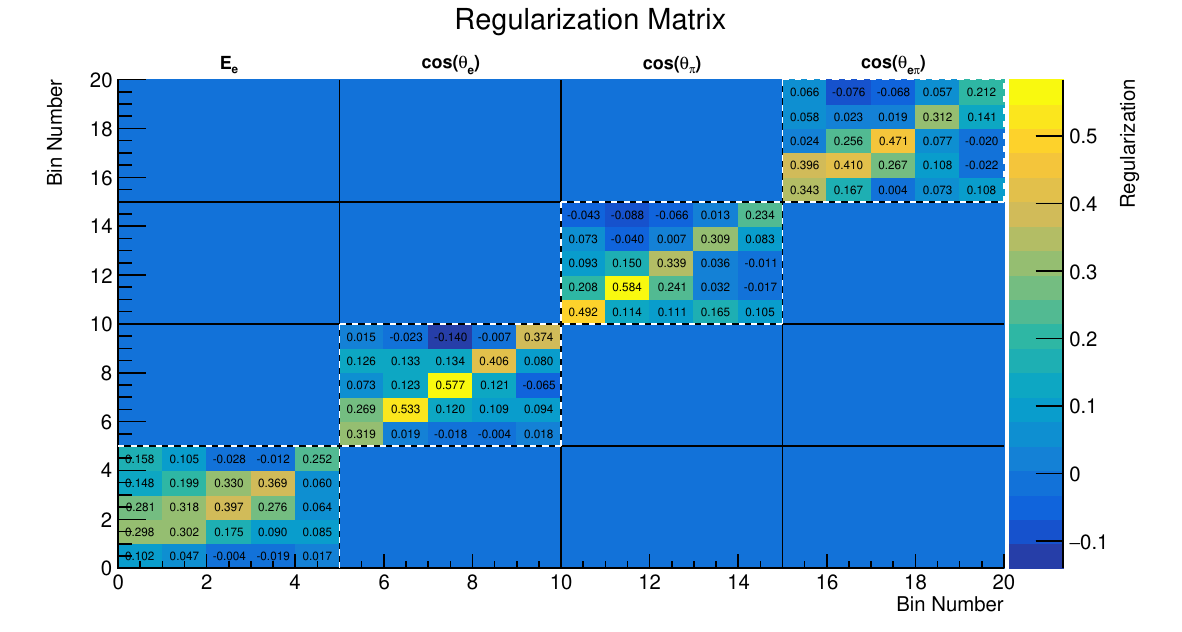}
\caption{Regularization matrix encoding the regularized truth space of the unfolded differential cross sections.}
\label{fig:RegularizationMatrix}
\end{figure}

Figure~\ref{fig:RegularizationMatrix} shows the the regularization matrix that encodes the regularized truth space of the unfolded differential cross sections. This is reported with the same block structure as Figures~\ref{fig:CovarianceMatrix} and \ref{fig:CorrelationMatrix}. Since the unfolding is performed separately for each block there is no regularization between blocks~\cite{Gardiner:2024gdy}. Therefore, the off-block-diagonal terms in the regularization matrix are zero and not shown.  

\clearpage

\section{Data release}

Tables~\ref{tab:UnfoldedResultsElectronEnergy}, \ref{tab:UnfoldedResultsElectronAngle}, \ref{tab:UnfoldedResultsPionAngle} and \ref{tab:UnfoldedResultsOpeningAngle} summarize the unfolded results in regularized truth space. To compare these results with a theoretical prediction, the prediction must first be transformed into this regularized space by applying the regularization matrix. Additionally, the prediction should be divided by bin width. The data release with the data results, the covariance matrix, and the regularization matrix is included in the DataRelease.root file. Instructions on how to use the data release and the description of the binning scheme are included in the README file.

\begin{table}[h]
\begin{tabular}{|c|c|c|c|c|c|}
\hline
\multicolumn{6}{|c|}{\textbf{Electron Energy, $E_e$}} \\
\hline
Block Bin \# & Global Bin \# & \makecell{Low Edge \\ \([ \text{GeV} ]\)} & \makecell{High Edge \\ \([ \text{GeV} ]\)} & \makecell{Cross Section \\ \([\)$\num{e-39}$cm$^2$/GeV/nucleon\(]\)} & \makecell{Uncertainty \\ \([\)$\num{e-39}$cm$^2$/GeV/nucleon\(]\)} \\
\hline
1 & 1 & 0.03 & 0.40 & 0.487 & 0.160 \\
2 & 2 & 0.40 & 0.65 & 0.680 & 0.221 \\
3 & 3 & 0.65 & 0.95 & 0.472 & 0.160 \\
4 & 4 & 0.95 & 1.45 & 0.211 & 0.076 \\
5 & 5 & 1.45 & 3.20 & 0.041 & 0.014 \\
\hline
\end{tabular}
\caption{Electron energy, $E_e$, unfolded cross section. The highest energy bin also serves as an overflow bin.}
\label{tab:UnfoldedResultsElectronEnergy}
\end{table}

\begin{table}[h]
\begin{tabular}{|c|c|c|c|c|c|}
\hline
\multicolumn{6}{|c|}{\textbf{Electron Angle, cos($\theta_e$})} \\
\hline
Block Bin \# & Global Bin \# & Low Edge & High Edge & \makecell{Cross Section \\ \([\)$\num{e-39}$cm$^2$/nucleon\(]\)} & \makecell{Uncertainty \\ \([\)$\num{e-39}$cm$^2$/nucleon\(]\)} \\
\hline
1 & 6 & -1.00 & 0.60 & 0.129 & 0.045 \\
2 & 7 & 0.60 & 0.80 & 0.650 & 0.252 \\
3 & 8 & 0.80 & 0.90 & 0.475 & 0.399 \\
4 & 9 & 0.90 & 0.95 & 1.897 & 0.658 \\
5 & 10 & 0.95 & 1.00 & 1.886 & 0.692 \\
\hline
\end{tabular}
\caption{Electron angle, $\theta_e$, unfolded cross section.}
\label{tab:UnfoldedResultsElectronAngle}
\end{table}

\begin{table}[h]
\begin{tabular}{|c|c|c|c|c|c|}
\hline
\multicolumn{6}{|c|}{\textbf{Pion Angle, cos($\theta_\pi$})} \\
\hline
Block Bin \# & Global Bin \# & Low Edge & High Edge & \makecell{Cross Section \\ \([\)$\num{e-39}$cm$^2$/nucleon\(]\)} & \makecell{Uncertainty \\ \([\)$\num{e-39}$cm$^2$/nucleon\(]\)} \\
\hline
1 & 11 & -1.00 & 0.00 & 0.119 & 0.057 \\
2 & 12 & 0.00 & 0.40 & 0.178 & 0.117 \\
3 & 13 & 0.40 & 0.65 & 0.365 & 0.157 \\
4 & 14 & 0.65 & 0.85 & 0.572 & 0.200 \\
5 & 15 & 0.85 & 1.00 & 0.583 & 0.227 \\
\hline
\end{tabular}
\caption{Pion angle, $\theta_\pi$, unfolded cross section.}
\label{tab:UnfoldedResultsPionAngle}
\end{table}

\begin{table}[h]
\begin{tabular}{|c|c|c|c|c|c|}
\hline
\multicolumn{6}{|c|}{\textbf{Electron-pion Opening Angle, cos($\theta_{e\pi}$})} \\
\hline
Block Bin \# & Global Bin \# & Low Edge & High Edge & \makecell{Cross Section \\ \([\)$\num{e-39}$cm$^2$/nucleon\(]\)} & \makecell{Uncertainty \\ \([\)$\num{e-39}$cm$^2$/nucleon\(]\)} \\
\hline
1 & 16 & -1.00 & 0.00 & 0.280 & 0.094 \\
2 & 17 & 0.00 & 0.40 & 0.273 & 0.136 \\
3 & 18 & 0.40 & 0.65 & 0.199 & 0.160 \\
4 & 19 & 0.65 & 0.85 & 0.322 & 0.129 \\
5 & 20 & 0.85 & 1.00 & 0.373 & 0.101 \\
\hline
\end{tabular}
\caption{Electron-pion opening angle, $\theta_{e\pi}$, unfolded cross section.}
\label{tab:UnfoldedResultsOpeningAngle}
\end{table}

\bibliography{supplemental.bib}